\documentstyle[preprint,aps,epsfig]{revtex}
\begin{document}
\draft
\tighten

\title{Four neutrino oscillation analysis of
the Superkamiokande atmospheric neutrino data}

\author{Osamu Yasuda\footnote{Email: yasuda@phys.metro-u.ac.jp}}
\address{Department of Physics, Tokyo Metropolitan University \\
Minami-Osawa, Hachioji, Tokyo 192-0397, Japan}

\date{June, 2000}
\preprint{
\parbox{5cm}{
TMUP-HEL-0010\\
hep-ph/0006319\\
}}

\maketitle

\begin{abstract}
Superkamiokande atmospheric neutrino data (contained and
upward going through $\mu$ events) for 990 days are analyzed in
the framework of four neutrinos (three active and one sterile neutrinos)
without imposing constraints of Big Bang Nucleosynthesis.  It
is shown that the wide range of the oscillation parameters is allowed
at 90\% confidence level ($0.1\lesssim |U_{s1}|^2+|U_{s2}|^2\le 1$), 
where the best fit point has
some contribution of $\nu_\mu\leftrightarrow\nu_s$ and
contribution of $\Delta m^2_{\mbox{\rm\scriptsize LSND}}$.
The case of pure $\nu_\mu\leftrightarrow\nu_s$ oscillation is excluded
at 99.7\%CL (3.0$\sigma$) which is consistent with
the recent analysis by the Superkamiokande group.
Combining this result with the analysis by Giunti, Gonzarez-Garcia
and Pe\~na-Garay, it is found that the Large Mixing Angle
and Vacuum Oscillation solutions of the solar neutrino problem
are also allowed.

\end{abstract}
\vskip 0.1cm

\section{Introduction}

There have been a few experiments which suggest neutrino oscillations:
the solar neutrino deficit
\cite{homestake,Kamsol,SKsol,sage,gallex,solar}
the atmospheric neutrino anomaly
\cite{Kamatm,Kamup,IMB,skatm,hasegawa,toshito,SKup,soudan2,macro,atmth}
and the LSND data \cite{lsnd}.  If we assume that all these three are
caused by neutrino oscillations then we need at least four species of
neutrinos and schemes with sterile neutrinos have been studied by many
people \cite{sterile,goswami,oy,bgg,bggs,fvy,gnpv,nus}.
By generalizing the discussion
on Big Bang Nucleosynthesis (BBN) from the two neutrino scheme
\cite{bbn} to four neutrino case, it has been shown \cite{oy}
\footnote{The result in \cite{oy} was refined later in \cite{bggs}
with more careful treatment.}
that the neutrino mixing angles are strongly constrained not only by
the reactor data \cite{goswami} but also by BBN if one demands that
the number $N_\nu$ of effective neutrinos be less than four.
In this case the $4\times 4$ MNS
matrix \cite{mns} splits approximately into two $2\times 2$ block diagonal
matrices, and the solar neutrino deficit is explained by
$\nu_e\leftrightarrow\nu_s$ oscillations with the Small Mixing Angle
(SMA) MSW solution \cite{msw} and the atmospheric neutrino anomaly is
accounted for by $\nu_\mu\leftrightarrow\nu_\tau$.  On the other hand,
some people have given conservative estimate for $N_\nu$ \cite{he4}
and if their estimate is correct then we no longer have strong
constraints on the mixing angles of sterile neutrinos. Recently
Giunti, Gonzalez-Garcia and Pe\~na-Garay \cite{ggp} have analyzed the
solar neutrino data in the four neutrino scheme without BBN
constraints.  They have shown that the scheme is reduced to the two
neutrino framework in which only one free parameter
$c_s\equiv|U_{s1}|^2+|U_{s2}|^2$ appears in the analysis.  Their
conclusion is that the SMA MSW solution exists for the entire region
of $0\le c_s
\le 1$, while the Large Mixing Angle (LMA) and Vacuum Oscillation (VO)
solutions survive only for $0\le c_s \lesssim 0.2$ and $0\le c_s \lesssim
0.4$, respectively.
In this paper the Superkamiokande atmospheric neutrino data
(contained and upward going through $\mu$ events) are analyzed
in the same scheme as in \cite{ggp}, i.e., in the four neutrino scheme
with all the constraints of accelerators and reactors and without
BBN constraints.
Zenith angle dependence of atmospheric neutrinos has
been analyzed by many theorists \cite{atmanalysis,fvy,gnpv,nus} as well as
by experimentalists \cite{Kamatm,Kamup,skatm,SKup,soudan2,macro}.

\section{Four neutrino scheme}
Here we adopt the notation in \cite{oy} for the $4\times 4$ MNS matrix:
\begin{eqnarray}
&{\ }&\left( \begin{array}{c} \nu_e  \\ \nu_{\mu} \\ 
\nu_{\tau}\\\nu_s \end{array} \right)
=U\left( \begin{array}{c} \nu_1  \\ \nu_2 \\ 
\nu_3\\\nu_4 \end{array} \right),\nonumber\\
U&\equiv&\left(
\begin{array}{cccc}
U_{e1} & U_{e2} &  U_{e3} &  U_{e4}\\
U_{\mu 1} & U_{\mu 2} & U_{\mu 3} & U_{\mu 4}\\
U_{\tau 1} & U_{\tau 2} & U_{\tau 3} & U_{\tau 4}\\
U_{s1} & U_{s2} &  U_{s3} &  U_{s4}
\end{array}\right) \nonumber\\
&\equiv&R_{34}({\pi \over 2}-\theta_{34})R_{24}(\theta_{24})
R_{23}({\pi \over 2})
e^{2i\delta_1\lambda_3}R_{23}(\theta_{23})
e^{-2i\delta_1\lambda_3}
e^{\sqrt{6}i\delta_3\lambda_{15}/2}\nonumber\\
&\times&R_{14}(\theta_{14})e^{-\sqrt{6}i\delta_3\lambda_{15}/2}
e^{2i\delta_2\lambda_8/\sqrt{3}}R_{13}(\theta_{13})
e^{-2i\delta_2\lambda_8/\sqrt{3}}
R_{12}(\theta_{12}),
\label{eqn:u}
\end{eqnarray}
where $c_{ij}\equiv\cos\theta_{ij}$, $s_{ij}\equiv\sin\theta_{ij}$ and
\begin{eqnarray}
R_{jk}(\theta)\equiv \exp\left(iT_{jk}\theta\right),
\end{eqnarray}
is a $4\times4$ orthogonal matrix with
\begin{eqnarray}
\left(T_{jk}\right)_{\ell m}=i\left(\delta_{j\ell}\delta_{km}
-\delta_{jm}\delta_{k\ell}\right),
\end{eqnarray}
and $2\lambda_3\equiv{\rm diag}(1,-1,0,0)$,
$2\sqrt{3}\lambda_8\equiv{\rm diag}(1,1,-2,0)$,
$2\sqrt{6}\lambda_{15}\equiv{\rm diag}(1,1,1,-3)$
are diagonal elements of the $SU(4)$ generators.

We can assume without loss of generality that
\begin{eqnarray}
m_1^2<m_2^2<m_3^2<m_4^2.
\label{eqn:inequality}
\end{eqnarray}
Three mass scales $\Delta m_\odot^2\sim{\cal O}(10^{-5}{\rm eV}^2)$ or
${\cal O}(10^{-10}{\rm eV}^2)$, $\Delta m_{\rm atm}^2\sim{\cal O}
(10^{-2}{\rm eV}^2)$, $\Delta m_{\rm LSND}^2\sim{\cal O}(1{\rm eV}^2)$
are necessary to explain the suppression of the ${}^7$Be solar
neutrinos \cite{solar}, the zenith angle dependence of the
atmospheric neutrino data \cite{Kamatm,skatm}, and the LSND data
\cite{lsnd}, so we assume that three independent mass squared
differences which are obtained from (\ref{eqn:inequality}) are $\Delta
m_\odot^2$, $\Delta m_{\rm atm}^2$, $\Delta m_{\rm LSND}^2$.  It has
been known \cite{oy,bgg} that schemes with three
degenerate masses and one distinct massive state do not work
to account for all the three neutrino anomalies, but
schemes with two degenerate massive states
($m_{1}^2 \simeq m_{2}^2 \ll m_{3}^2 \simeq m_{4}^2$,
where (a) $(\Delta m_{21}^2, \Delta m_{43}^2)=
(\Delta m_\odot^2,\Delta m_{\rm atm}^2)$ or (b)
$(\Delta m_{21}^2, \Delta m_{43}^2)=
(\Delta m_{\rm atm}^2,\Delta m_\odot^2)$)
do.  As far as the analyses of atmospheric
neutrinos and solar neutrinos are concerned,
the two cases (a) and (b) can be treated in the same manner,
so we assume $\Delta m_{21}^2=\Delta m_\odot^2$,
$\Delta m_{43}^2=\Delta m_{\rm atm}^2$ in the following.

For the range of the $\Delta m^{2}$
suggested by the LSND data, which is given by 0.2
eV$^2~\lesssim \Delta m^{2}_{\mbox{\rm{\scriptsize LSND}}}
\lesssim$ 2 eV$^2$ when
combined with the data of Bugey \cite{bugey} and E776 \cite{e776},
the constraint by the Bugey data is very stringent and
\begin{eqnarray}
|U_{e3}|^2+|U_{e4}|^2\lesssim 10^{-2},
\end{eqnarray}
has to be satisfied \cite{goswami,oy,bgg}, therefore we put
$U_{e3}=U_{e4}=0$ for simplicity in the following discussions.
Also in the analysis of atmospheric neutrinos,
$|\Delta m_\odot^2L/4E|\ll 1$ is satisfied for typical values of
the neutrino path length $L$ and the neutrino energy $E$ for
atmospheric neutrinos, so we assume $\Delta m_\odot^2= 0$ for simplicity
throughout this paper.

In the following analysis we will consider the situation where
non-negligible contributions from the largest mass squared difference
$\Delta m^{2}_{\mbox{\rm{\scriptsize LSND}}}$ appear in the
oscillation probability $P(\nu_\mu\rightarrow\nu_\mu)$.
To avoid contradiction with the negative result
of the CDHSW disappearing experiment on
$\nu_\mu\rightarrow\nu_\mu$ \cite{cdhsw}, we will
take $\Delta m^{2}_{\mbox{\rm{\scriptsize LSND}}}$=0.3eV$^2$
as a reference value.

Having assumed $U_{e3}=U_{e4}=0$ and $\Delta m_{21}^2= 0$,
we have only mixings among $\nu_\mu$, $\nu_\tau$, $\nu_s$
in the analysis of atmospheric neutrinos, and the
Schr\"odinger equation we have to consider is
\begin{eqnarray}
i {d \over dx} \left( \begin{array}{c} \nu_{\mu}(x) \\ 
\nu_{\tau}(x) \\ \nu_s (x)
\end{array} \right) = 
\left[ \widetilde U {\rm diag} \left(-\Delta E_{32} ,0,\Delta E_{43}
\right) \widetilde U^{-1}
+{\rm diag} \left(0,0,A(x) \right) \right]
\left( \begin{array}{c}
\nu_{\mu}(x) \\ \nu_{\tau}(x) \\ \nu_s (x)
\end{array} \right),
\label{eqn:sch}
\end{eqnarray}
where
\begin{eqnarray}
\widetilde U&\equiv&\left(
\begin{array}{ccc}
 U_{\mu 2} & U_{\mu 3}&U_{\mu 4} \\
 U_{\tau 2} & U_{\tau 3}&U_{\tau 4} \\
 U_{s2} &  U_{s3}&U_{s4} 
\end{array}\right)\nonumber\\
&=&\left(
\begin{array}{rrr}
-c_{24}s_{23}e^{i\delta_1}
& c_{23}c_{24}
& s_{24}\\
-c_{23}s_{34}+c_{34}s_{23}s_{24}e^{i\delta_1}
& -c_{23}c_{34}s_{24}-s_{23}s_{34}e^{-i\delta_1}
& c_{24}c_{34}\\
c_{23}c_{34}+s_{23}s_{24}s_{34}e^{i\delta_1}
& -c_{23}s_{24}s_{34}+c_{34}s_{23}e^{-i\delta_1}
& c_{24}s_{34}
\end{array}
\right),
\label{eqn:mns3}
\end{eqnarray}
$\Delta E_{ij}\equiv\Delta m_{ij}^2/2E$ 
and
$A(x)\equiv G_F N_n(x)/\sqrt{2}$ stands for the effect due to the
neutral current interactions between $\nu_\mu$, $\nu_\tau$ and matter in the
Earth \cite{nr} after adding the unit matrix
${\rm diag} \left(A(x),A(x),A(x) \right)$ to the right hand side of
(\ref{eqn:sch}).  Since $\nu_e$ does not oscillate with any other
neutrinos, the only oscillation probability which is required in the analysis
of atmospheric neutrinos is $P(\nu_\mu\rightarrow\nu_\mu)$.
In the present case, we have mass hierarchy $|\Delta E_{32}|\gg
|\Delta E_{43}|, |A(x)|$ and to the leading order in $|\Delta
E_{43}|/|\Delta E_{32}|$ and $|A(x)|/|\Delta E_{32}|$ we can obtain
the analytical expression for the oscillation probability
$P(\nu_\mu\rightarrow\nu_\mu)$ in adiabatic approximation, i.e.,
assuming that the derivative $|dA(x)/dx|$ is not large compared to
$\Delta E_{jk}$.  As we will see in the appendix,
with such approximation we can show that
the oscillation probability $P(\nu_\mu\rightarrow\nu_\mu)$
is invariant under $\theta_{34}\rightarrow-\theta_{34}$ and
$\delta_1\rightarrow\pi-\delta_1$ to the leading order in $|\Delta
E_{43}|/|\Delta E_{32}|$ and $|A(x)|/|\Delta E_{32}|$.  So
we will study our scheme for the range
$0\le\theta_{24}\le\pi/2$, $0\le\theta_{23}\le\pi/2$,
$-\pi/2\le\theta_{34}\le\pi/2$, $0\le\delta_1\le\pi/2$.
Note that all the three mixing angles would lie in the first
quadrant \cite{pdg} if there were no matter effects.

\section{Analysis of the atmospheric neutrino data}
We calculate the disappearance probability
$P(\nu_\mu\rightarrow\nu_\mu)$ by solving (\ref{eqn:sch})
numerically, and evaluate the number of events:
\begin{eqnarray}
\displaystyle
N(\mu)&=&(1-r_\mu) N_0(\mu)+r_\mu N_0(e),\nonumber\\
N(e)&=& (1-r_e)N_0(e)+r_e N_0(\mu),\nonumber\\
N_0(\mu)
&=& n_T
\int dE
\int dq
\int d\cos\Theta
\int d\cos\theta
\int d\varphi~\epsilon_\mu (q)\nonumber\\
&\times&
{d^3F_\mu (E,\theta) \over dE~d\cos\theta~d\varphi}
\cdot{ d^2\sigma_\mu (E,q) \over dq~d\cos\psi }
\cdot{d\cos\psi \over d\cos\theta}
{\ }P(\nu_\mu\rightarrow\nu_\mu\,; E, \theta),\nonumber\\
N_0(e)
&=& n_T
\int dE
\int dq
\int d\cos\Theta
\int d\cos\theta
\int d\varphi~\epsilon_e (q)\nonumber\\
&\times&
{d^3F_e (E,\theta) \over dE~d\cos\theta~d\varphi}
\cdot{ d^2\sigma_e (E,q) \over dq~d\cos\psi }
\cdot{d\cos\psi \over d\cos\theta},
\end{eqnarray}
where $d^3F_\alpha /dEd\cos\theta d\varphi$ ($\alpha=\mu, e$) is the
flux of atmospheric neutrinos $\nu_\alpha$ ($\alpha=\mu, e$) with
energy $E$ from the zenith angle $\theta$ \cite{hkkm}, $n_T$ is the
effective number of target nucleons, $\epsilon_\alpha (q)$ is the
detection efficiency function for charged leptons $\alpha$
($\alpha=\mu, e$) \cite{kaneyuki}, $d\sigma_\alpha/dqd\cos\psi$
($\alpha=\mu, e$) is the differential cross section of the interaction
$\nu_\alpha N \rightarrow \alpha X$ ($\alpha$ = $e$ or $\mu$; for
sub-GeV events quasi-elastic scatterings $\nu_{\alpha} N \to \alpha
N'$ are dominant and the cross-section given in \cite{gaisser} is
used, while for multi-GeV events the inclusive cross-section for
$\nu_{\alpha} N \to \alpha X$ given in \cite{barger} is used), and
$\Theta$ is the zenith angle of the direction from which the charged
lepton $\alpha$ comes (See Fig. 1 of \cite{fvy}), $r_\mu$ and $r_e$
stand for ratios of contamination due to misidentification of
$\mu$-like and $e$-like events and we have put $r_\mu$=0.03,
$r_e$=0.06 for sub-GeV events, $r_\mu$=0.007, $r_e$=0.12 for fully
contained multi-GeV events \cite{hasegawa}, $r_\mu$=0 for partially
contained multi-GeV and upward going through $\mu$ events.

$\chi^2$ is defined as
\begin{eqnarray}
\chi^2=\chi_{\rm sub-GeV}^2+\chi_{\rm multi-GeV}^2
+\chi_{\rm through}^2
\label{eqn:chi}
\end{eqnarray}
where
\begin{eqnarray}
\displaystyle\chi_{\rm sub-GeV}^2&=&
{\beta_s^2 \over \sigma_{\beta s}^2}
+\sum_{j=1}^{10}\left\{
{\left[ \alpha (1-{\beta_s \over 2})N_j^{\rm s}(e)
-n_j^{\rm s}(e)\right]^2
\over n_j^{\rm s}(e)}
+{\left[ \alpha (1+{\beta_s \over 2})N_j^{\rm s}(\mu)
-n_j^{\rm s}(\mu)\right]^2
\over n_j^{\rm s}(\mu)}\right\},\nonumber\\
\displaystyle\chi_{\rm multi-GeV}^2&=&
{\beta_m^2 \over \sigma_{\beta m}^2}
+\sum_{j=1}^{10}\left\{
{\left[ \alpha (1-{\beta_m \over 2})N_j^{\rm m}(e)
-n_j^{\rm m}(e)\right]^2
\over n_j^{\rm m}(e)}
+{\left[ \alpha (1+{\beta_m \over 2})N_j^{\rm m}(\mu)
-n_j^{\rm m}(\mu)\right]^2
\over n_j^{\rm m}(\mu)}\right\},\nonumber\\
\displaystyle\chi_{\rm through}^2&=&
{\alpha_{\rm th}^2 \over \sigma_{\alpha\,{\rm th}}^2}
+\sum_{j=1}^{10}
{\left[ \alpha_{\rm th} N_j^{\rm t}(\mu)
-n_j^{\rm t}(\mu)\right]^2
\over n_j^{\rm t}(\mu)}
\end{eqnarray}
are $\chi^2$ for sub-GeV, multi-GeV, and upward going through
$\mu$ events, respectively, the summation on $j$ runs
over the ten bins for each $\chi^2$, $N_j^a(\alpha)$ and
$n_j^a(\alpha)$ ($a$=s, m, th) stand for theoretical predictions and data
for the numbers of sub-GeV, multi-GeV, and upward going through
$\mu$ events,
and it is understood that
$\chi^2$ is minimized with respect to all the normalization
factors $\alpha$, $\beta_s$, $\beta_m$, $\alpha_{\rm th}$.
We have put $\sigma_s$=0.08, $\sigma_m$=0.12,
$\sigma_{\alpha\,{\rm th}}$=0.22 and we have assumed that
the overall flux normalization $\alpha$ in
the contained events is a free parameter as in \cite{skatm}, and
we have omitted the uncertainties of $E_\nu$ spectral index,
relative normalization between PC and FC and up-down
correlation for simplicity.

We have evaluated $\chi^2$ for $\theta_{24}= (25+5j)^\circ$
($j=0,\cdots,7$), $\theta_{34}= 15j^\circ$ ($j=-6,\cdots,6$),
$\theta_{23}= 10j^\circ$ ($j=0,\cdots,4$),
$\delta_1 = 0^\circ, 45^\circ, 90^\circ$,
$\Delta m^2_{43}= 10^{-4+j/10}$eV$^2$ ($j=5,\cdots,20$)
and it is found that $\chi^2$ has the minimum value
\begin{eqnarray}
\chi^2_{\rm min}=43.1 \qquad(\chi_{\rm sub-GeV}^2=19.0,
\chi_{\rm multi-GeV}^2=13.2,
\chi_{\rm through}^2=11.6)
\end{eqnarray}
for
\begin{eqnarray}
&{ }&\Delta m_{43}^2=10^{-2.9}{\rm eV}^2=
1.3\times 10^{-3}{\rm eV}^2,~~
(\theta_{24},\theta_{34},\theta_{23})=
(35^\circ,15^\circ,20^\circ)
\label{eqn:bestfit}
\end{eqnarray}
for 45 degrees of freedom.
For pure $\nu_\mu\leftrightarrow\nu_\tau$
($\theta_{34}=\theta_{23}=0$),
the best fit is obtained
\begin{eqnarray}
\chi^2_{\rm min}(\nu_\mu\leftrightarrow\nu_\tau)=48.3
~~(\chi_{\rm sub-GeV}^2=19.8,
\chi_{\rm multi-GeV}^2=17.0,
\chi_{\rm through}^2=10.6)
\end{eqnarray}
for
\begin{eqnarray}
&{ }&\Delta m_{43}^2=
2.0\times 10^{-3}{\rm eV}^2,~~
(\theta_{24},\theta_{34},\theta_{23})=
(40^\circ,0^\circ,0^\circ).
\label{eqn:mutau}
\end{eqnarray}
The allowed regions at 1$\sigma$CL, 90\%CL, 99\%CL
are obtained by
$\chi^2\le \chi^2_{\rm min}+\Delta \chi^2$,
where $\Delta \chi^2$=5.9, 9.2, 15.1 for five degrees
of freedom, respectively,
and they are depicted in Fig. 1 for various values of
$\theta_{24}$ and for $\delta_1=0$ (1(a)),
$\delta_1=\pi/4$ (1(b)) and $\delta_1=\pi/2$ (1(c)).
At 99\% confidence level we find
$27^\circ\lesssim\theta_{24}\lesssim 58^\circ$.
The difference of the best fit point among the all
parameter space and the best fit for pure $\nu_\mu\leftrightarrow\nu_\tau$
($\Delta m_{43}^2=10^{-27/10}$eV$^2$, $\theta_{24}=40^\circ$,
$\theta_{34}=0^\circ$, $\theta_{23}=0^\circ$)
is $\Delta \chi^2$=5.1 which corresponds to 0.84$\sigma$, so
this difference is not significant.
Zenith angle dependence for contained
events and upward going through muon events for
three sets of the parameters including the case (\ref{eqn:bestfit})
and (\ref{eqn:mutau}) is given in Fig. 2(a) and 2(b).
Zenith angle dependence for no oscillation case that we obtained
is reasonably in good agreement with the Superkamiokande
result \cite{toshito} and this puts confidence to the present
analysis.
In general, the reason that the best fit point is
slightly away from pure $\nu_\mu\leftrightarrow\nu_\tau$
case is because a better fit to the multi-GeV contained events
compensates a worse fit to the upward going through $\mu$
events, and in total the case of hybrid oscillations of
$\nu_\mu\leftrightarrow\nu_\tau$ and
$\nu_\mu\leftrightarrow\nu_s$ fits better to the data
(See Figs. 2(a) and 2(b)).

\section{discussions}

(1) Pure $\nu_\mu\leftrightarrow\nu_s$ oscillation is
obtained only for $\theta_{23}=0$, $\theta_{34}=\pm 90^\circ$,
and in this case $\chi^2$ satisfies
$\chi^2-\chi^2_{\rm min}\ge$17.9
for any value of $\Delta m_{43}^2$, $\theta_{24}$ and $\delta_1$
and it implies that
pure $\nu_\mu\leftrightarrow\nu_s$ oscillation is excluded
at 99.7\%CL (3.0$\sigma$CL).  This is consistent with
the recent claim by Superkamiokande group \cite{hasegawa,toshito}.

(2) For some $\Delta m_{43}^2$, $\theta_{24}$ $\delta_1$
(e.g., $\Delta m_{43}^2$=1.3$\times 10^{-3}$eV$^2$,
$\theta_{24}=50^\circ$, $\theta_{23}=30^\circ$, $\delta_1=0$),
$\theta_{34}=90^\circ$ is allowed at 90\%CL, and one might wonder
how such a solution can give a reasonable fit to the data.
For this set of the oscillation parameters,
we find that the oscillation probabilities without
matter effects are given by
\begin{eqnarray}
P(\nu_\mu\rightarrow\nu_\tau;{\mbox{without matter effects}})&=&
{3 \over 8}\cos^2\left({5\pi \over 18}\right)\nonumber\\
P(\nu_\mu\rightarrow\nu_s;{\mbox{without matter effects}})&=&
\sin^2\left({5\pi \over 9}\right)\left[{1 \over 32}+{3 \over 4}
\sin^2\left({\Delta m^2_{43}L \over 4E}\right)\right],
\label{eqn:hybrid}
\end{eqnarray}
where $(3/8)\cos^2(5\pi/18) \simeq 0.15$, $\sin^2(5\pi/9)\simeq 0.97$,
and we have averaged over rapid oscillations which come from
$\Delta m_{32}^2$.
From (\ref{eqn:hybrid}) we observe that zenith angle dependence with
a non-maximal coefficient comes
solely from $\nu_\mu\leftrightarrow\nu_s$ oscillation
and both $\nu_\mu\leftrightarrow\nu_\mu$ and
$\nu_\mu\leftrightarrow\nu_s$ oscillations have constant
contribution in the oscillation probabilities.
As was shown in \cite{fvy}, the zenith angle dependence of
contained events is explained well also by $\nu_\mu\leftrightarrow\nu_s$
oscillation, but the fit of $\nu_\mu\leftrightarrow\nu_s$ to
upward going $\mu$ events is poor \cite{hasegawa,toshito}.
In the present case, we have $\chi_{\rm sub-GeV}^2$=19.7,
$\chi_{\rm multi-GeV}^2$=15.5,
$\chi_{\rm through}^2$=15.0 and we observe that this solution is
allowed at 90\%CL because a poor fit to the upward going through $\mu$
events is compensated by
a good fit to the multi-GeV contained events.

(3) One of the remarkable features of our result is
that there is a region where relatively small value
of $\theta_{24}(\sim 30^\circ)$ is allowed at 90\%CL for
$\theta_{23}\simeq 20^\circ$, i.e., there exists a solution
in which all the mixing angles are relatively small
$|\theta_{jk}|\lesssim \pi/6$,
and such situation does not
occur for $\theta_{23}=0$.  This phenomena is
reminiscent of the work \cite{dgh} on three flavor
analysis of the atmospheric neutrino data,
in which it was argued that
relatively small mixing angles could account for the data
because of matter effects
if there were no CHOOZ constraint \cite{chooz}.
However, in the present case it turns out that the matter effect
is not so important, i.e., even if we put the density of
the matter to zero the fit to the data is still reasonable.
The reason that the fit to the data is good for $\theta_{24}\sim 30^\circ$
is because the disappearance probability behaves as
$1-P(\nu_\mu\leftrightarrow\nu_\mu) = A + B\sin^2(\Delta m_{43}^2L/4E)$
($A, B$ are constant) and even if the coefficient $B$
is relatively small the presence of the constant term $A$
compensates the goodness of the fit to a certain extent.
This argument applies both to the multi-GeV $\mu$-like
contained events and the upward going through muon events.

(4) To combine the present result with the analysis
of the solar neutrinos in
\cite{ggp}, it is necessary to obtain the value
of $c_s\equiv |U_{s1}|^2+|U_{s2}|^2$ for each point.
In our parametrization (\ref{eqn:u}) we have
\begin{eqnarray}
U_{s1}&=&-s_{12}(c_{23}c_{34}+
s_{23}s_{24}s_{34}e^{i\delta_1})\nonumber\\
U_{s2}&=&~~c_{12}(c_{23}c_{34}+
s_{23}s_{24}s_{34}e^{i\delta_1})
\end{eqnarray}
so that
\begin{eqnarray}
c_s&\equiv& |U_{s1}|^2+|U_{s2}|^2\nonumber\\
&=&|c_{23}c_{34}+
s_{23}s_{24}s_{34}e^{i\delta_1}|^2.
\end{eqnarray}
The contours of $c_s=$0.2, 0.4, 0.6, 0.8 are plotted together with
the allowed region for various $\Delta m_{43}^2$
in Fig. 3(a) ($\delta_1=0$), 3(b) ($\delta_1=\pi/4$) and
3(c) ($\delta_1=\pi/2$).  For each value of $\delta_1$
some point in the allowed region
satisfies $c_s\lesssim 0.2$ for
$40^\circ\lesssim \theta_{24}\lesssim50^\circ$.
We find that the allowed region at 90\% confidence level satisfies
$0.15\lesssim c_s\le 1$ for $\delta_1=0$,
$0.10\lesssim c_s\le 1$ for $\delta_1=\pi/4$,
$0.05\lesssim c_s\le 1$ for $\delta_1=\pi/2$, respectively.
Hence combination of the present result with the analysis
in \cite{ggp} suggests that the LMA and VO solutions
as well as SMA solution of the solar neutrino problem
are possible for some region in the parameter space.
Recently Superkamiokande group has announced \cite{suzuki} that
the SMA and VO solutions of the solar neutrino problem
are disfavored at 95\% confidence level.  If Nature
is described by a four neutrino scenario, therefore,
the present scheme with $c_s\lesssim 0.2$ may be the right
solution for all the oscillation data.

\section{Conclusions}

We have shown in the framework of four neutrino oscillations
without assuming the BBN constraints
that the Superkamiokande atmospheric neutrino data
are explained by wide range of the oscillation parameters
which implies hybrid oscillations of $\nu_\mu\leftrightarrow\nu_\tau$
and $\nu_\mu\leftrightarrow\nu_s$ as well as hybrid
oscillations with $\Delta m_{\mbox{\rm\scriptsize atm}}^2$ and
$\Delta m_{\mbox{\rm\scriptsize LSND}}^2$.
The case of pure $\nu_\mu\leftrightarrow\nu_s$ is excluded
at 3.0$\sigma$CL in good agreement with the Superkamiokande
analysis.  It is found by combining the analysis on the solar
neutrino data by Giunti, Gonzalez-Garcia and Pe\~na-Garay
that the LMA and VO solutions
as well as SMA solution of the solar neutrino problem
are allowed.  This gives us another possibility
in phenomenology of neutrino oscillations
and such scenarios deserve further study.

\section*{Acknowledgments}
I would like to thank Alex Friedland for showing
me how to obtain interpolation in 3 dimensions by Mathematica.
This
research was supported in part by a Grant-in-Aid for Scientific
Research of the Ministry of Education, Science and Culture,
\#12047222, \#10640280.

\section*{Appendix}
With hierarchy $|\Delta E_{32}|\gg |\Delta E_{43}|, |A(x)|$ we can
obtain the analytical expression for the oscillation probability
$P(\nu_\mu\rightarrow\nu_\mu)$ to the leading order in $|\Delta
E_{43}|/|\Delta E_{32}|$ and $|A(x)|/|\Delta E_{32}|$ in adiabatic
approximation.
The purpose of this Appendix is to show that
the oscillation probability $P(\nu_\mu\leftrightarrow\nu_\mu)$
is invariant under $\theta_{34}\rightarrow-\theta_{34}$ and
$\delta_1\rightarrow\pi-\delta_1$
to the leading order in $|\Delta E_{43}/\Delta E_{32}|$
and $|A(x)/\Delta E_{32}|$.
To do that, it is convenient to use different parametrization:
\begin{eqnarray}
V\equiv R_{13}\left({\pi \over 2}\right){\cal D}
R_{23}(\varphi_{23})
R_{13}(\varphi_{13})
{\cal D}^{-1}
R_{12}(\varphi_{12}),
\label{eqn:v}
\end{eqnarray}
where ${\cal D}\equiv$ diag($e^{i\delta'}$,1,1).
To relate this parametrization with (\ref{eqn:mns3}),
we have to multiply diagonal phase matrices
\begin{eqnarray}
V= {\rm diag}\left(e^{i\omega_1},e^{i\omega_2},
e^{-i(\omega_1+\omega_2)}\right)
{\widetilde U}
{\rm diag}\left(e^{i\gamma_1},e^{i\gamma_2},
e^{-i(\gamma_1+\gamma_2)}\right),
\label{eqn:vu}
\end{eqnarray}
where
\begin{eqnarray}
\omega_1&=&{1 \over 3}\arg U_{s2}+{1 \over 3}\arg U_{s3}
-{1 \over 3}\arg U_{\mu 4},~~
\omega_2={1 \over 3}\arg U_{s2}+{1 \over 3}\arg U_{s3}
+{2 \over 3}\arg U_{\mu 4}\nonumber\\
\gamma_1&=&-{1 \over 3}\arg U_{s2}+{2 \over 3}\arg U_{s3}
+{1 \over 3}\arg U_{\mu 4},~~
\gamma_2={2 \over 3}\arg U_{s2}-{1 \over 3}\arg U_{s3}
+{1 \over 3}\arg U_{\mu 4}.
\end{eqnarray}
Comparing the both hand sides of (\ref{eqn:vu}), we obtain
\begin{eqnarray}
\tan\varphi_{12}&=&{|U_{s3}| \over |U_{s2}|},~~~
\tan\varphi_{13}={c_{24}s_{34} \over
\sqrt{c_{24}^2c_{34}^2+s_{24}^2}}\nonumber\\
\tan\varphi_{23}&=&{c_{24}c_{34} \over s_{24}},~~
\arg \left(-V_{e3}\right)\equiv\pi-\delta'=
-\arg U_{\mu 4}-\arg U_{s 2}-\arg U_{s 3}.
\end{eqnarray}
Using this parametrization, we have
\begin{eqnarray}
&{\ }&V^{-1}{\rm diag}\left(0,0,A\right)V
+\left(0,0,\Delta E_{43}\right)\nonumber\\
&=&R_{12}(\varphi_{12})^{-1}{\cal D}
R_{13}(\varphi_{13}-\varphi_{13}^M)^{-1}
{\rm diag}\left(t_-,0,t_+\right)
R_{13}(\varphi_{13}-\varphi_{13}^M)
{\cal D}^{-1}
R_{12}(\varphi_{12})\nonumber\\
&=&{\rm diag}\left(C_{12}^2(t_-{\tilde C}^2+t_+{\tilde S}^2),
S_{12}^2(t_-{\tilde C}^2+t_+{\tilde S}^2),
t_+{\tilde C}^2+t_-{\tilde S}^2\right)\nonumber\\
&{\ }&+{1 \over 2}\sin2\varphi_{12}(t_-{\tilde C}^2+t_+{\tilde S}^2)\lambda_1
+{1 \over 2}C_{12}(t_--t_+)\sin2{\tilde\varphi}
(\lambda_4\cos\delta'-\lambda_5\sin\delta')\nonumber\\
&{\ }&+{1 \over 2}S_{12}(t_--t_+)\sin2{\tilde\varphi}
(\lambda_6\cos\delta'-\lambda_7\sin\delta'),
\end{eqnarray}
where $\lambda_j (j=1,\cdots, 8)$ are the $3\times3$ Gell-Mann
matrices with normalization $tr(\lambda_j \lambda_k)=2\delta_{jk}$,
$S_{12}\equiv\sin\varphi_{12},C_{12}\equiv\cos\varphi_{12}$,
$\tilde S\equiv\sin(\varphi_{13}-\varphi_{13}^M),
\tilde C\equiv\cos(\varphi_{13}-\varphi_{13}^M)$
and $\varphi_{13}^M$ is the effective mixing angle in matter
given by
\begin{eqnarray}
\tan2\varphi_{13}^M\equiv {\Delta E_{43}\sin2\varphi_{13}
\over \Delta E_{43}\cos2\varphi_{13}- A},
\end{eqnarray}
and $t_\pm$ are the eigenvalues defined by
\begin{eqnarray}
t_\pm\equiv{1 \over 2}\left(
A+\Delta E_{43}\pm\sqrt{(\Delta E_{43}\cos2\varphi_{13}-A)^2
+(\Delta E_{43}\sin2\varphi_{13})^2}\right).
\end{eqnarray}
Putting
\begin{eqnarray}
\Lambda\equiv {1 \over 2\Delta E_{32}}\left[\sin2\varphi_{12}
(t_-{\tilde C}^2+t_+{\tilde S}^2)\lambda_1
+C_{12}(t_--t_+)\sin2{\tilde\varphi}
(\lambda_4\cos\delta'-\lambda_5\sin\delta')\right],
\end{eqnarray}
we have
\begin{eqnarray}
&{\ }&e^{-i\Lambda}\left[{\rm diag}\left(-\Delta E_{32},0,0\right)
+V^{-1}{\rm diag}\left(0,0,A\right)V
+{\rm diag}\left(0,0,\Delta E_{43}\right)
\right]e^{i\Lambda}\nonumber\\
&=&{\rm diag}\left(-\Delta E_{32},0,0\right)
+(C_{12}^2(t_-{\tilde C}^2+t_+{\tilde S}^2)I_3\nonumber\\
&{\ }&+{1 \over 2}\left[(t_-{\tilde C}^2+t_+{\tilde S}^2)
(S_{12}^2-2C_{12}^2)+t_+{\tilde C}^2+t_-{\tilde S}^2\right]
{\rm diag}(0,1,1)\nonumber\\
&{\ }&+{1 \over 2}\left[
S_{12}^2(t_-{\tilde C}^2+t_+{\tilde S}^2)
-t_+{\tilde C}^2-t_-{\tilde S}^2\right]\widetilde\lambda_8
+{1 \over 2}S_{12}(t_--t_+)\sin2{\tilde\varphi}
(\lambda_6\cos\delta'-\lambda_7\sin\delta'),
\label{eqn:diag}
\end{eqnarray}
where $I_3\equiv{\rm diag}(1,1,1),
\widetilde\lambda_8\equiv{\rm diag}(0,1,-1)$.
The last two lines in (\ref{eqn:diag}) can be
diagonalized by noting
\begin{eqnarray}
&{\ }&
{\cal F}\widetilde\lambda_8+{\cal G}
(\lambda_6\cos\delta'-\lambda_7\sin\delta')
\nonumber\\
&=&{\cal F}\widetilde\lambda_8+{\cal G}
e^{{i \over 2}\delta'\widetilde\lambda_8}\,
\lambda_6\,e^{-{i \over 2}\delta'\widetilde\lambda_8}
=e^{{i \over 2}\delta'\widetilde\lambda_8}
\left({\cal F}\widetilde\lambda_8
+{\cal G}\lambda_6\right)
e^{-{i \over 2}\delta'\widetilde\lambda_8}\nonumber\\
&=&e^{{i \over 2}\delta'\widetilde\lambda_8}
e^{-i\psi\lambda_7}
\sqrt{{\cal F}^2+{\cal G}^2}\,{\widetilde\lambda_8}\,
e^{i\psi\lambda_7}
e^{-{i \over 2}\delta'\widetilde\lambda_8},
\end{eqnarray}
where
\begin{eqnarray}
{\cal F}&\equiv&{1 \over 2}\left[
S_{12}^2(t_-{\tilde C}^2+t_+{\tilde S}^2)
-t_+{\tilde C}^2-t_-{\tilde S}^2\right]\nonumber\\
{\cal G}&\equiv&{1 \over 2}S_{12}(t_--t_+)\sin2{\tilde\varphi}
\end{eqnarray}
and the effective mixing angle $\psi$ in matter
is given by
\begin{eqnarray}
\tan2\psi\equiv {{\cal G} \over {\cal F}}.
\end{eqnarray}
Putting everything together, we obtain
\begin{eqnarray}
&{\ }&V
{\rm diag}\left(-\Delta E_{32},0,\Delta E_{43}\right)
V^{-1}
+{\rm diag}\left(0,0,A\right)\nonumber\\
&=&Ve^{{i \over 2}\delta'\widetilde\lambda_8}
e^{-i\psi\lambda_7}e^{i\Lambda}\left\{
{\rm diag}\left(-\Delta E_{32},0,0\right)
+(C_{12}^2(t_-{\tilde C}^2+t_+{\tilde S}^2)I_3
\right.\nonumber\\
&{\ }&\left.+{1 \over 2}\left[(t_-{\tilde C}^2+t_+{\tilde S}^2)
(S_{12}^2-2C_{12}^2)+t_+{\tilde C}^2+t_-{\tilde S}^2\right]
{\rm diag}(0,1,1)+\sqrt{{\cal F}^2+{\cal G}^2}\,{\widetilde\lambda_8}
\right\}\nonumber\\
&{\ }&\times e^{-i\Lambda}e^{i\psi\lambda_7}
e^{-{i \over 2}\delta'\widetilde\lambda_8}V^{-1},
\end{eqnarray}
to first order in $|\Delta E_{43}/\Delta E_{32}|$ and
$|A(x)/\Delta E_{32}|$.
The effective MNS matrix in matter
is given by $V e^{i\delta'\widetilde\lambda_8/2}
e^{-i\psi\lambda_7}$ 
to zeroth order, and the three eigenvalues in matter are $-\Delta
E_{32}$, $\xi_+$, $\xi_-$, where
$\xi_\pm\equiv(t_-{\tilde C}^2+t_+{\tilde S}^2)
(S_{12}^2-2C_{12}^2)+t_+{\tilde C}^2+t_-{\tilde S}^2 \pm\sqrt{{\cal
F}^2+{\cal G}^2}$, to first order in $|\Delta E_{43}/\Delta E_{32}|$
and $|A(x)/\Delta E_{32}|$, after subtracting the contribution from
$C_{12}^2(t_-{\tilde C}^2+t_+{\tilde S}^2)I_3$.  The oscillation
probability $P(\nu_\mu\rightarrow\nu_\mu)$ is given by
\begin{eqnarray}
P(\nu_\mu\rightarrow\nu_\mu)
&=&1-4|V_{\mu 1}|^2(1-|V_{\mu 1}|^2)\sin^2\left(
{\Delta E_{32}L \over 2}\right)\nonumber\\
&{\ }&-4|e^{{i \over 2}\delta'}C_\psi V_{\mu 2}
+e^{-{i \over 2}\delta'}S_\psi V_{\mu 3}|^2
|-e^{{i \over 2}\delta'}S_\psi V_{\mu 2}
+e^{-{i \over 2}\delta'}C_\psi V_{\mu 3}|^2
\sin^2\left(
{(\xi_+-\xi_-)L \over 2}\right)\nonumber\\
\label{eqn:pmm}
\end{eqnarray}
to the leading order in $|\Delta
E_{43}|/|\Delta E_{32}|$ and $|A(x)|/|\Delta E_{32}|$,
where $C_\psi\equiv\cos\psi$, $S_\psi\equiv\sin\psi$ and
$V_{\mu j}$ are the matrix elements given by (\ref{eqn:v}):
\begin{eqnarray}
V_{\mu 1}&\equiv&-S_{12}C_{23}-C_{12}S_{23}S_{13}e^{i\delta'}\nonumber\\
V_{\mu 2}&\equiv&C_{12}C_{23}-S_{12}S_{23}S_{13}e^{i\delta'}\nonumber\\
V_{\mu 3}&\equiv&S_{23}C_{13}
\end{eqnarray}
with $S_{jk}\equiv\sin\varphi_{jk}$, $C_{jk}\equiv\cos\varphi_{jk}$.
We observe that (\ref{eqn:pmm}) is invariant under
$\theta_{34}\rightarrow-\theta_{34}$ and $\delta_1\rightarrow\pi-\delta_1$
since under these transformation the mixing angles in $V$ behave as
$\varphi_{12}\rightarrow\varphi_{12}$,
$\varphi_{13}\rightarrow-\varphi_{13}$
($\tilde\varphi_{13}\rightarrow-\tilde\varphi_{13}$),
$\varphi_{23}\rightarrow\varphi_{23}$,
$\arg U_{sj}\rightarrow-\arg U_{sj}$ ($j$=2,3),
$\arg U_{\mu 4}\rightarrow\pi-\arg U_{\mu 4}$,
$\delta'\rightarrow\pi-\delta'$,
$\psi\rightarrow-\psi$,
$V_{\mu j}\rightarrow V_{\mu j}^\ast$ ($j$=1,2),
$V_{\mu 3}\rightarrow V_{\mu 3}$,
$|e^{{i \over 2}\delta'}C_\psi V_{\mu 2}
+e^{-{i \over 2}\delta'}S_\psi V_{\mu 3}|^2
\rightarrow|e^{{i \over 2}\delta'}C_\psi V_{\mu 2}
+e^{-{i \over 2}\delta'}S_\psi V_{\mu 3}|^2$,
$|-e^{{i \over 2}\delta'}S_\psi V_{\mu 2}
+e^{-{i \over 2}\delta'}C_\psi V_{\mu 3}|^2
\rightarrow|-e^{{i \over 2}\delta'}S_\psi V_{\mu 2}
+e^{-{i \over 2}\delta'}C_\psi V_{\mu 3}|^2$.
Therefore, the oscillation probability is invariant
under $\theta_{34}\rightarrow-\theta_{34}$,
$\delta_1\rightarrow\pi-\delta_1$
to zeroth order in $|\Delta E_{43}/\Delta E_{32}|$
and $|A(x)/\Delta E_{32}|$.


\newpage
\noindent
{\Large{\bf Figures}}

\begin{description}
\item[Fig.1 (a), (b), (c)] The allowed region in the $(\theta_{34},~
\theta_{23})$ plane for various values of $\theta_{24}=27^\circ,
30^\circ,\cdots,58^\circ$ and (a) $\delta_1=0$ ($\Delta m_{43}^2=
10^{-29/10}$eV$^2$), (b) $\delta_1=\pi/4$ ($\Delta m_{43}^2=
10^{-27/10}$eV$^2$), (c) $\delta_1=\pi/2$ ($\Delta m_{43}^2=
10^{-29/10}$eV$^2$), respectively.  The solid, dashed and dotted lines
represent 68\%, 90\%, 99\% confidence level, respectively.  The
asterisk in Fig. 1 (a) stands for the best fit point.  The value of
$\Delta m_{43}^2$ for each set of figures is that of the best fit point
with GIVEN value of $\delta_1$.  The best fit with a fixed value
of $\delta_1=0, \pi/4, \pi/2$ is obtained for $\Delta m_{43}^2=
10^{-29/10}$eV$^2$, $10^{-27/10}$eV$^2$, $10^{-29/10}$eV$^2$,
respectively.

\item[Fig.2 (a), (b)] The zenith angles dependence of (a) contained
events and (b) upward going through muon events for
$\Delta m_{43}^2=10^{-29/10}$eV$^2$, $\theta_{24}=50^\circ$,
$\theta_{34}=90^\circ$, $\theta_{23}=30^\circ$
(fine dotted line),
$\Delta m_{43}^2=10^{-29/10}$eV$^2$, $\theta_{24}=35^\circ$,
$\theta_{34}=15^\circ$, $\theta_{23}=20^\circ$ (dashed line;
best fit case)
$\Delta m_{43}^2=10^{-27/10}$eV$^2$, $\theta_{24}=40^\circ$,
$\theta_{34}=0^\circ$, $\theta_{23}=0^\circ$ 
(coarse dotted line; best fit case among pure
$\nu_\mu\leftrightarrow\nu_\tau$ oscillations),
no oscillation case (solid line), respectively.
$\delta_1=0$ for all the three cases.
The zenith angle dependence of the multi-GeV $\mu$-like events
is different for the three sets of the oscillation parameters,
but that of the upward going muon events is almost
similar for the best fit case and the pure
$\nu_\mu\leftrightarrow\nu_\tau$ case.

\item[Fig.3 (a), (b), (c)] The shadowed area is the
allowed region projected on the $(\theta_{34},~
\theta_{23})$ plane for various values of $\Delta m_{43}^2$
($10^{-3.5}$eV$^2\le\Delta m_{43}^2\le 10^{-2}$eV$^2$)
for each value of $\theta_{24}=27^\circ,
30^\circ,\cdots,58^\circ$ and for (a) $\delta_1=0$,
(b) $\delta_1=\pi/4$, (c) $\delta_1=\pi/2$, respectively,
and the thin solid lines are boundary of the allowed region
for various values of $\Delta m_{43}^2$.
The solid, dashed, coarse dotted and fine dotted lines
stand for the contours of
$c_s\equiv|U_{s1}|^2+|U_{s2}|^2$
=$|c_{23}c_{34}+s_{23}s_{24}s_{34}e^{i\delta_1}|^2=0.2,
0.4, 0.6, 0.8$,
respectively.  Solutions with $c_s\lesssim 0.2$ exist
for $40^\circ\lesssim\theta_{24}\lesssim 50^\circ$
and for each value of $\delta_1$ 
and they can have Large Mixing Angle solutions of the solar
neutrino problem.

\end{description}
\newpage
\pagestyle{empty}
\vglue -4.5cm
\hglue -5.5cm 
\epsfig{file=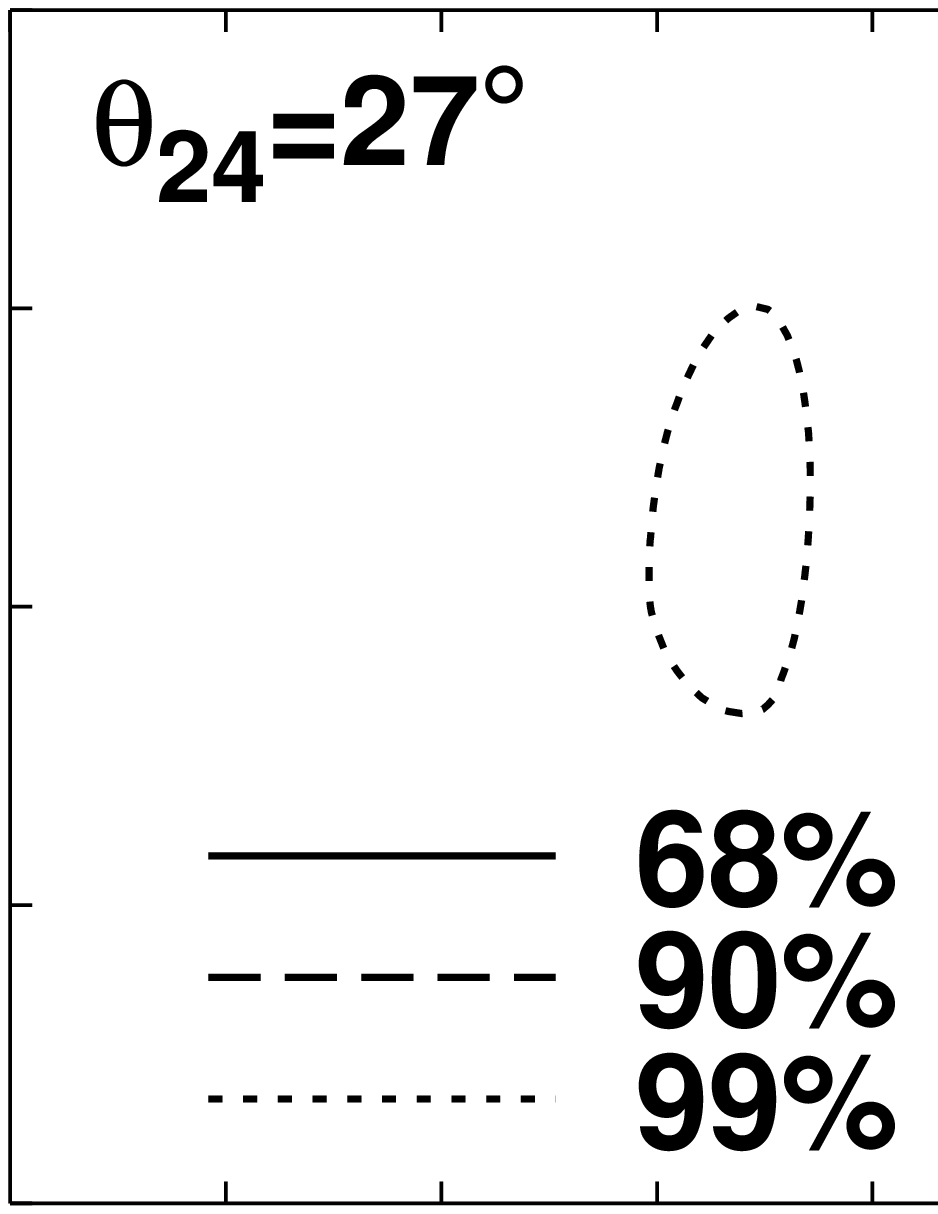,width=10cm}
\vglue -10.1cm \hglue 0.6cm \epsfig{file=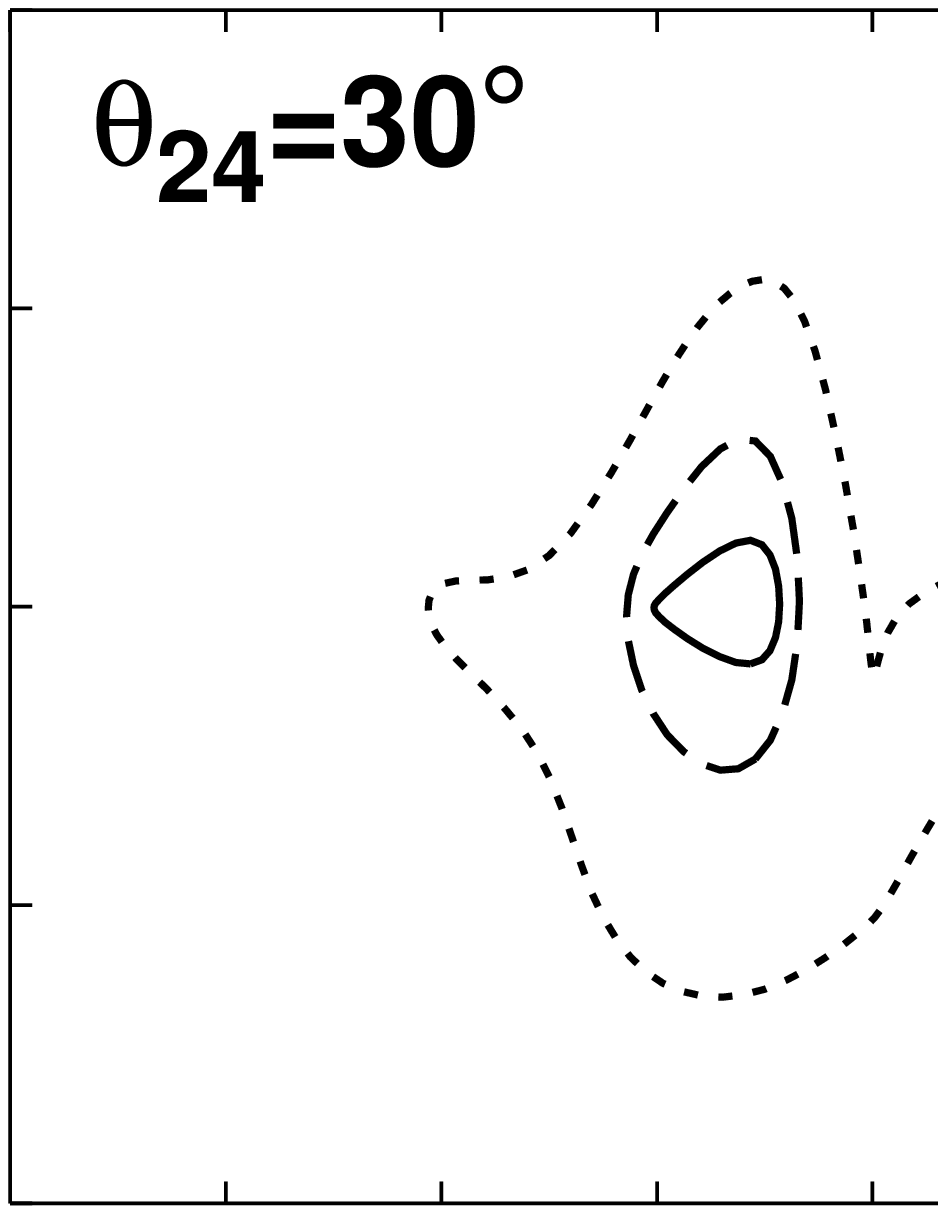,width=10cm}

\vglue -4.3cm
\hglue -5.5cm 
\epsfig{file=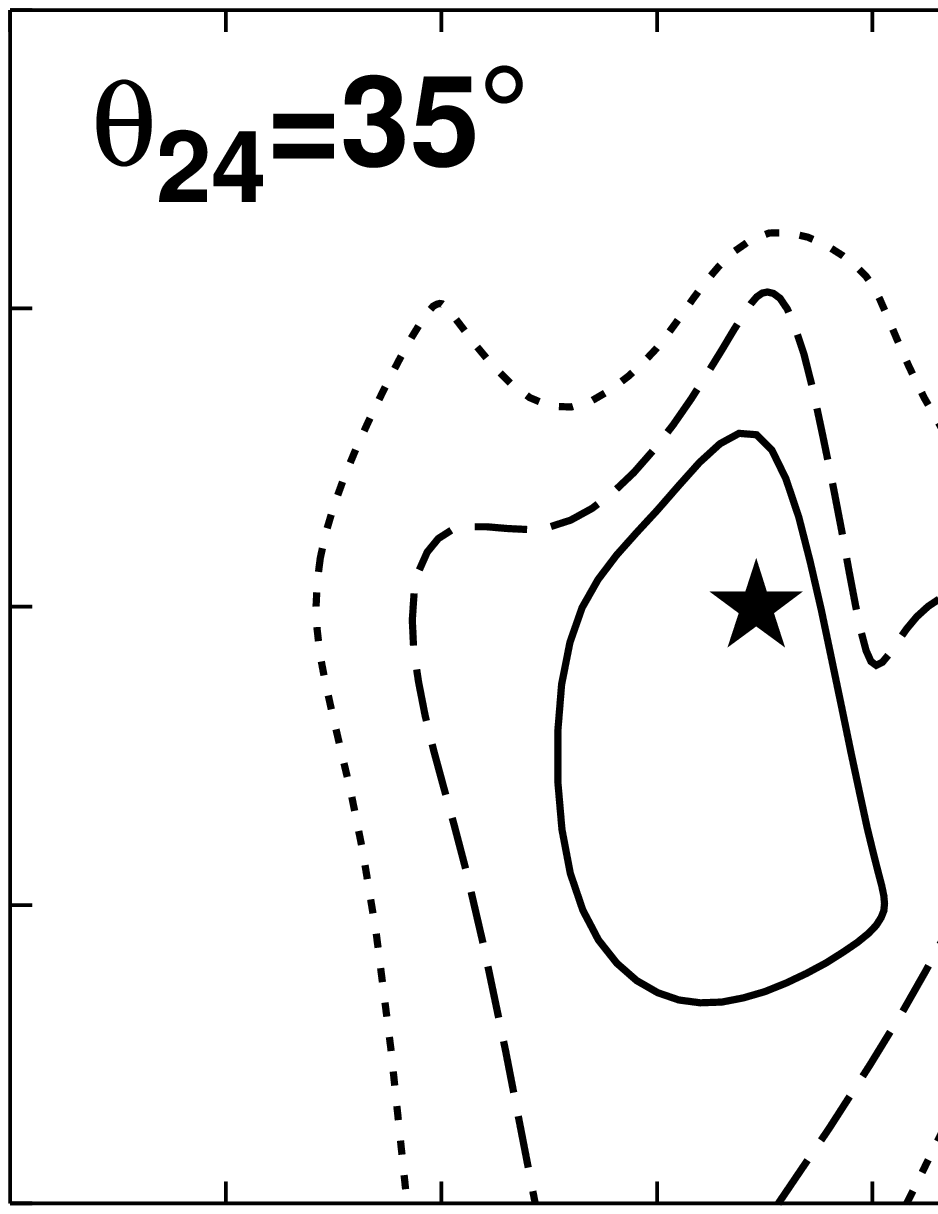,width=10cm}
\vglue -10.1cm \hglue 0.6cm \epsfig{file=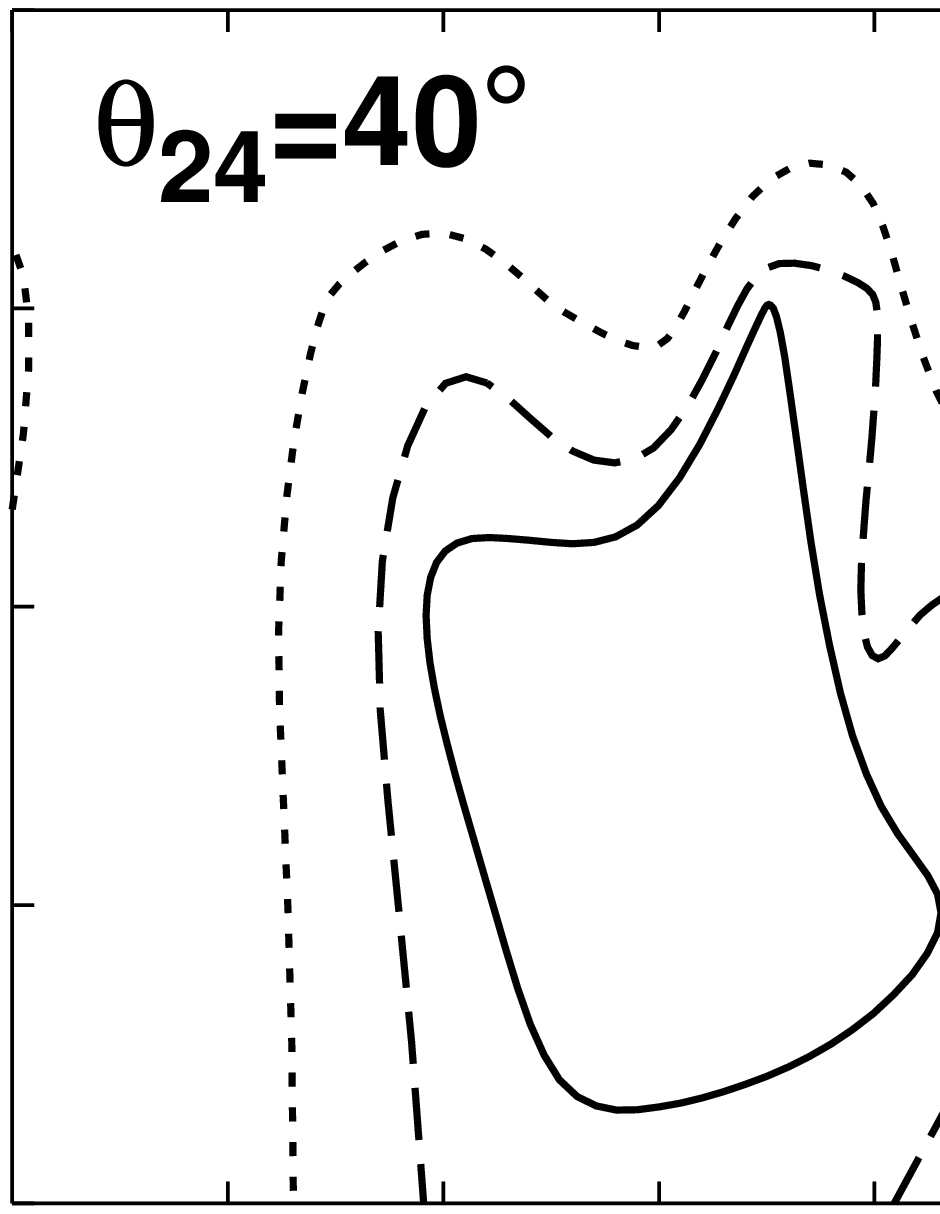,width=10cm}

\vglue -4.3cm
\hglue -5.5cm 
\epsfig{file=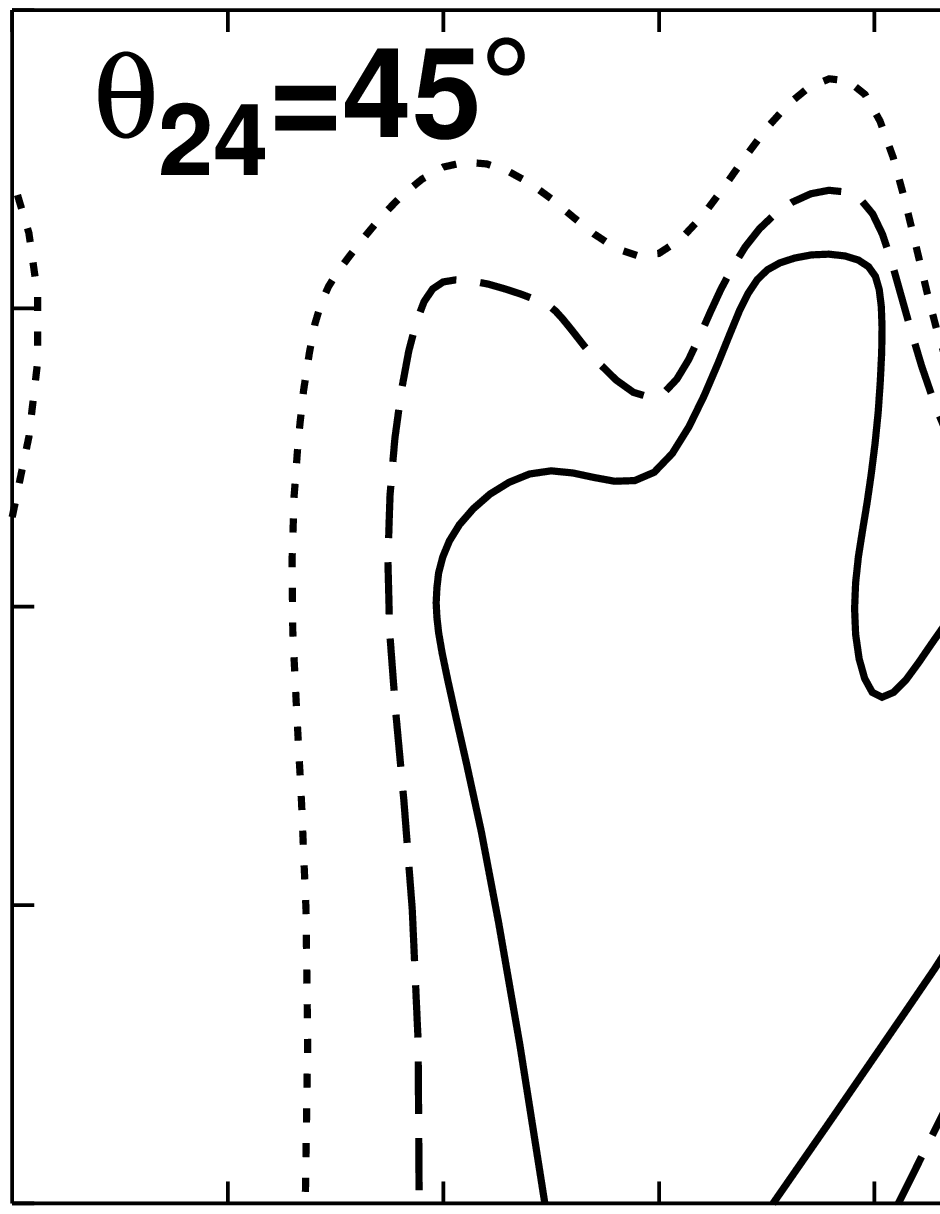,width=10cm}
\vglue -10.1cm \hglue 0.6cm \epsfig{file=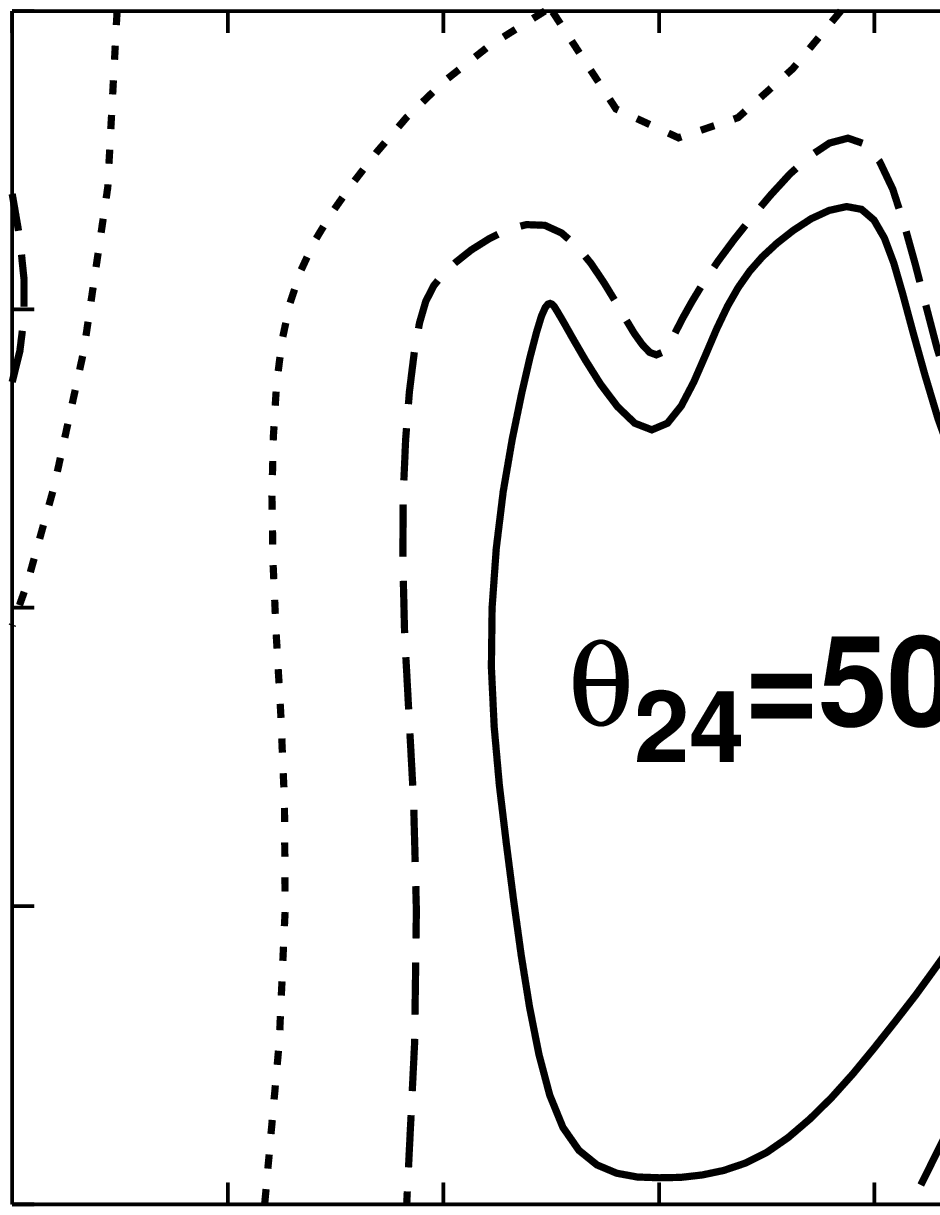,width=10cm}

\vglue -4.3cm
\hglue -5.5cm 
\epsfig{file=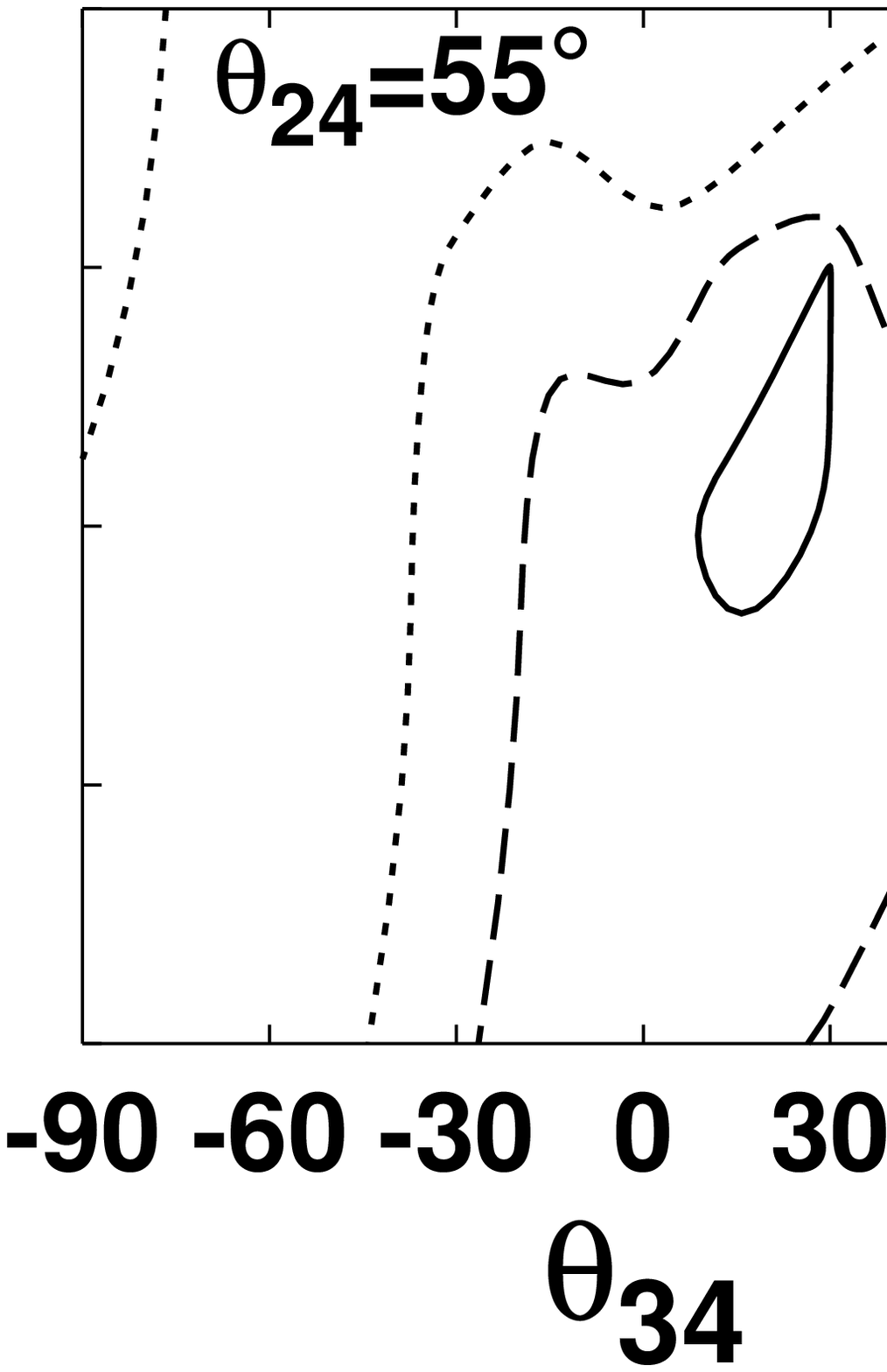,width=10cm}
\vglue -10.1cm \hglue 0.6cm \epsfig{file=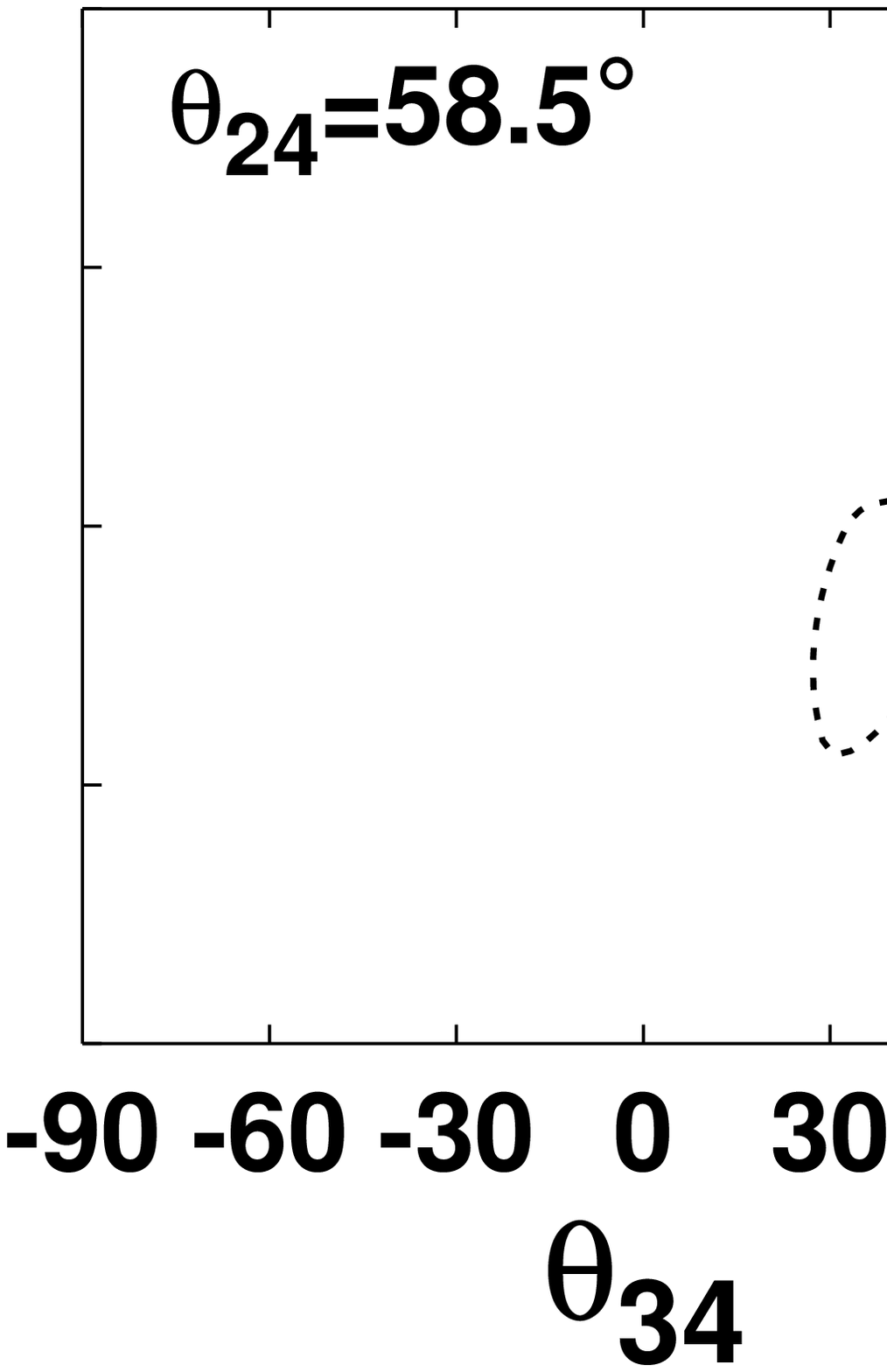,width=10cm}

\vglue -1.0cm
\hglue 6.0cm \epsfig{file=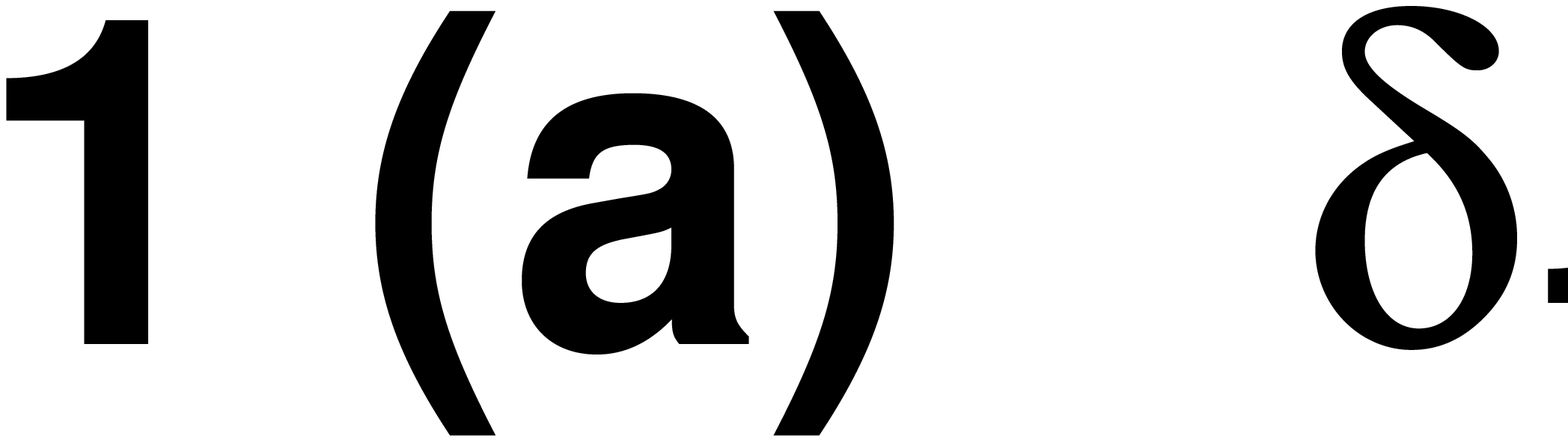,width=3cm}
\newpage
\pagestyle{empty}
\vglue -4.5cm
\hglue -5.5cm 
\epsfig{file=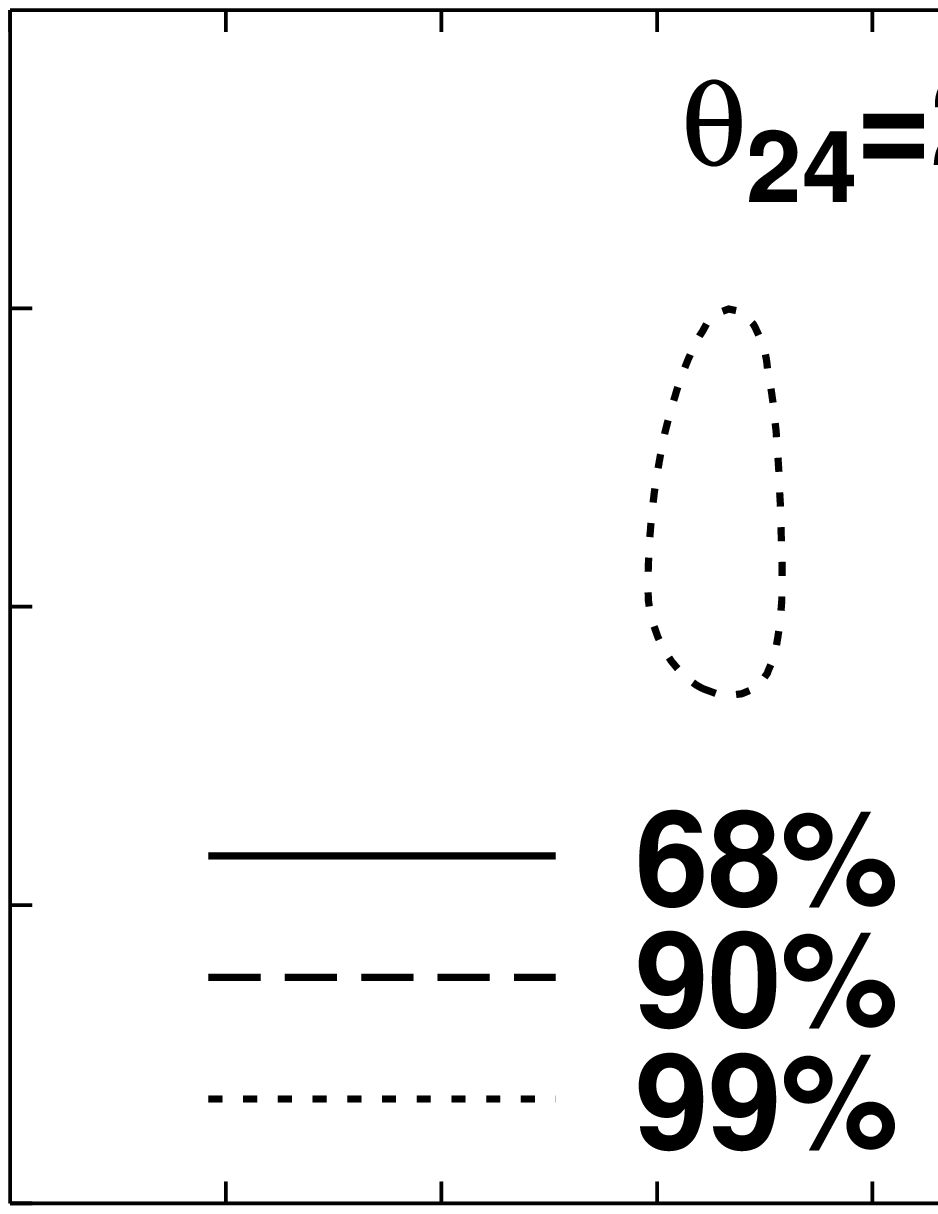,width=10cm}
\vglue -10.1cm \hglue 0.6cm \epsfig{file=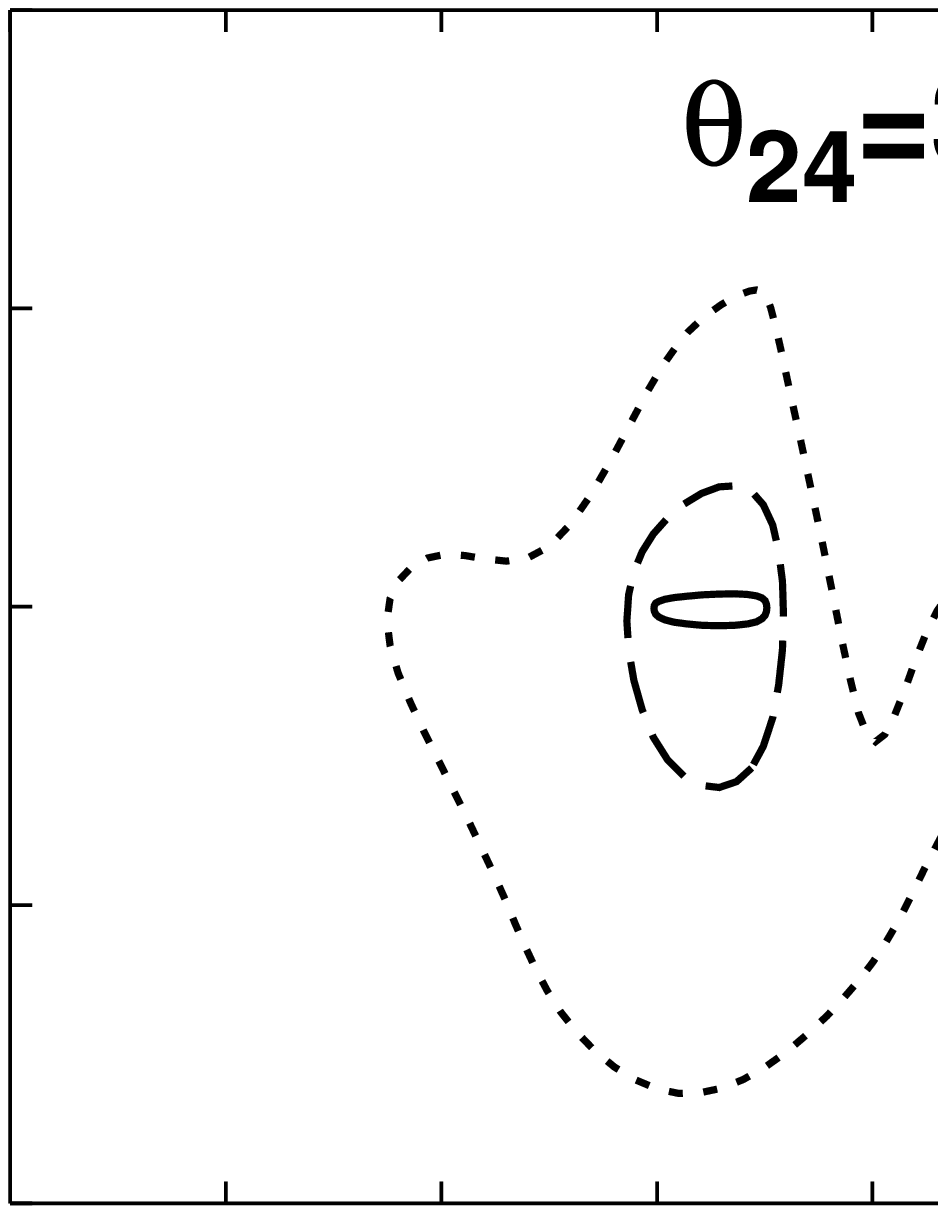,width=10cm}

\vglue -4.3cm
\hglue -5.5cm 
\epsfig{file=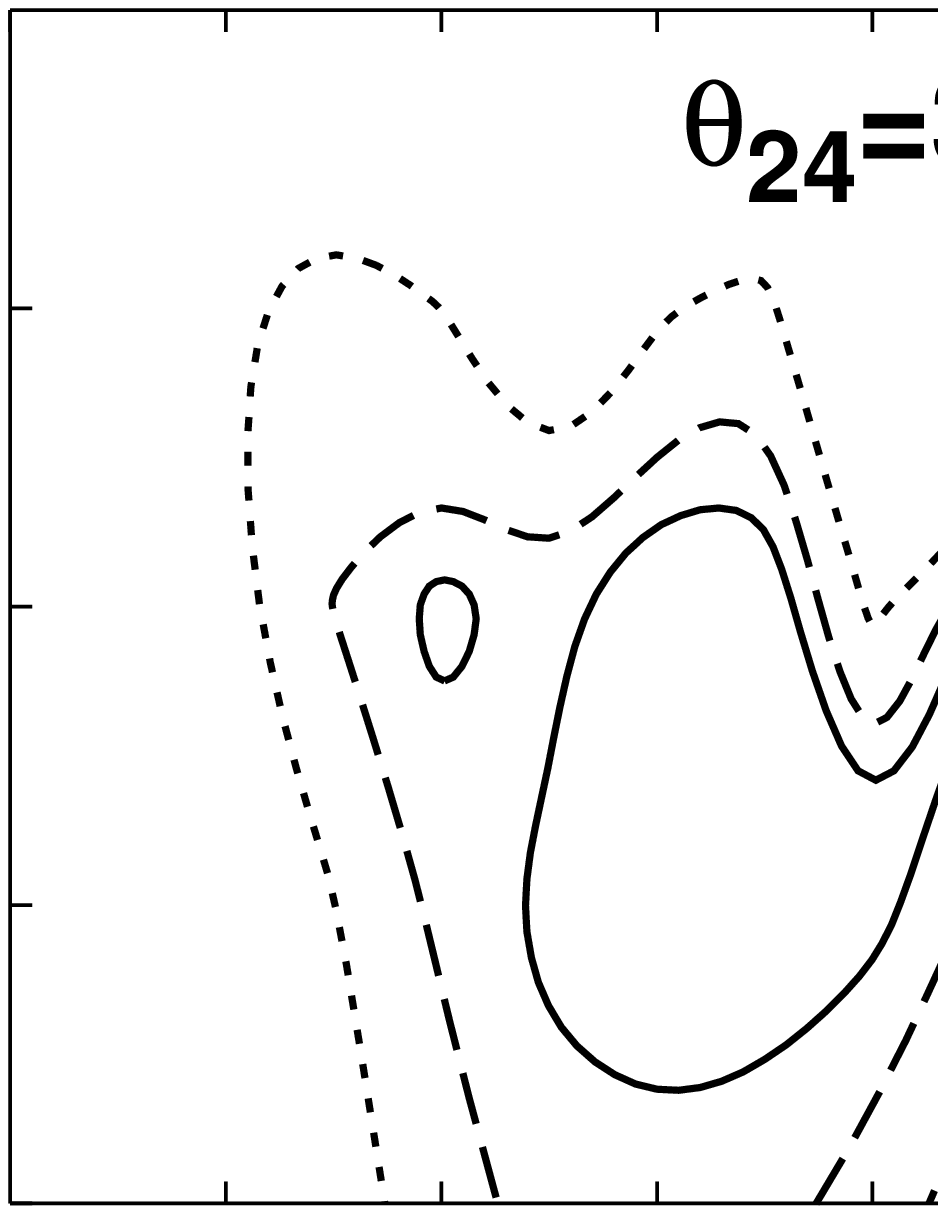,width=10cm}
\vglue -10.1cm \hglue 0.6cm \epsfig{file=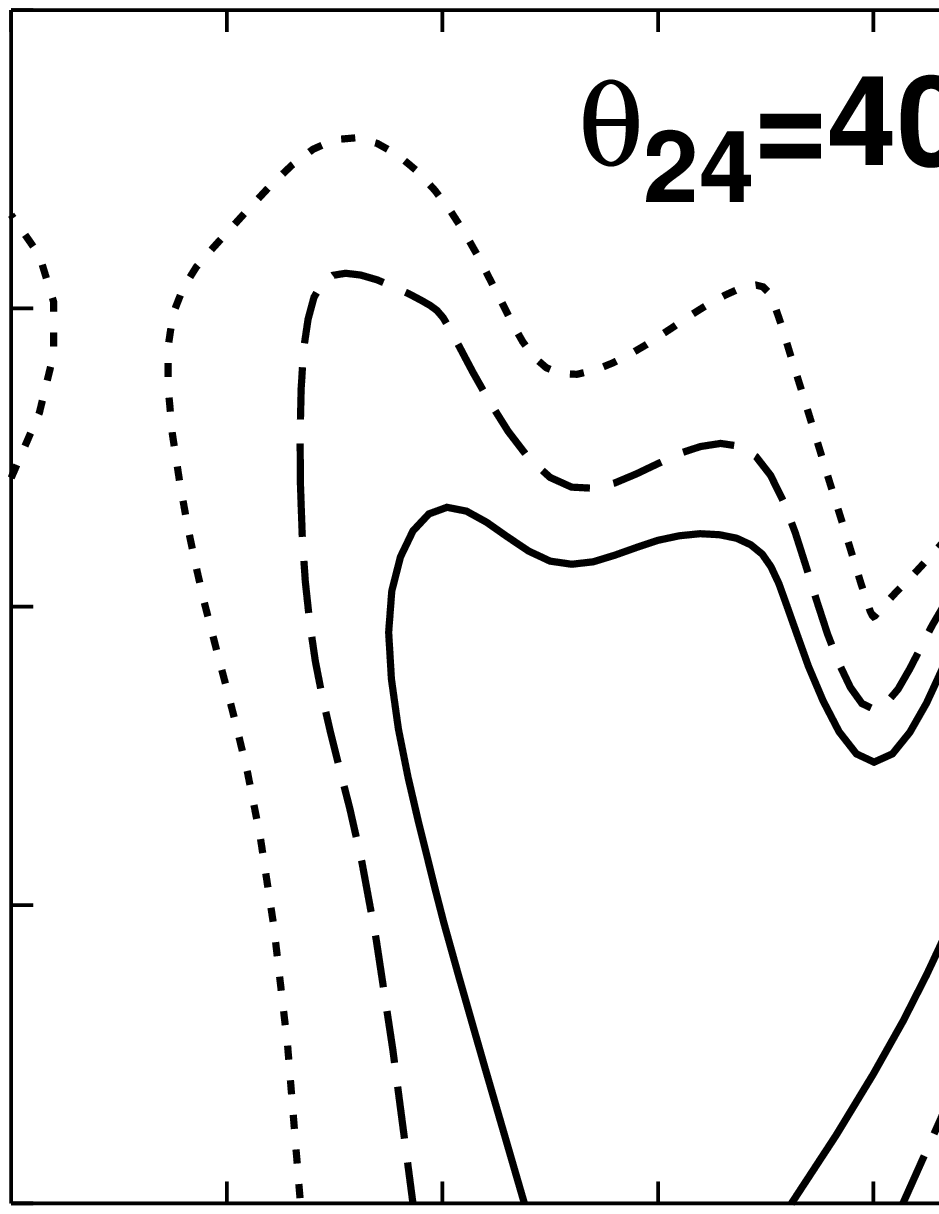,width=10cm}

\vglue -4.3cm
\hglue -5.5cm 
\epsfig{file=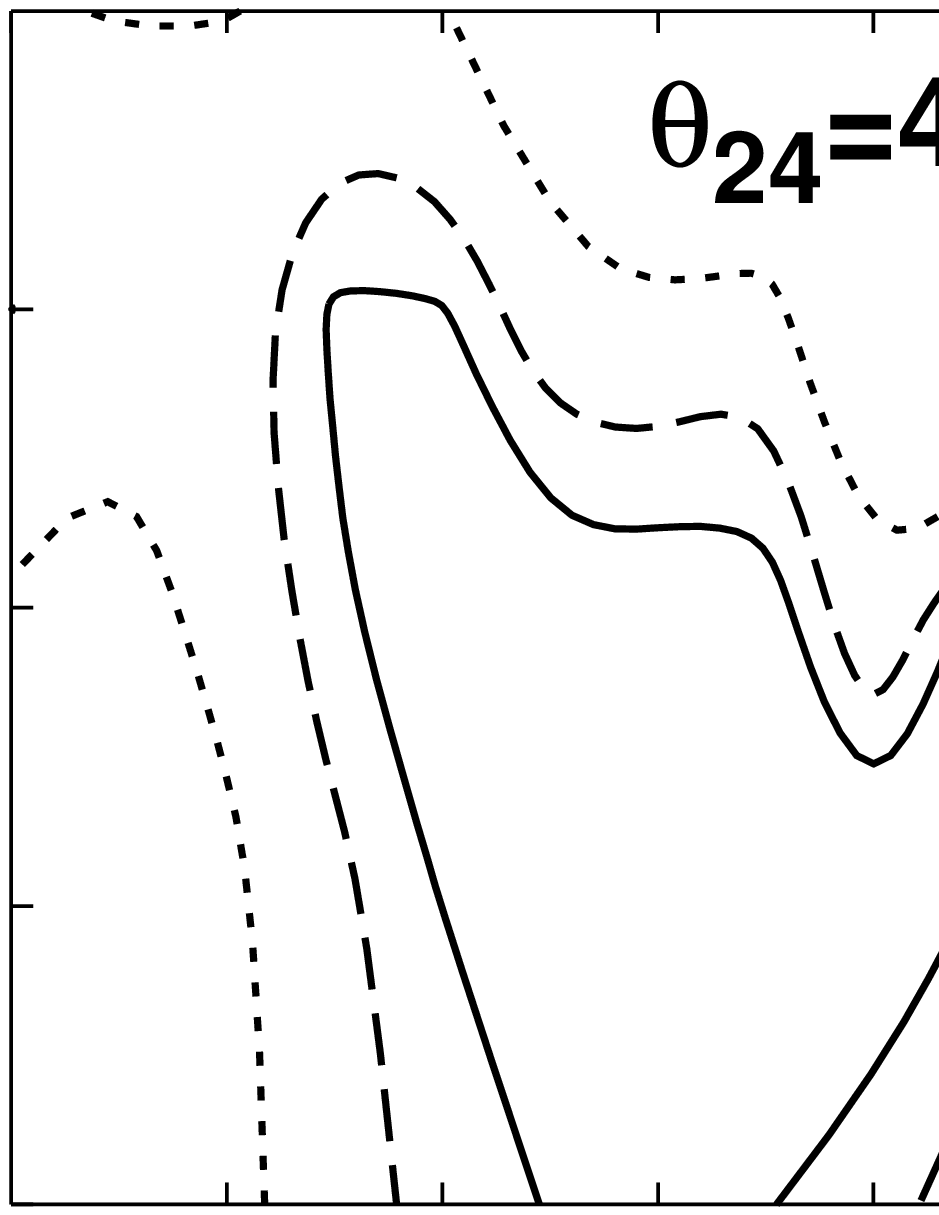,width=10cm}
\vglue -10.1cm \hglue 0.6cm \epsfig{file=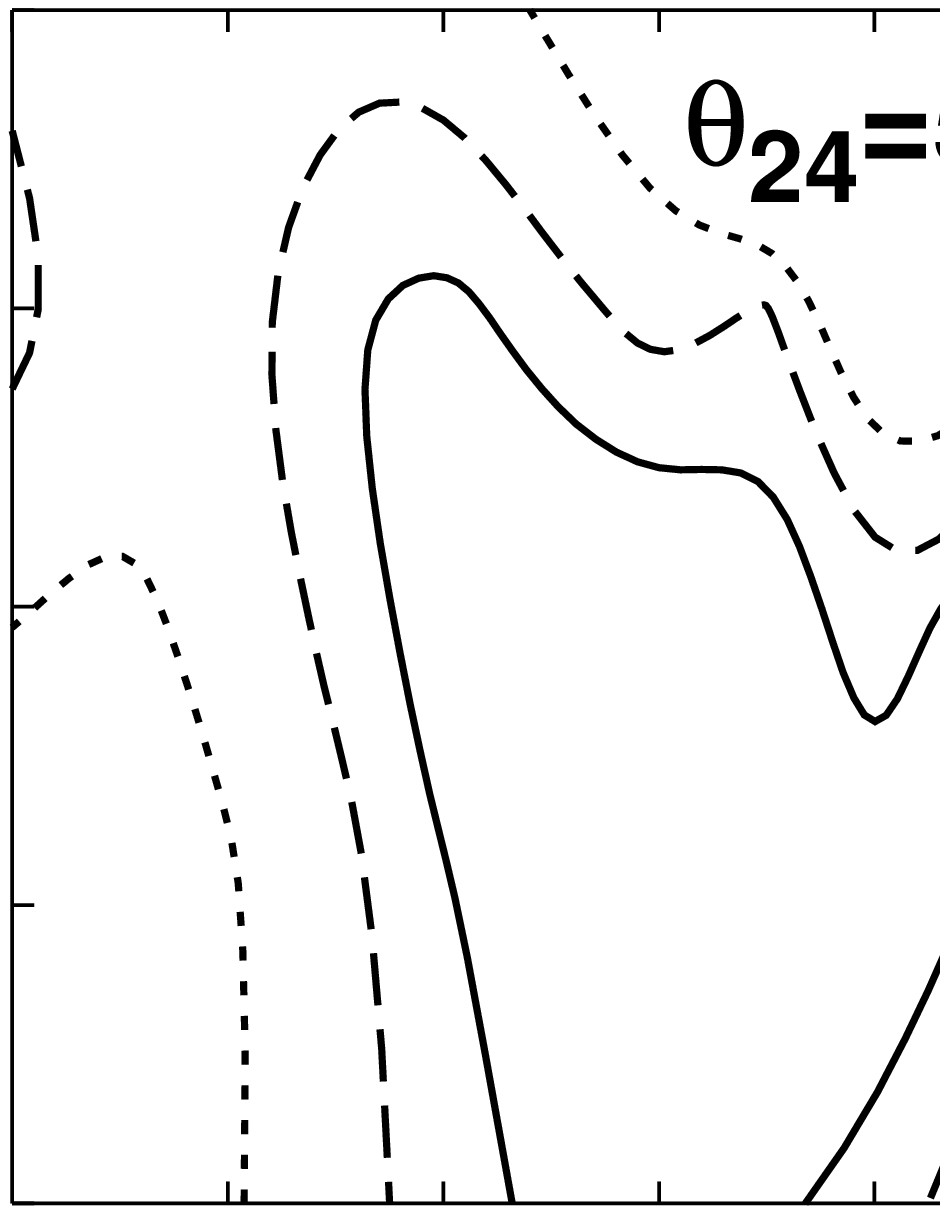,width=10cm}

\vglue -4.3cm
\hglue -5.5cm 
\epsfig{file=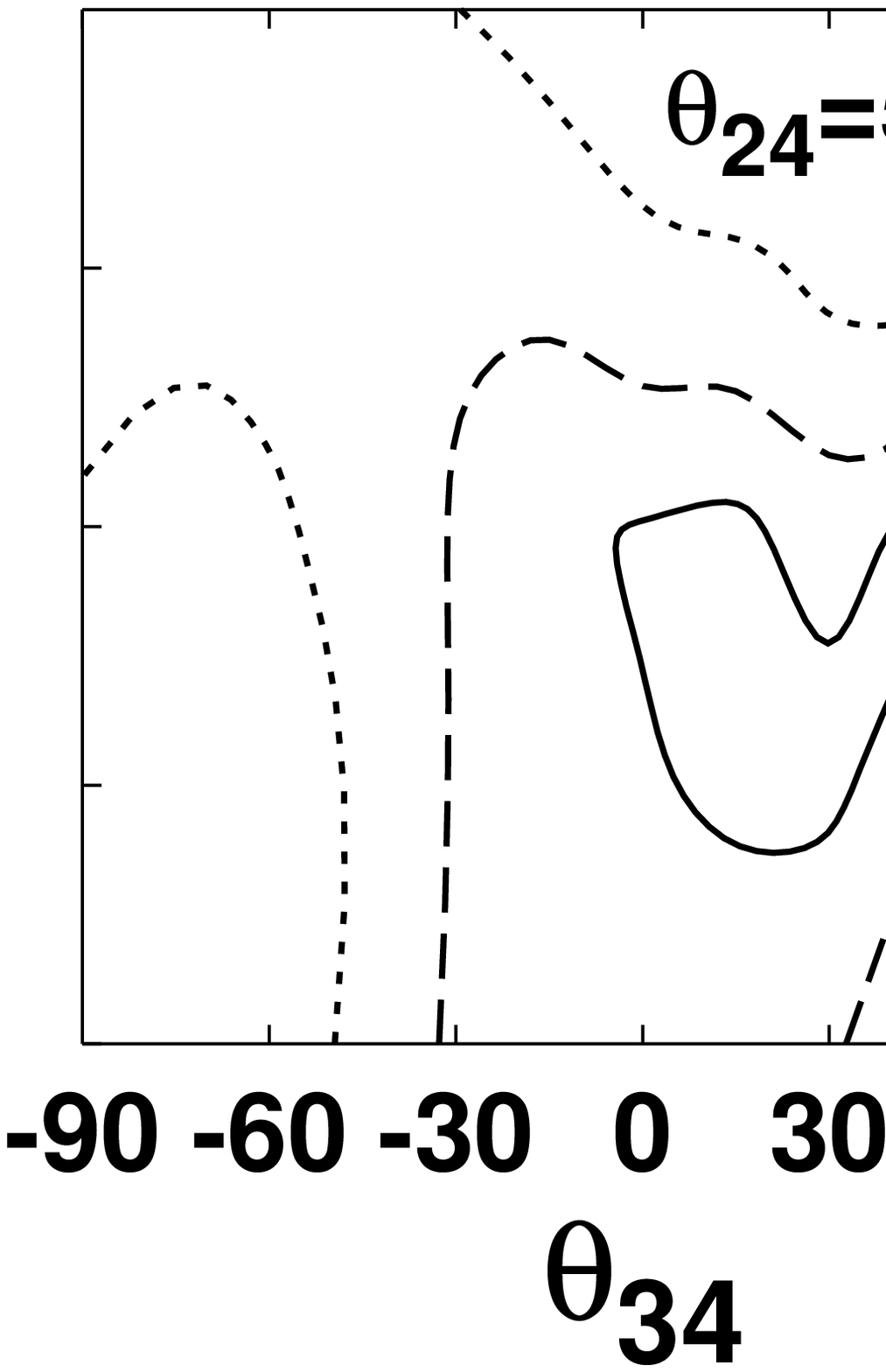,width=10cm}
\vglue -10.1cm \hglue 0.6cm \epsfig{file=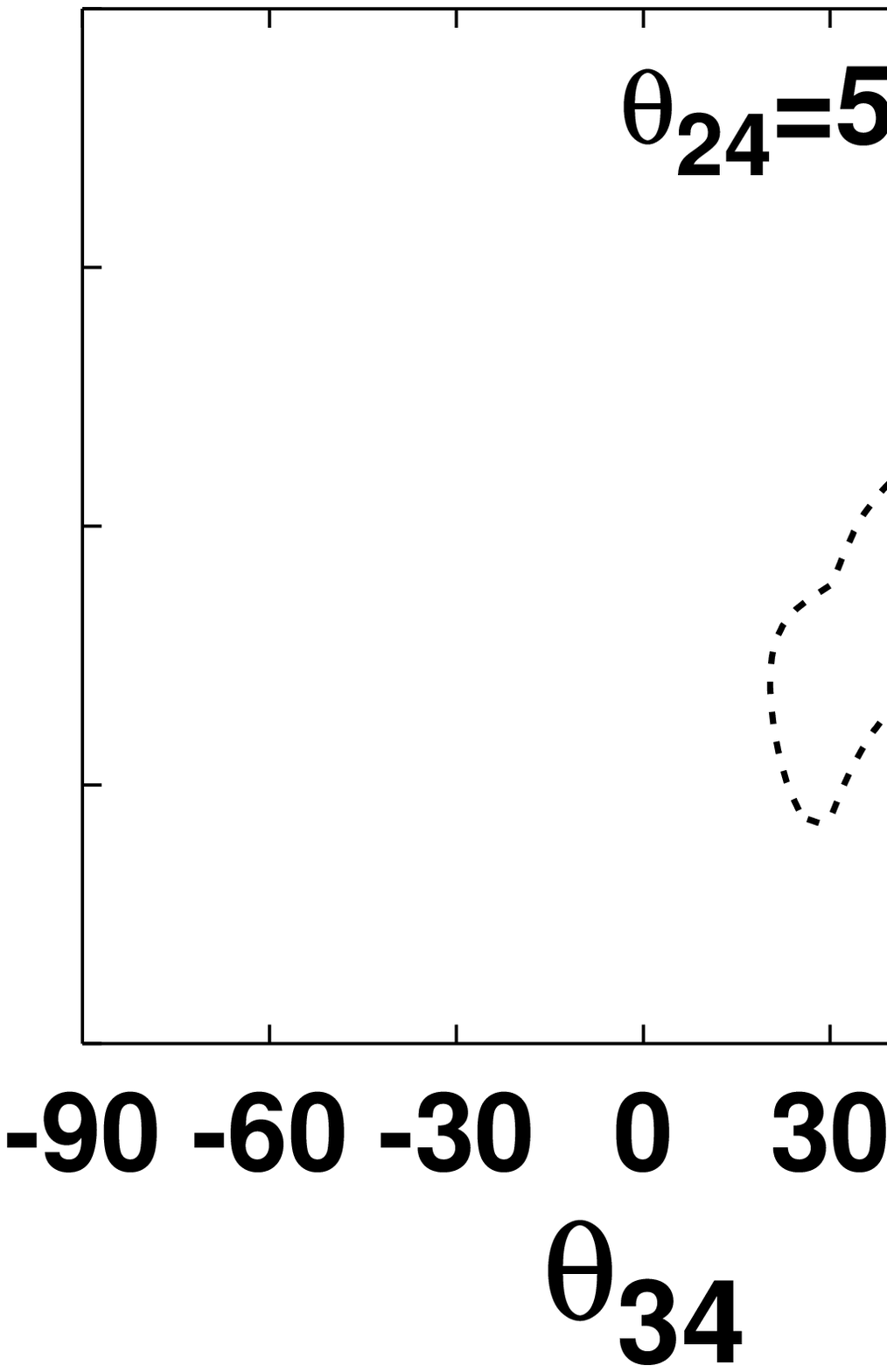,width=10cm}

\vglue -1.0cm
\hglue 6.0cm \epsfig{file=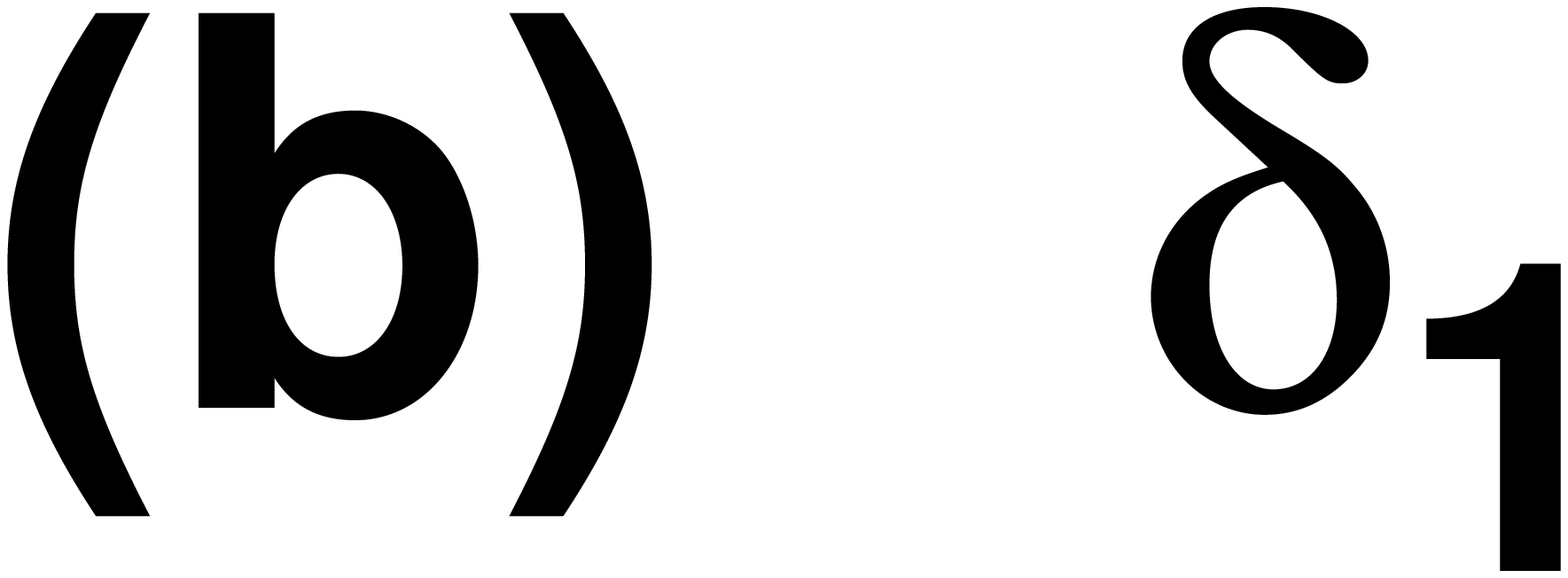,width=3cm}
\newpage
\pagestyle{empty}
\vglue -4.5cm
\hglue -5.5cm 
\epsfig{file=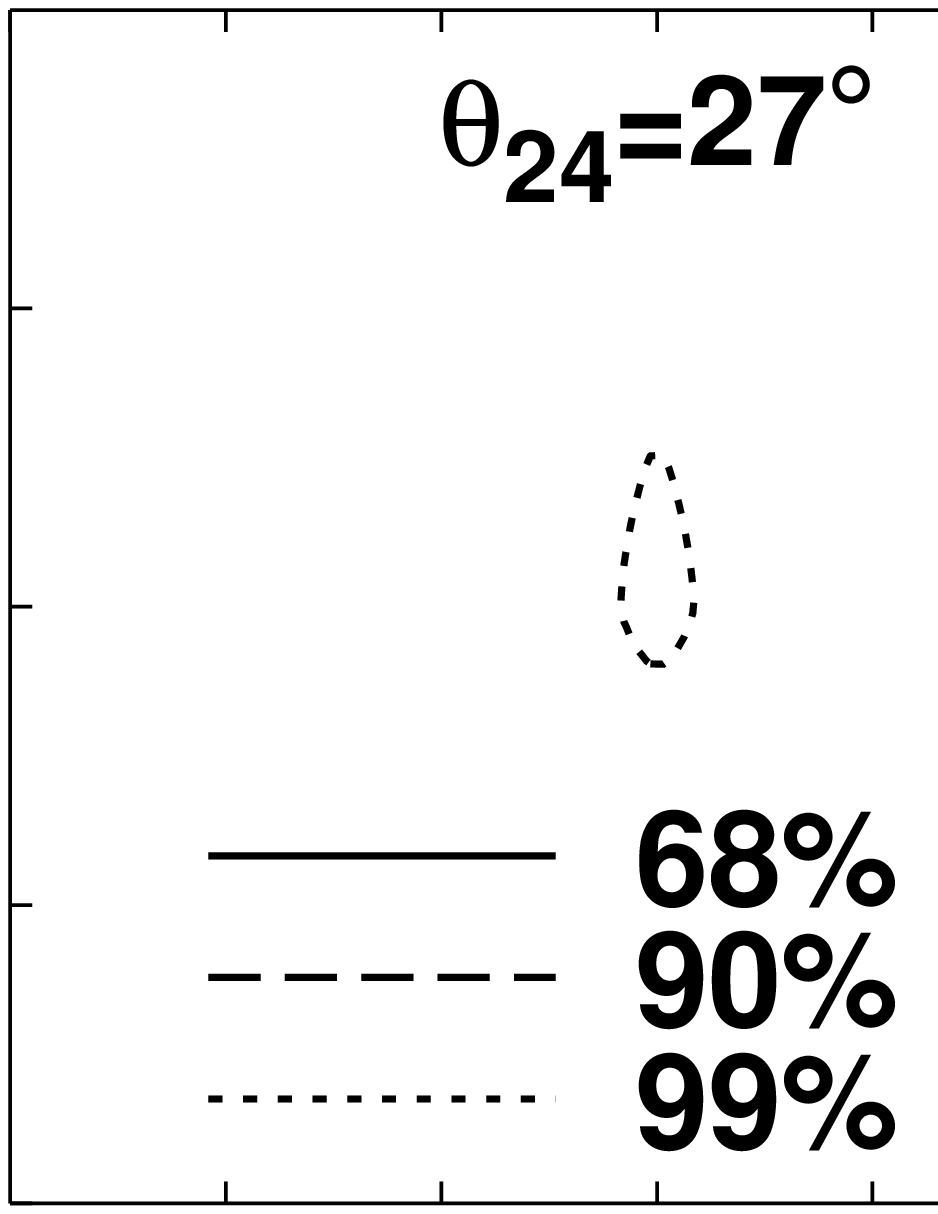,width=10cm}
\vglue -10.1cm \hglue 0.6cm \epsfig{file=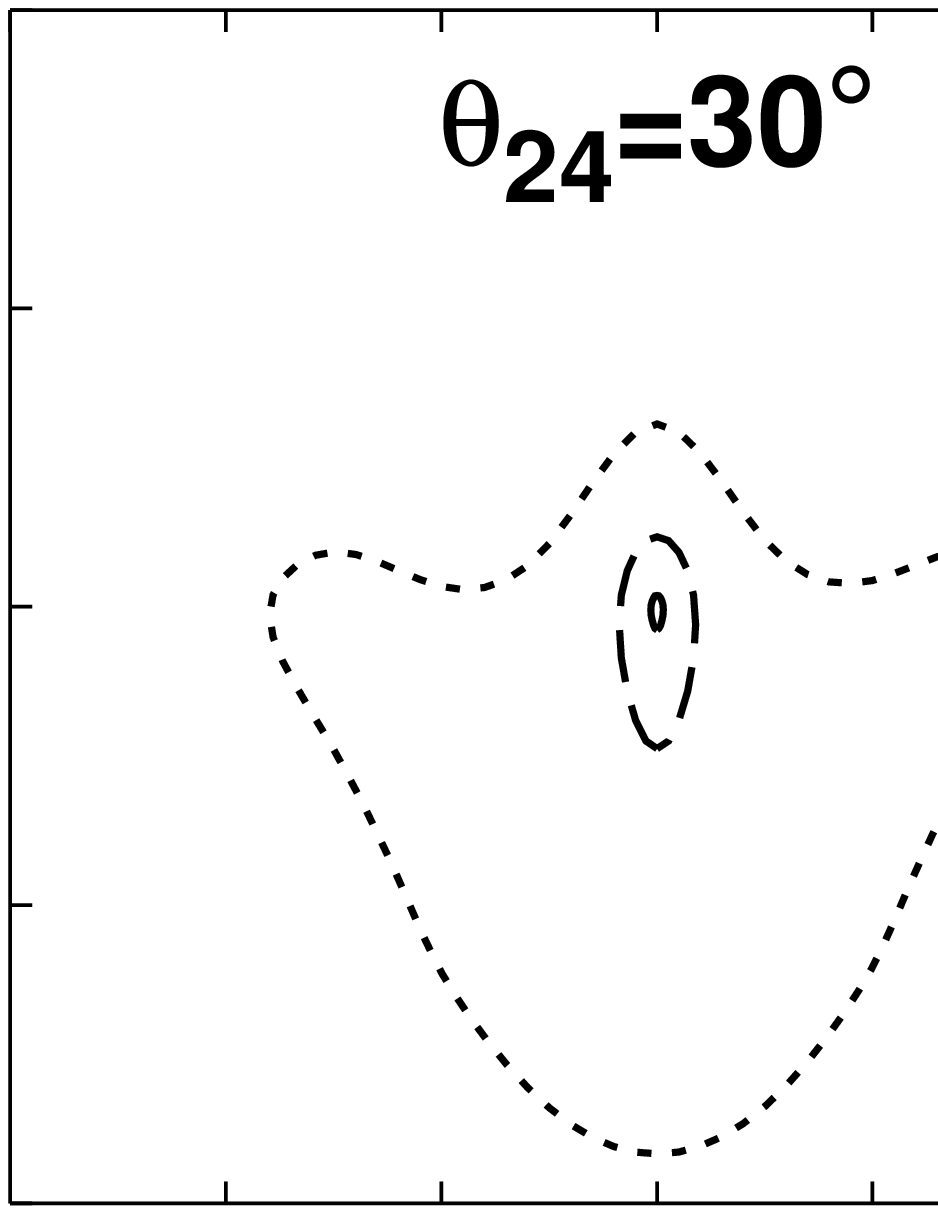,width=10cm}

\vglue -4.3cm
\hglue -5.5cm 
\epsfig{file=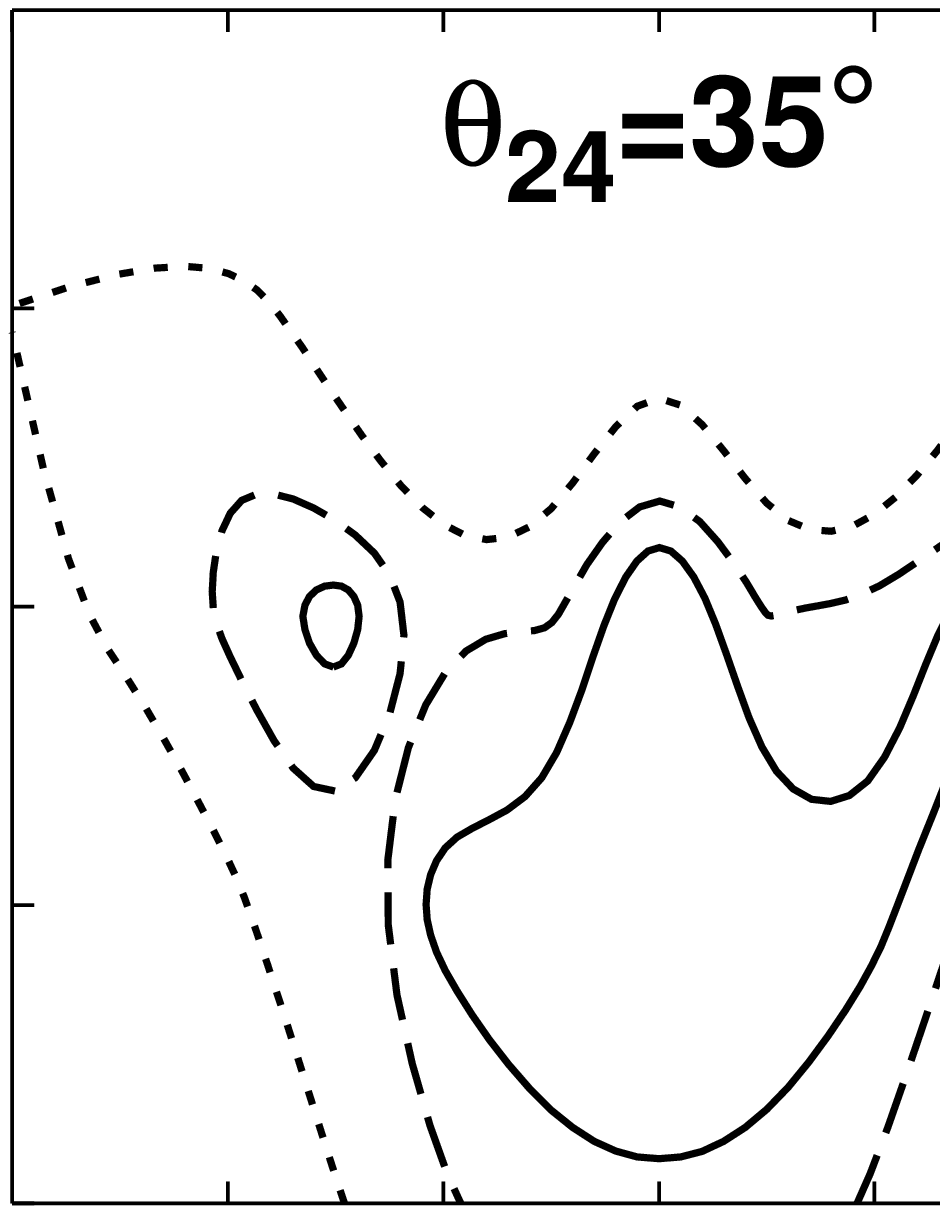,width=10cm}
\vglue -10.1cm \hglue 0.6cm \epsfig{file=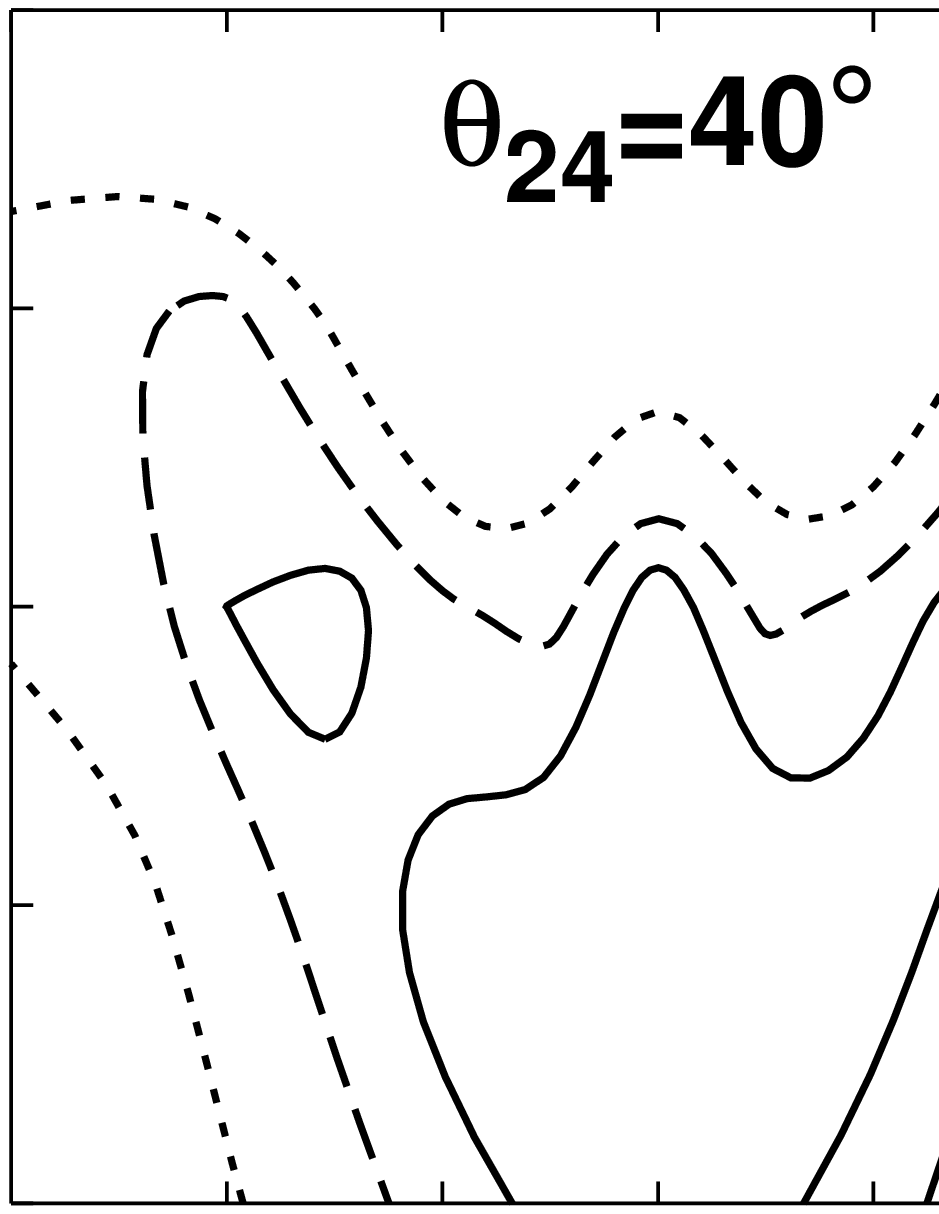,width=10cm}

\vglue -4.3cm
\hglue -5.5cm 
\epsfig{file=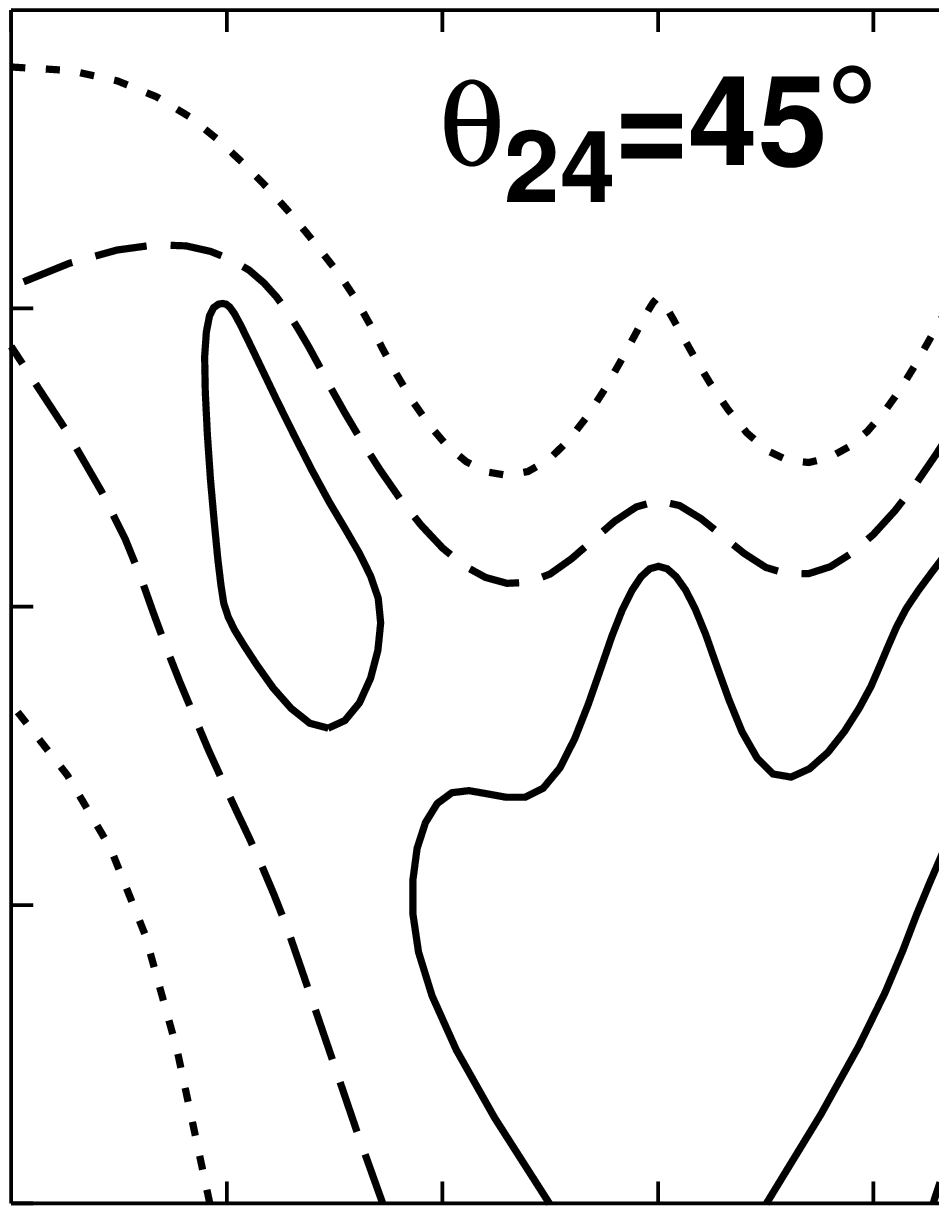,width=10cm}
\vglue -10.1cm \hglue 0.6cm \epsfig{file=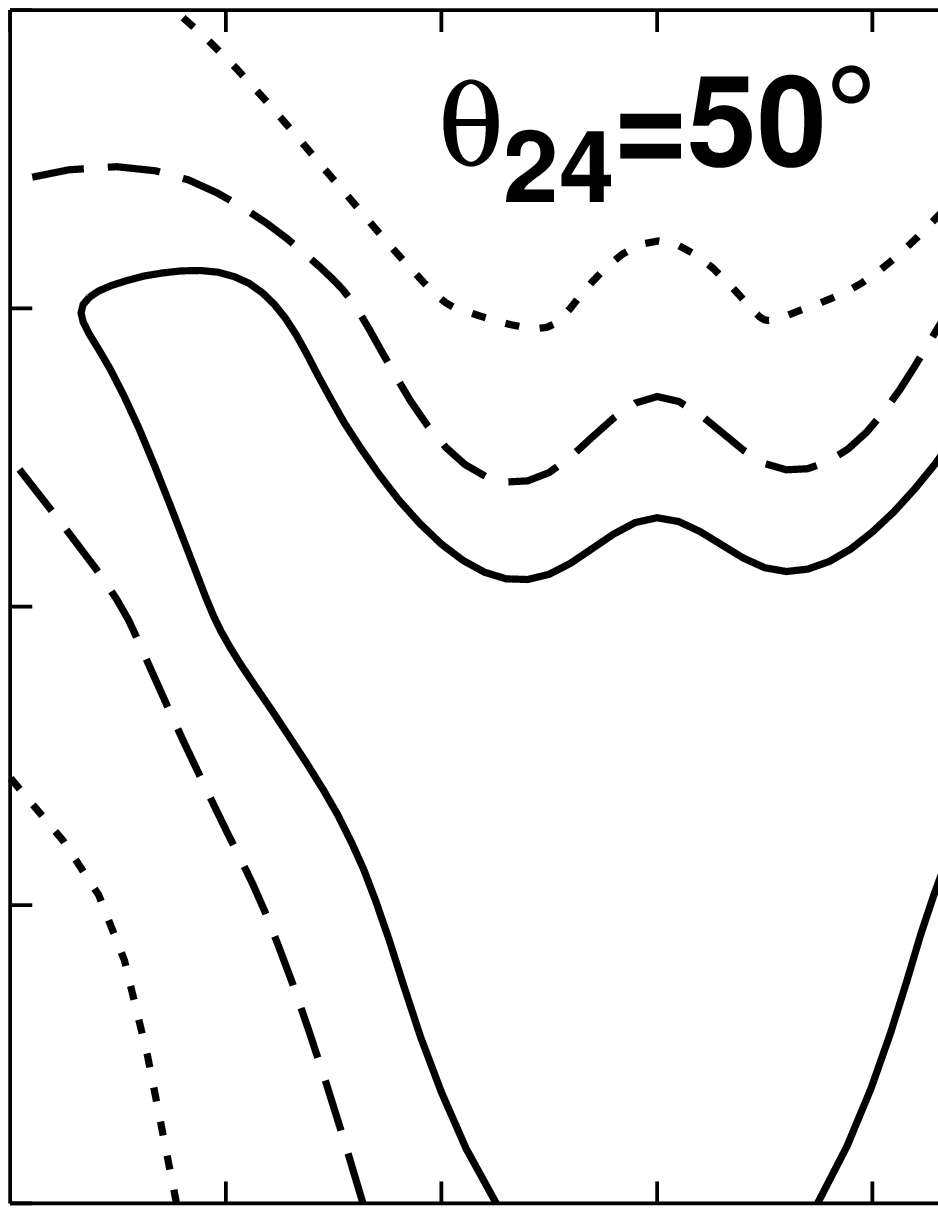,width=10cm}

\vglue -4.3cm
\hglue -5.5cm 
\epsfig{file=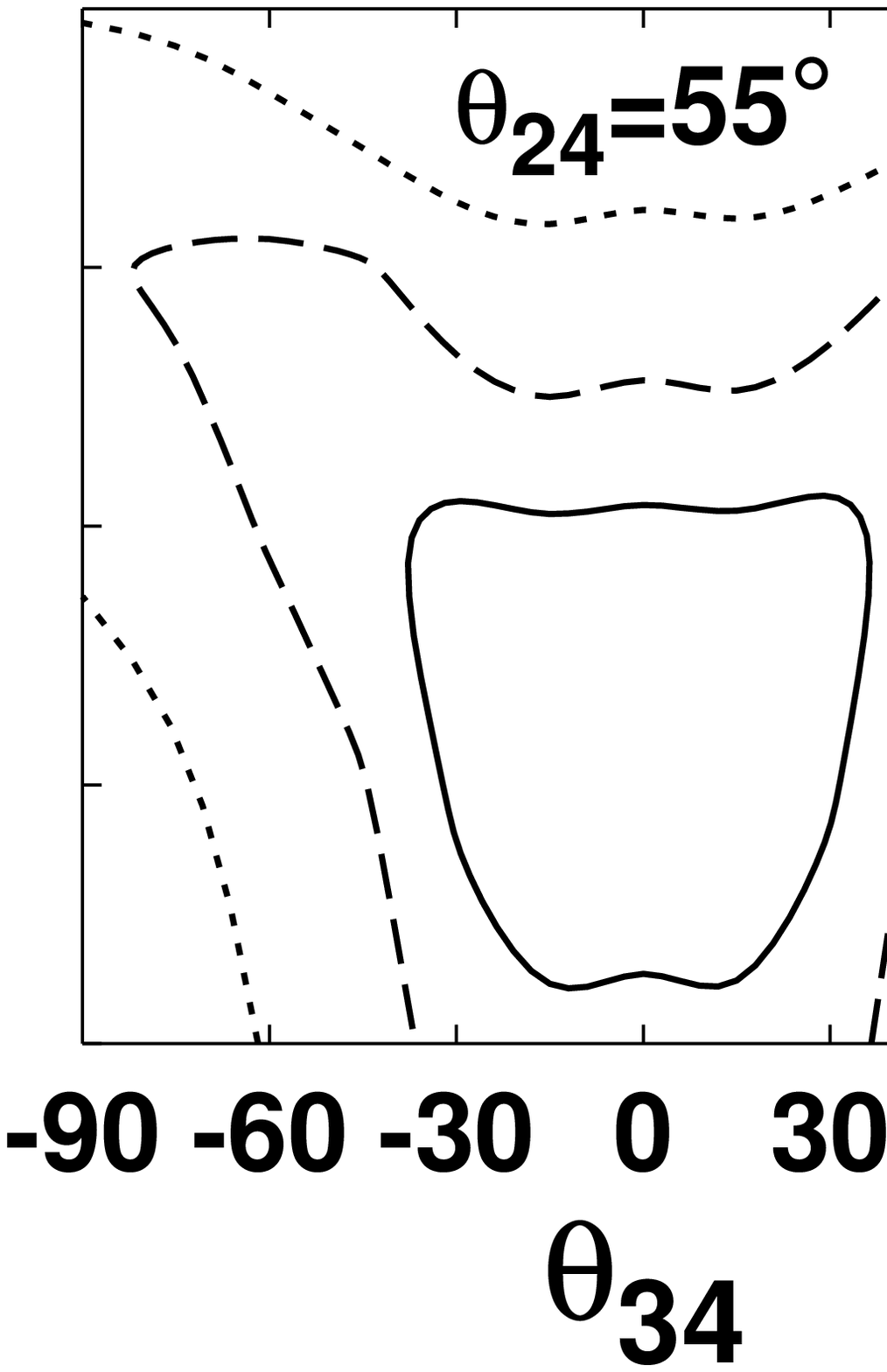,width=10cm}
\vglue -10.1cm \hglue 0.6cm \epsfig{file=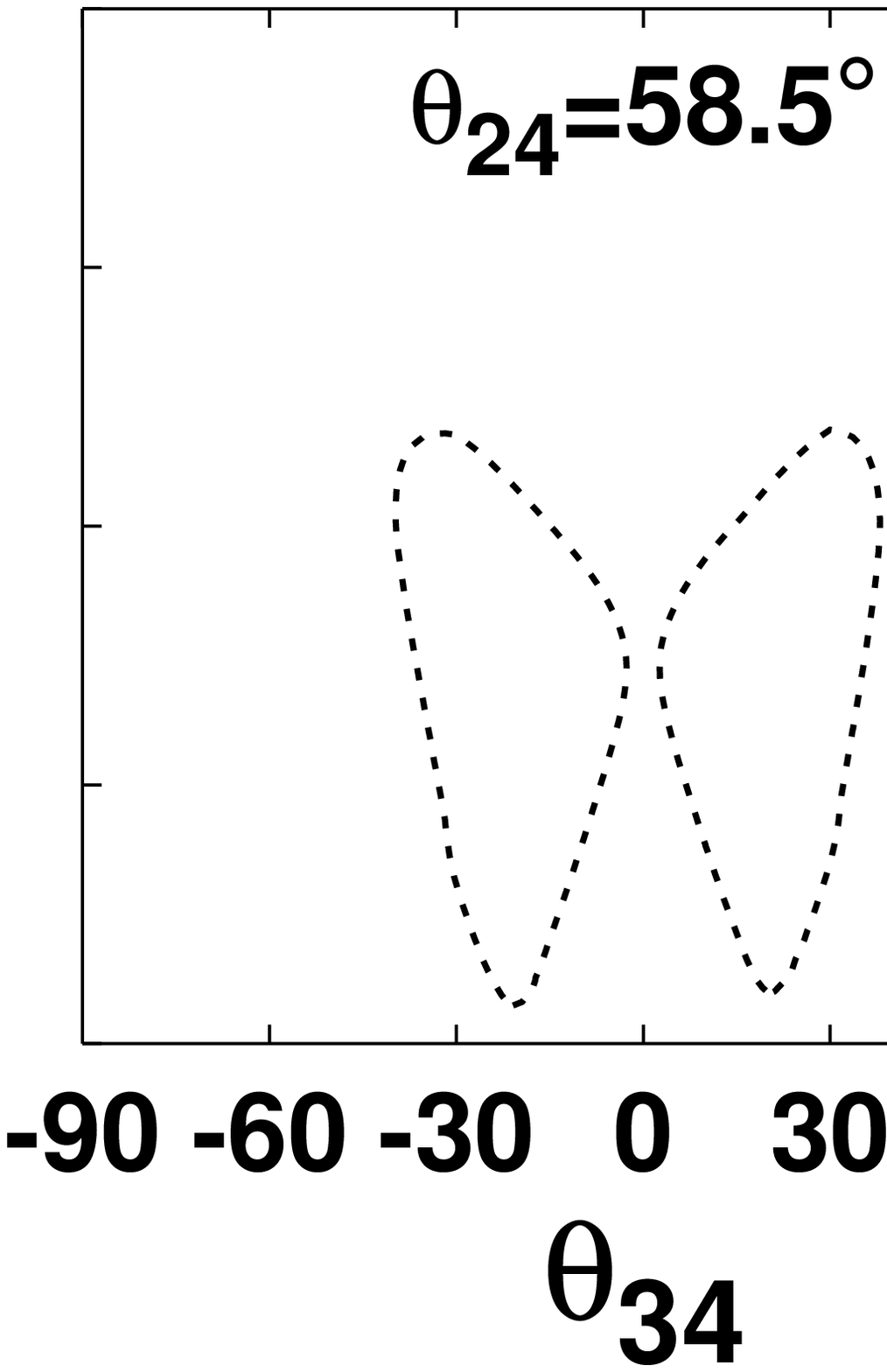,width=10cm}

\vglue -1.0cm
\hglue 6.0cm \epsfig{file=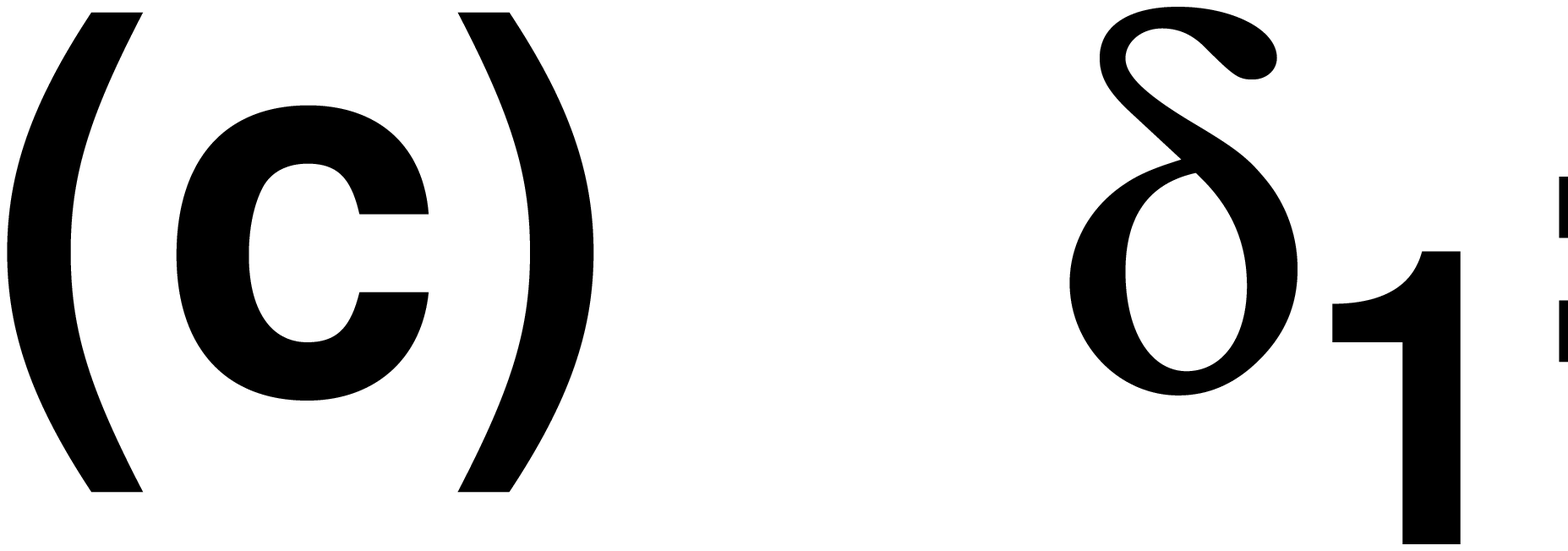,width=3cm}
\newpage
\vglue 2.10cm \hglue 2.5cm
\epsfig{file=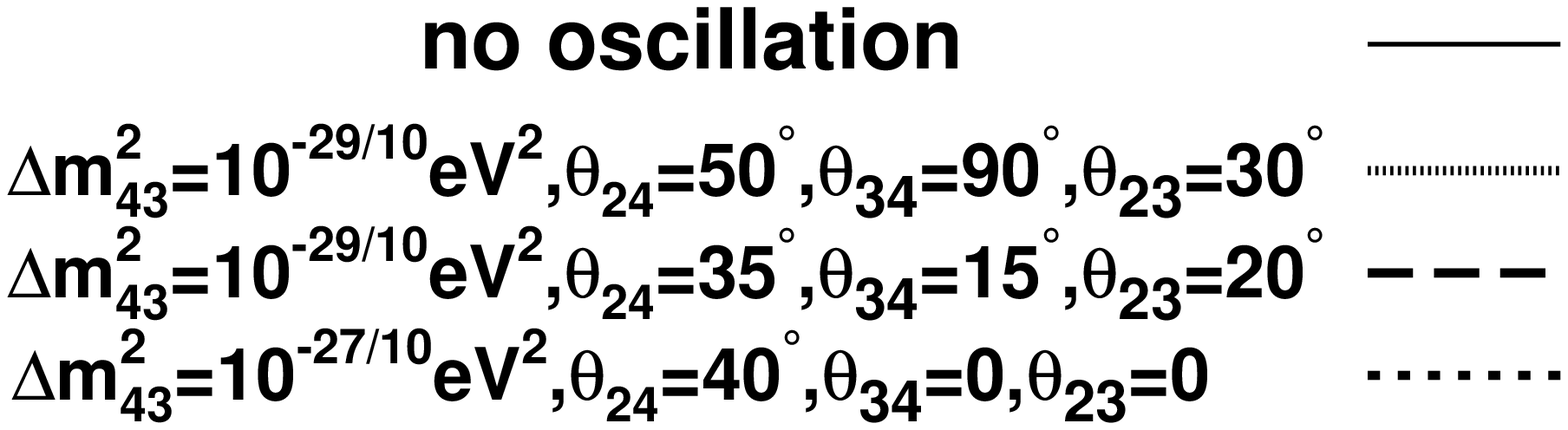,width=10cm}
\vglue -11.0cm \hglue -0.5cm
\epsfig{file=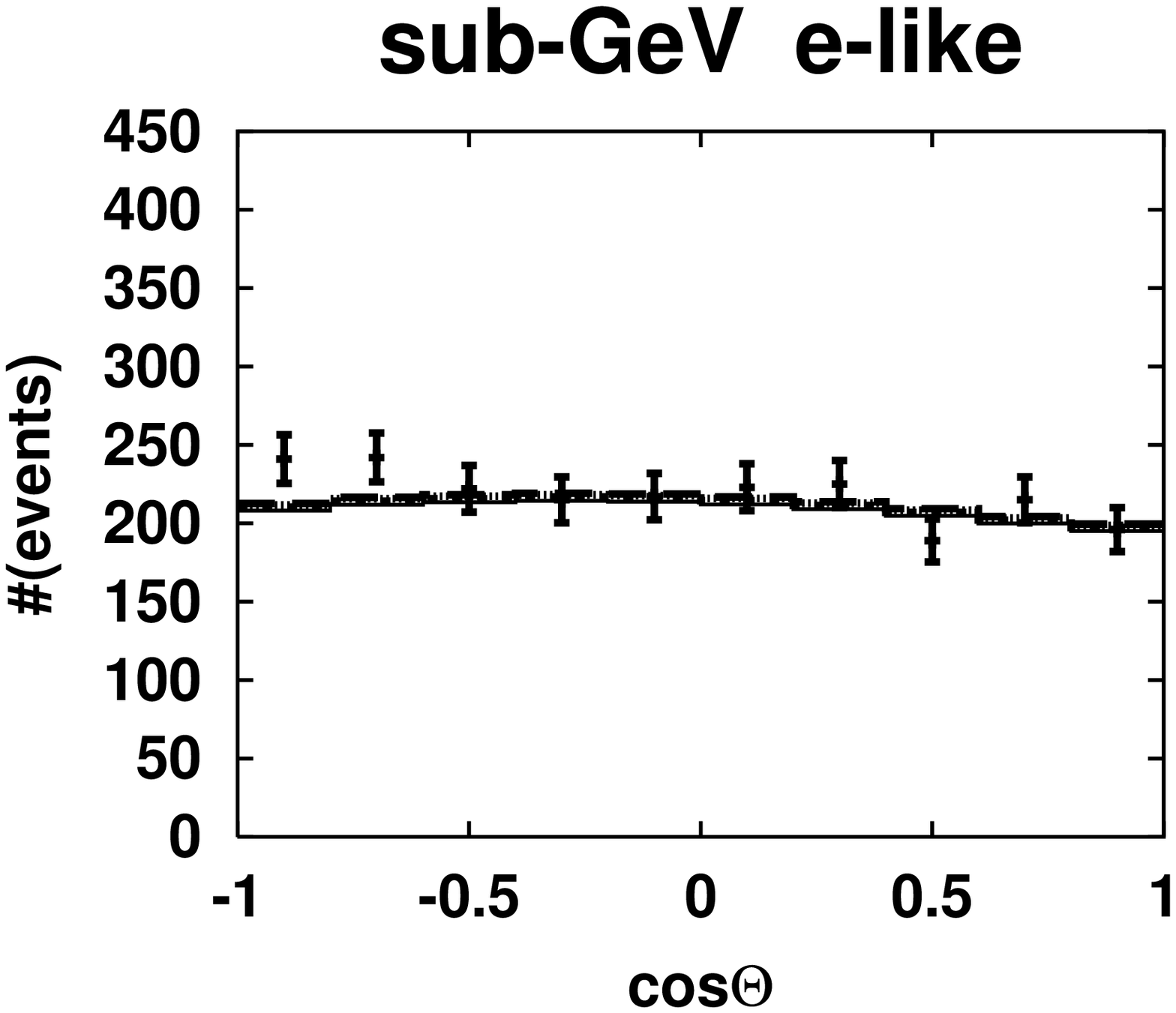,width=8cm}
\vglue -9.25cm \hglue 7.8cm
\epsfig{file=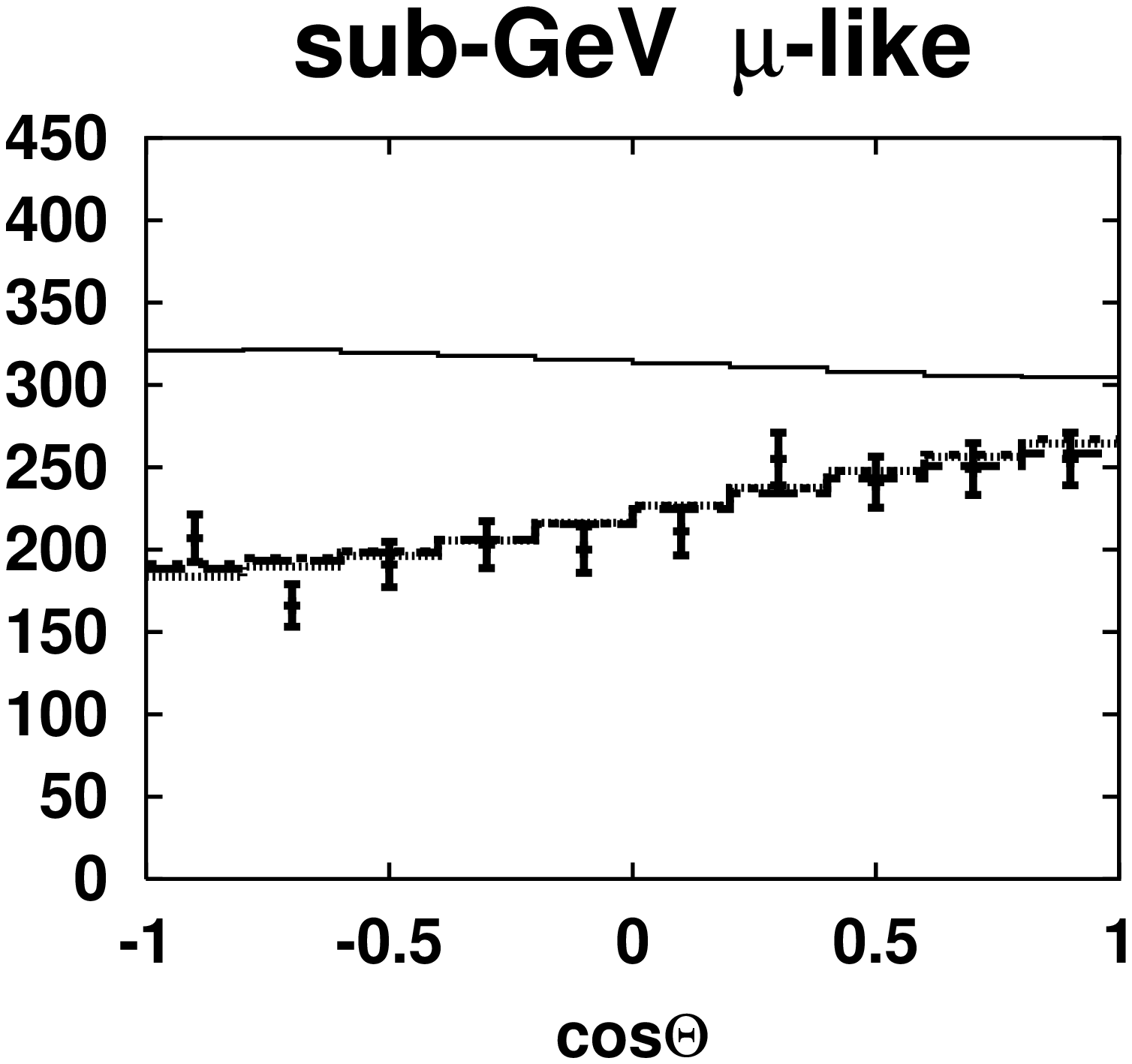,width=8cm}
\vglue -0.5cm \hglue -0.5cm
\epsfig{file=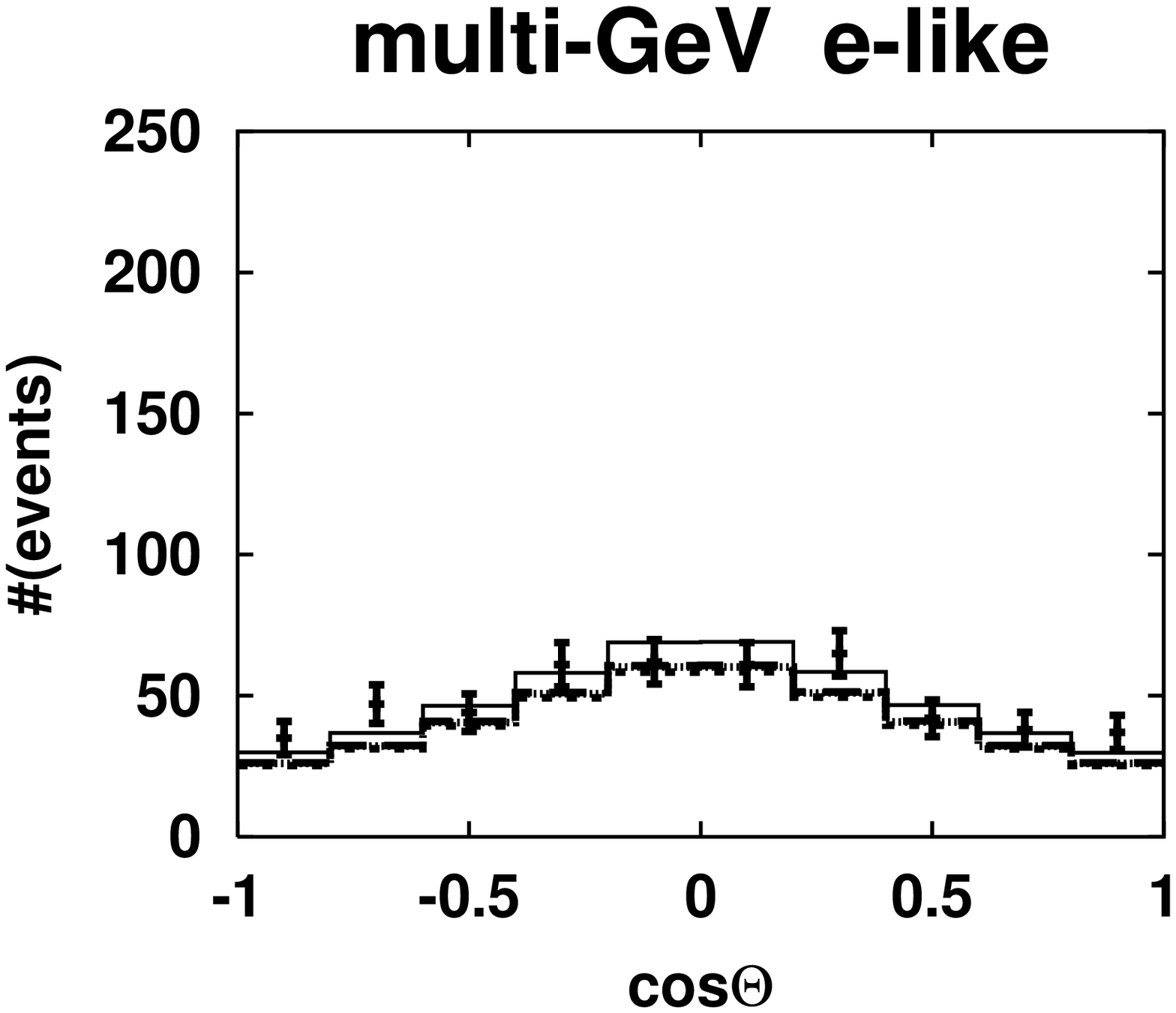,width=8cm}
\vglue -9.25cm \hglue 7.8cm
\epsfig{file=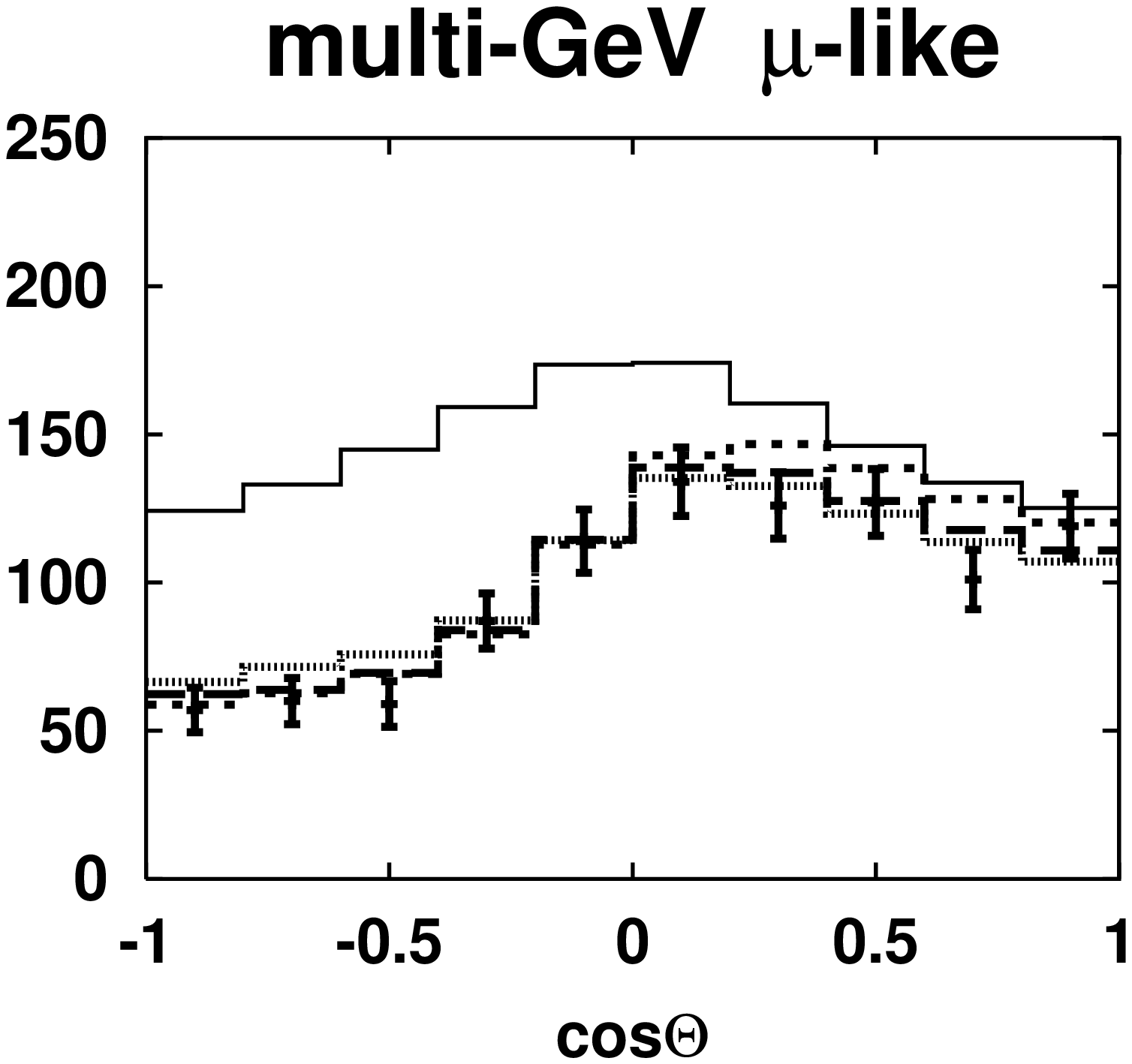,width=8cm}
\vglue 0.0cm
\hglue 6.0cm \epsfig{file=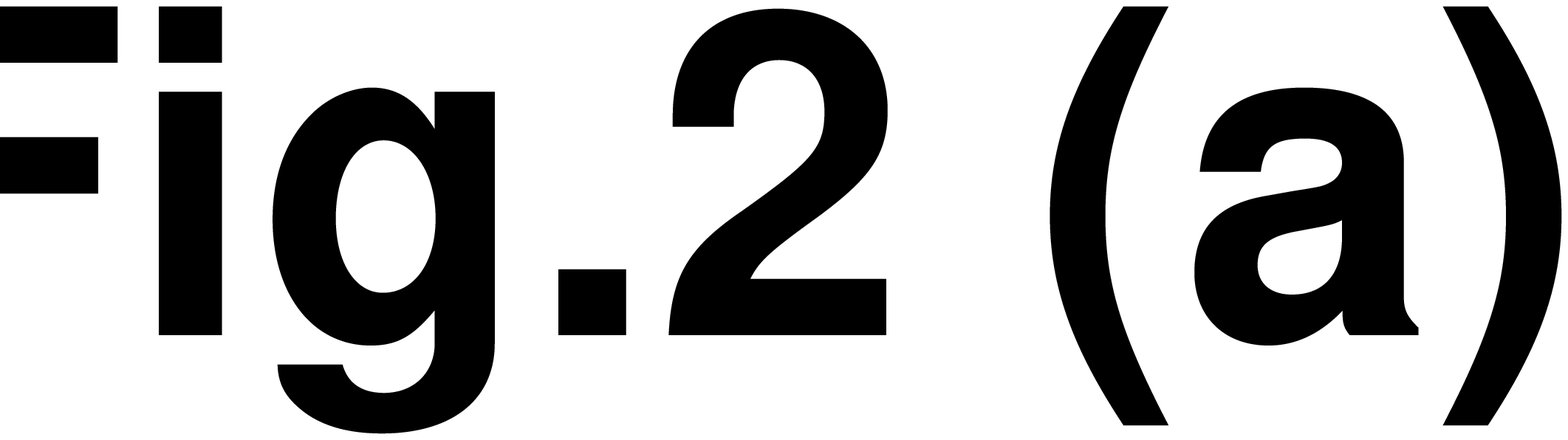,width=3cm}
\newpage
\vglue 6.10cm \hglue 2.0cm
\epsfig{file=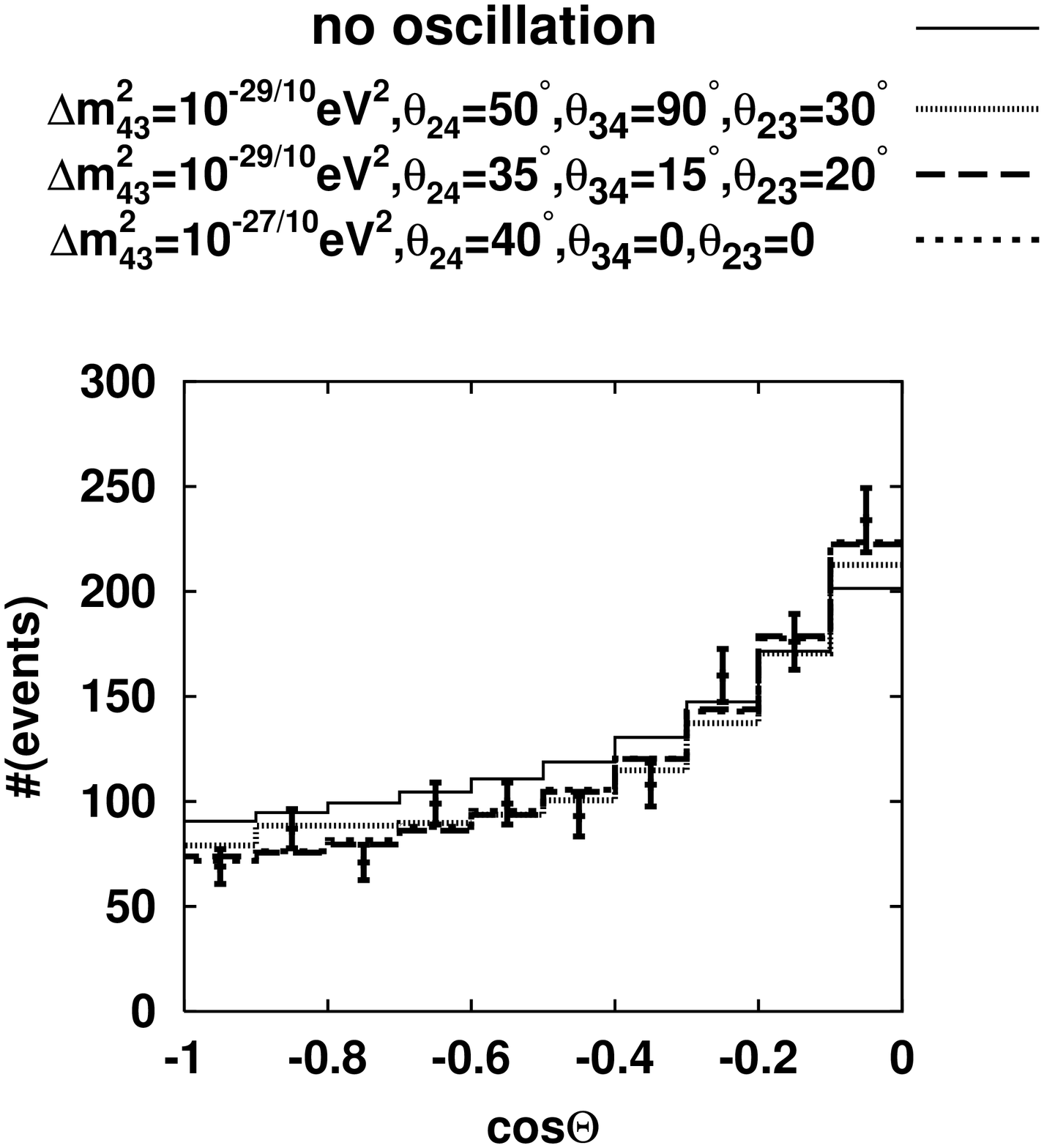,width=12cm}
\vglue 0.5cm
\hglue 6.0cm \epsfig{file=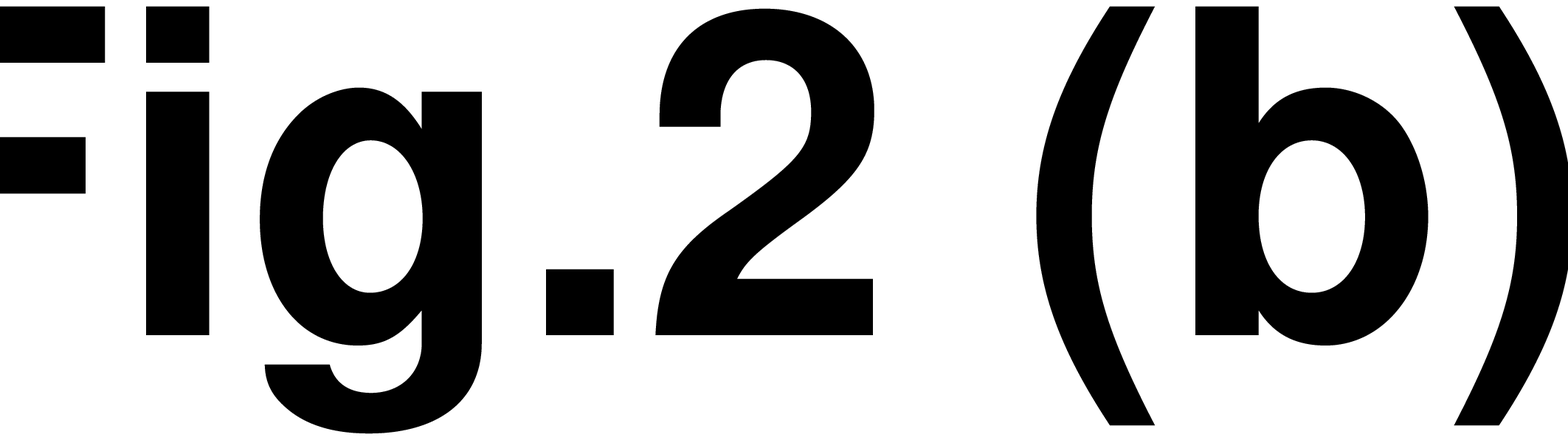,width=3cm}
\newpage
\pagestyle{empty}
\vglue -2.5cm
\hglue -6cm 
\epsfig{file=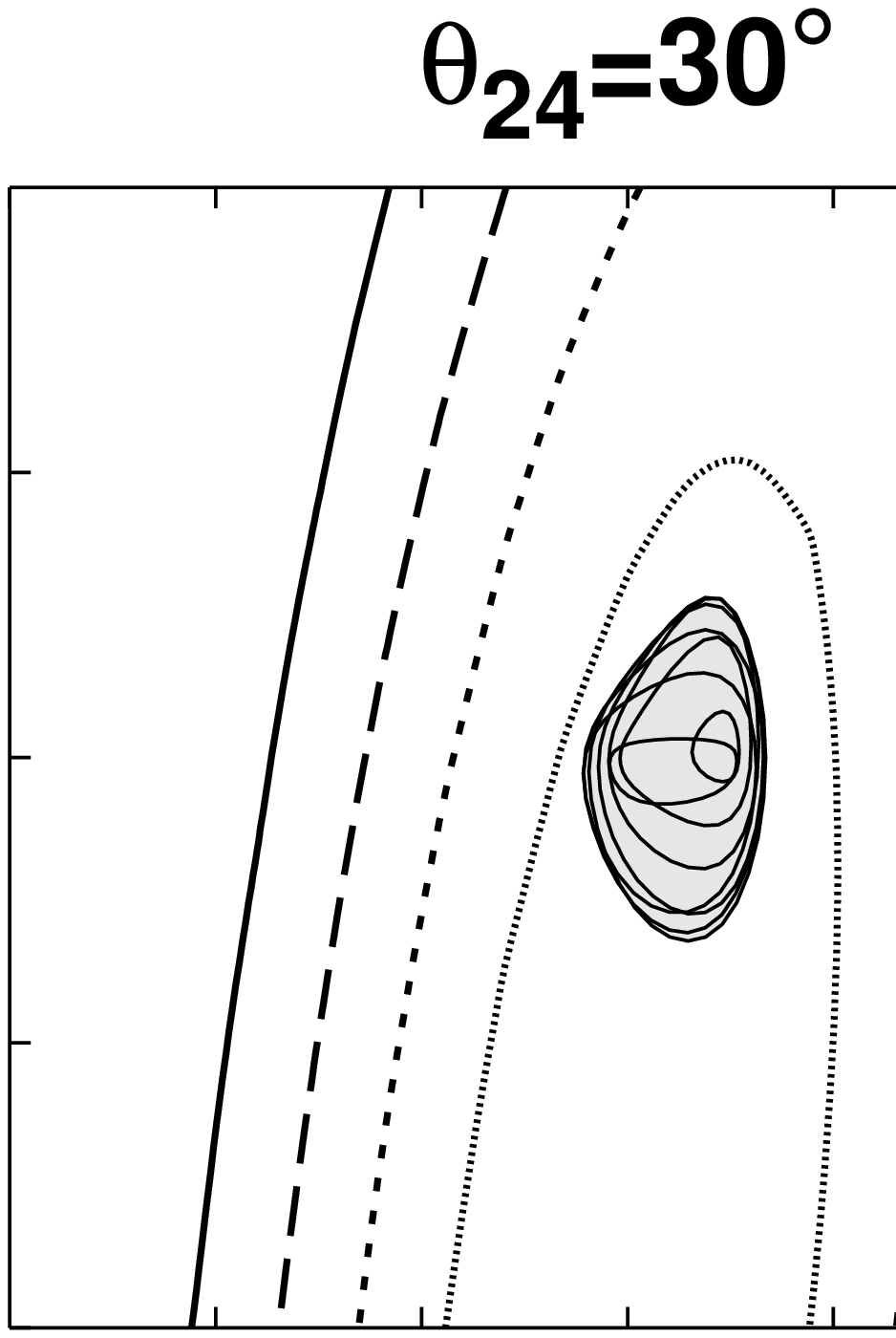,width=10cm}
\vglue -10.1cm \hglue 1cm \epsfig{file=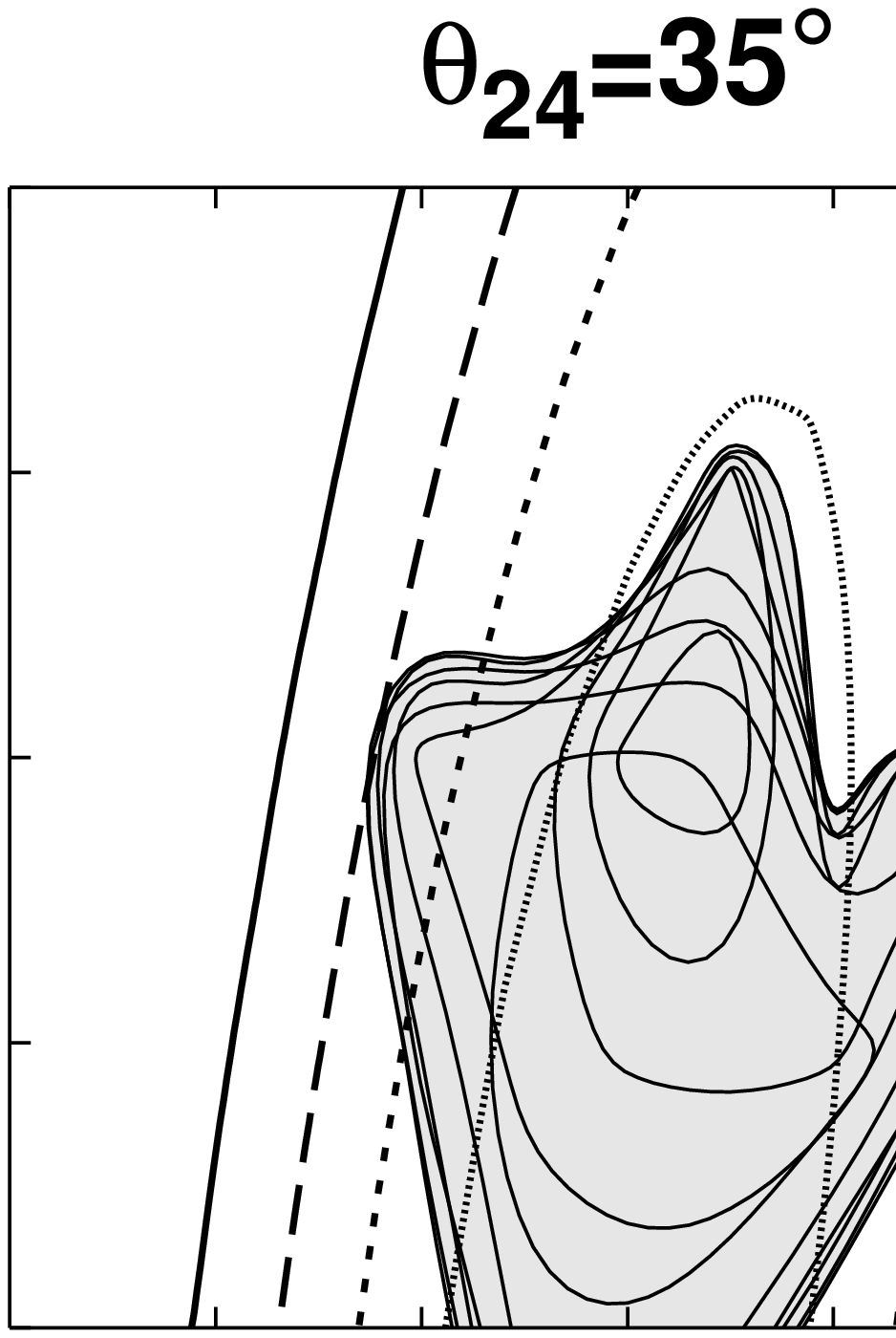,width=10cm}

\vglue -3.8cm
\hglue -6cm 
\epsfig{file=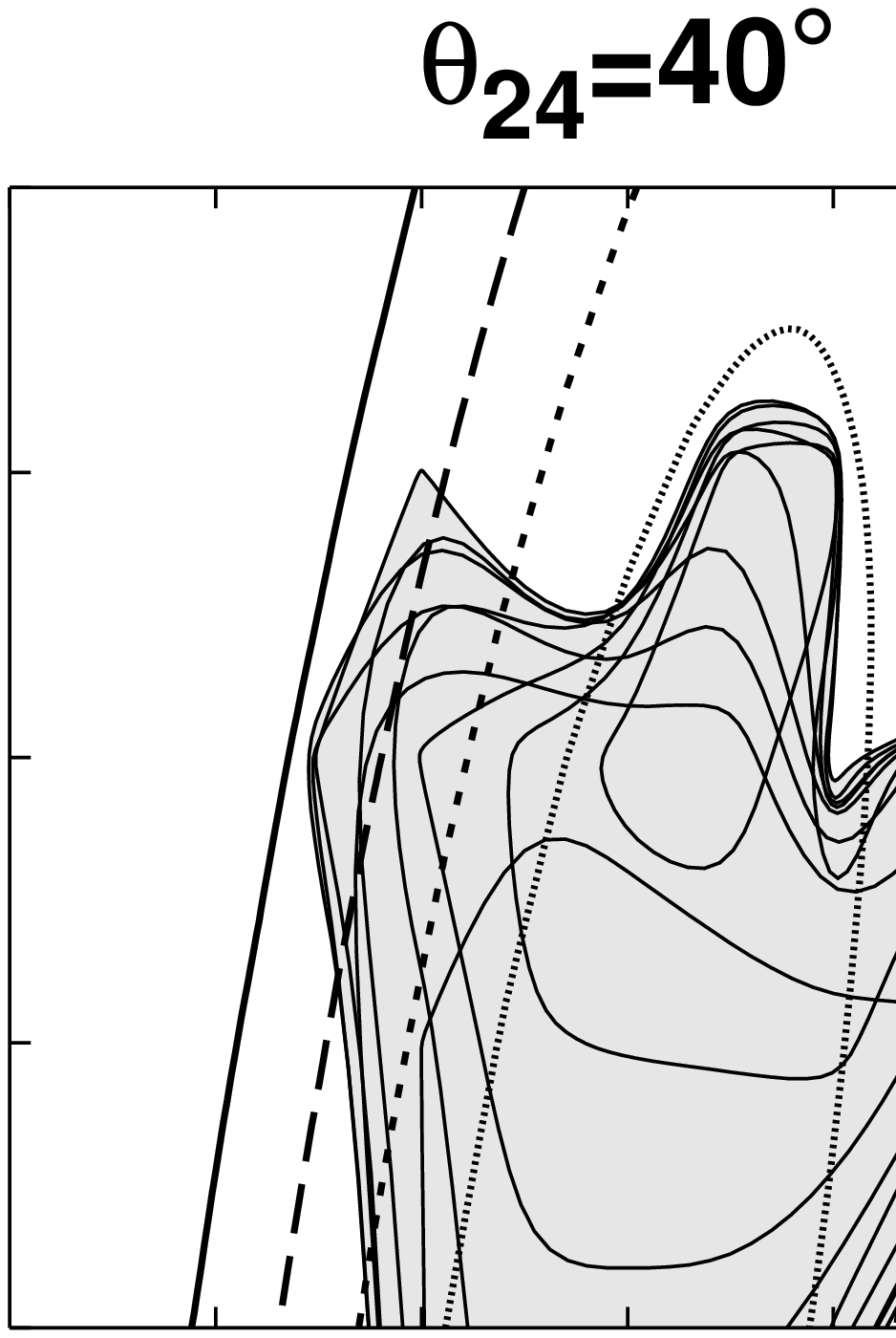,width=10cm}
\vglue -10cm \hglue 1cm \epsfig{file=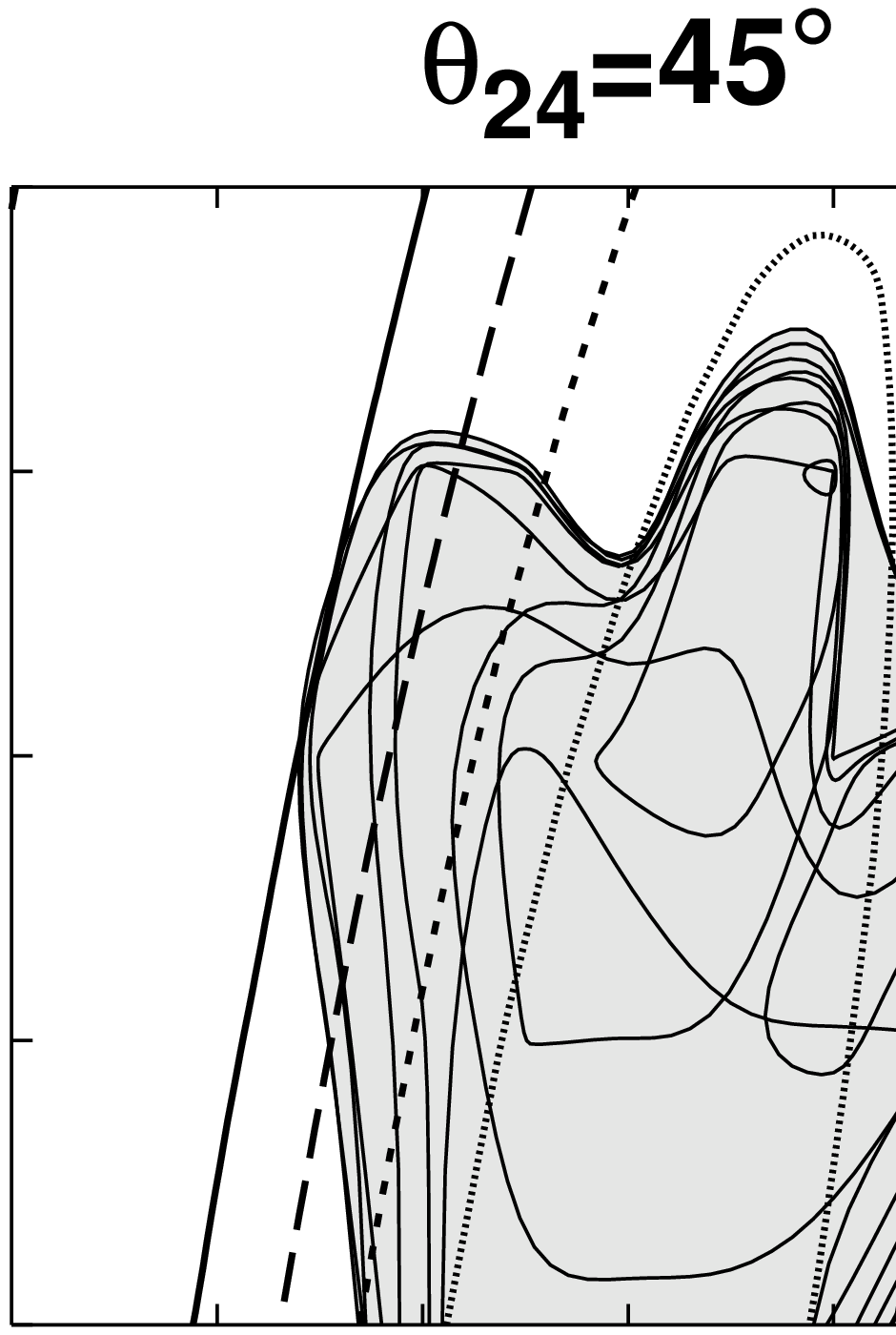,width=10cm}

\vglue -3.8cm
\hglue -6cm 
\epsfig{file=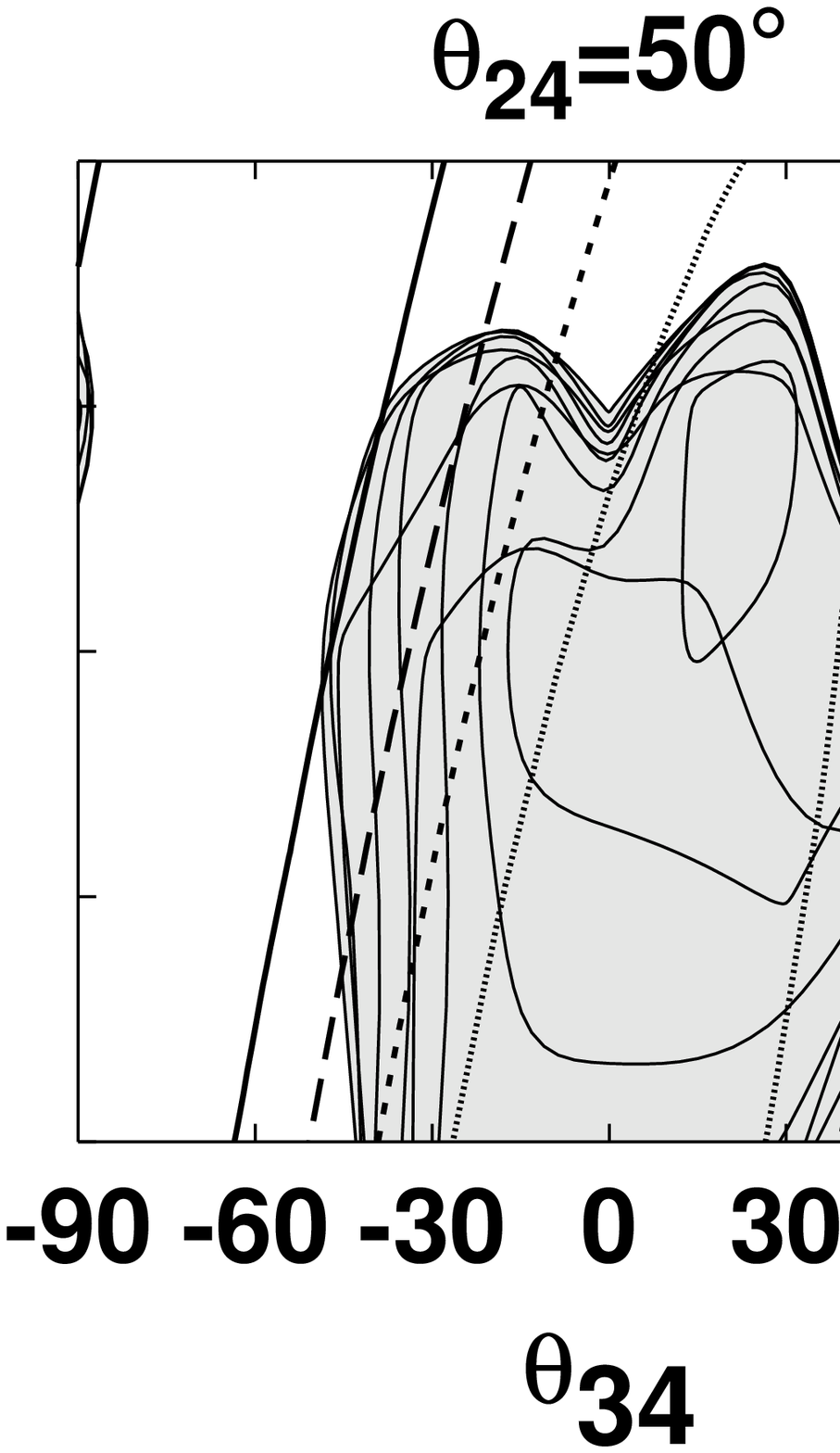,width=10cm}
\vglue -10cm \hglue 1cm \epsfig{file=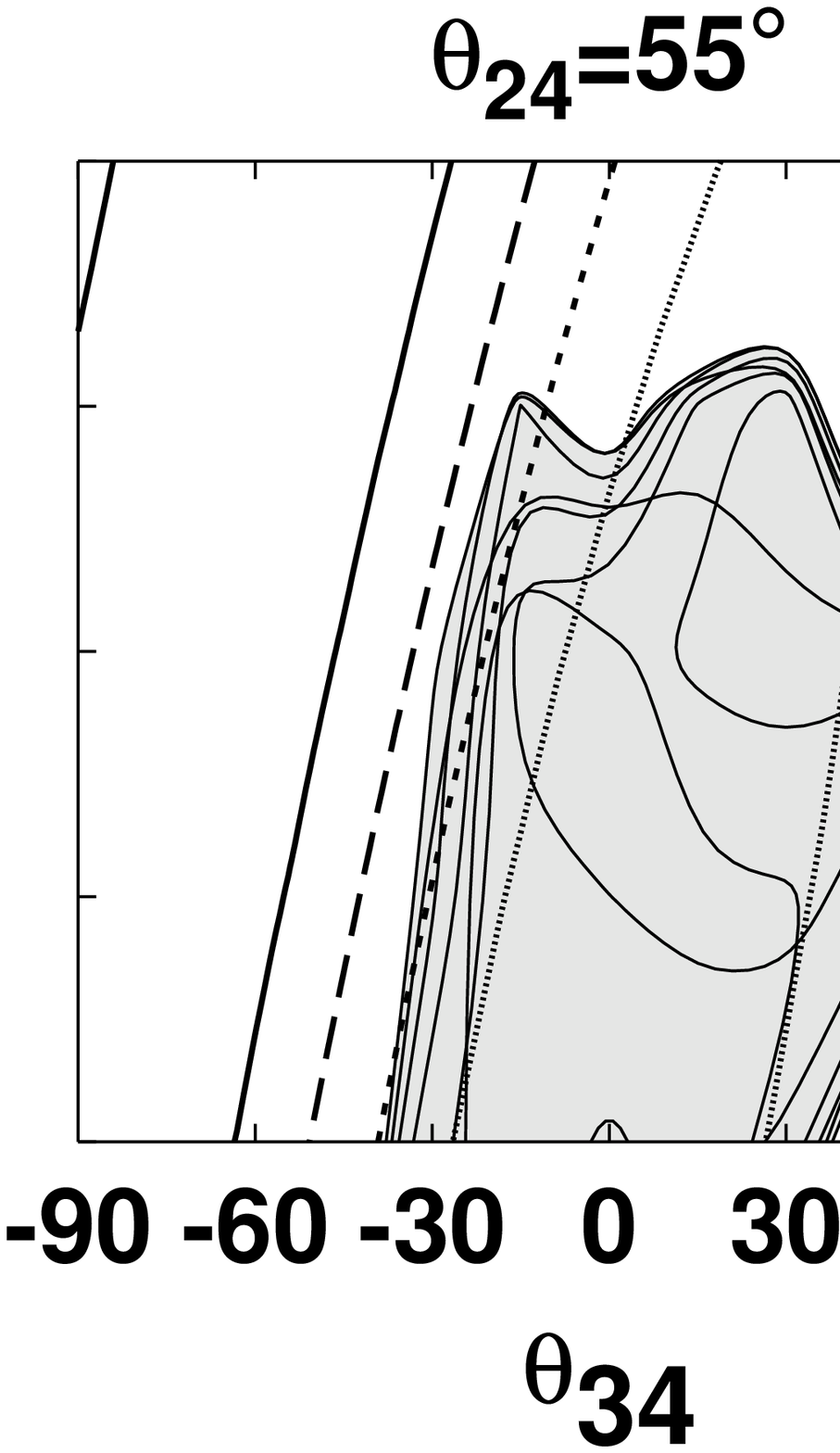,width=10cm}
\vglue -2.0cm \hglue -1.5cm \epsfig{file=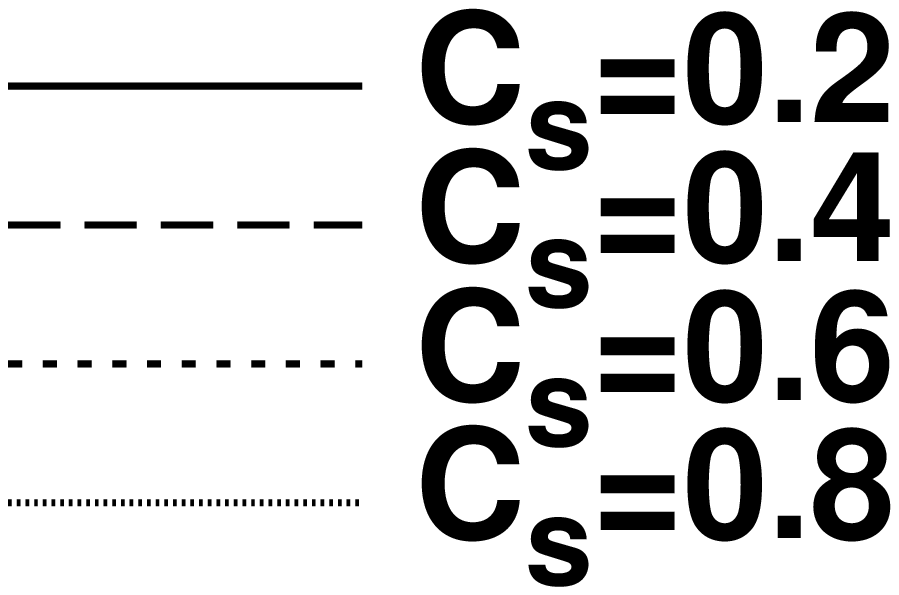,width=10cm}

\vglue -7.cm
\hglue 6.0cm \epsfig{file=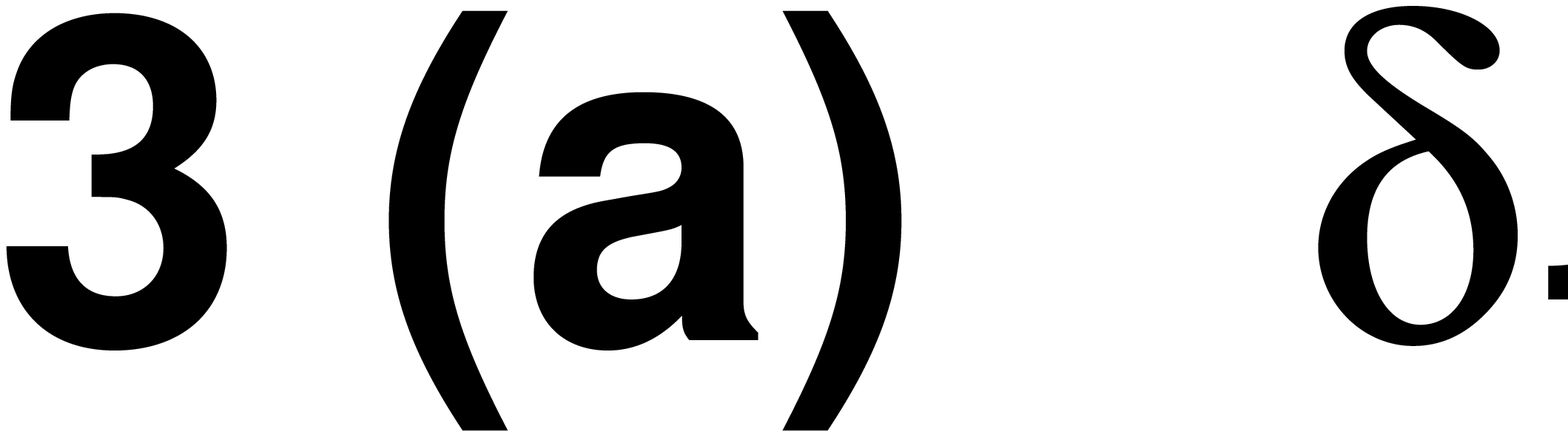,width=3cm}
\newpage
\pagestyle{empty}
\vglue -2.5cm
\hglue -6cm 
\epsfig{file=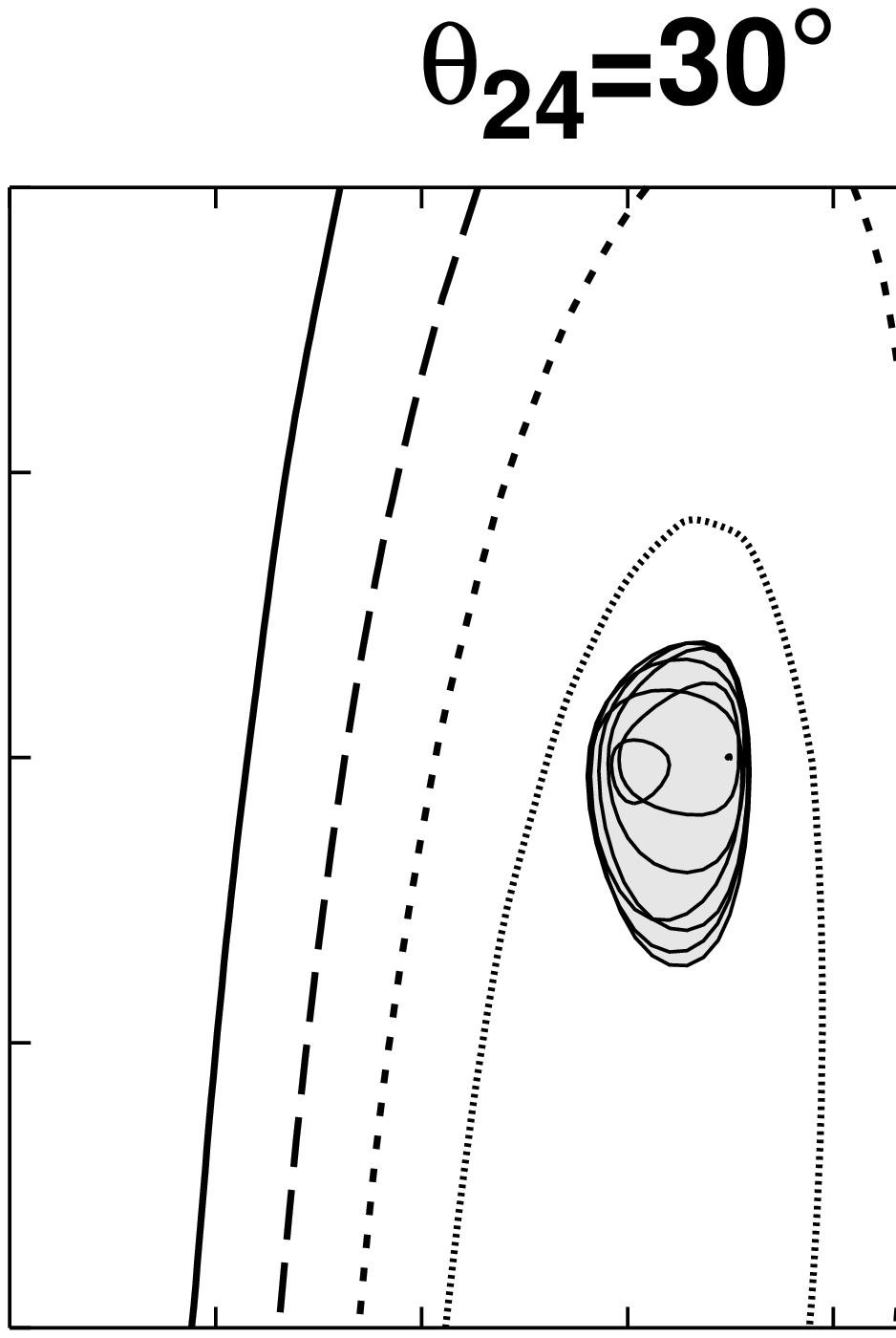,width=10cm}
\vglue -10.1cm \hglue 1cm \epsfig{file=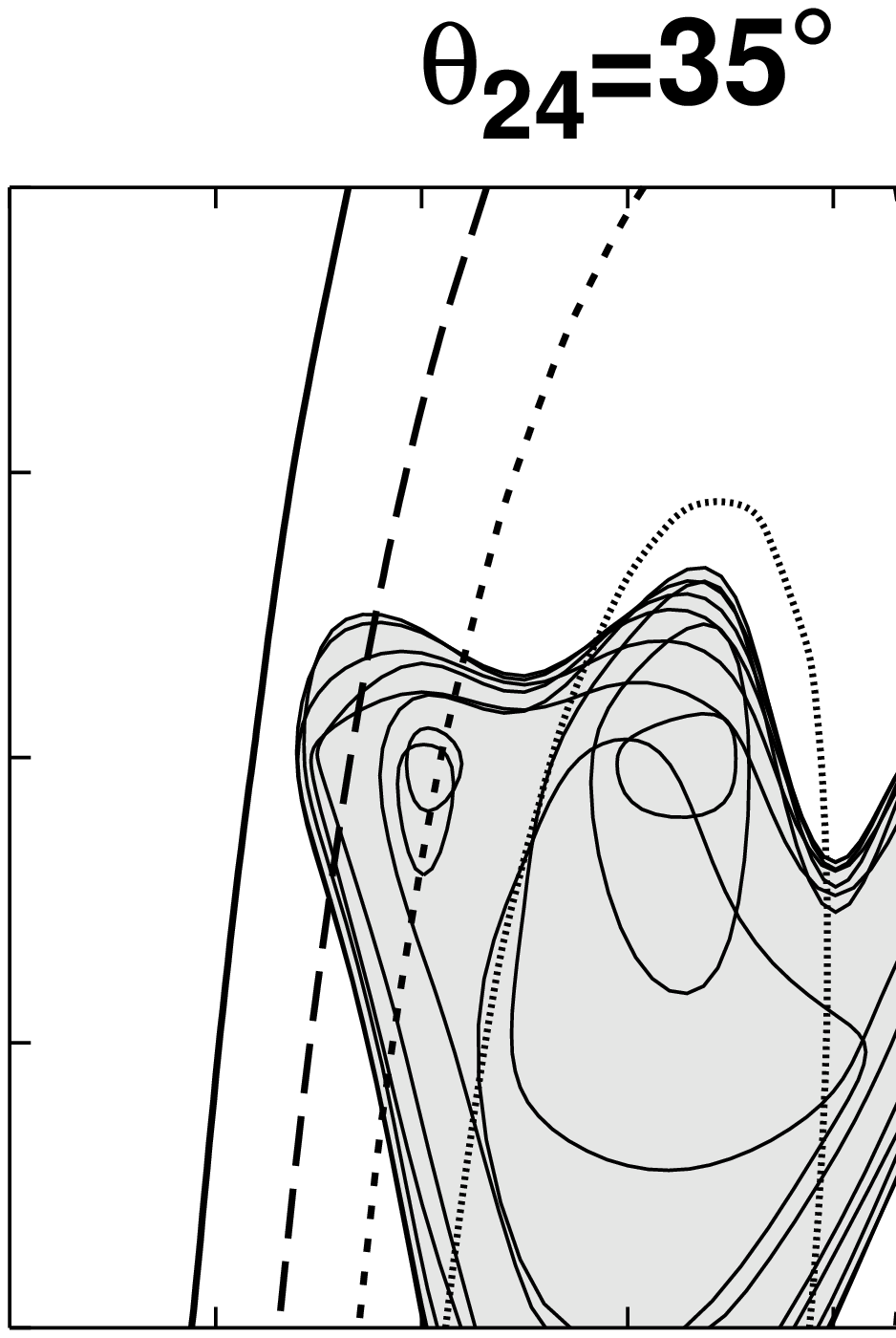,width=10cm}

\vglue -3.8cm
\hglue -6cm 
\epsfig{file=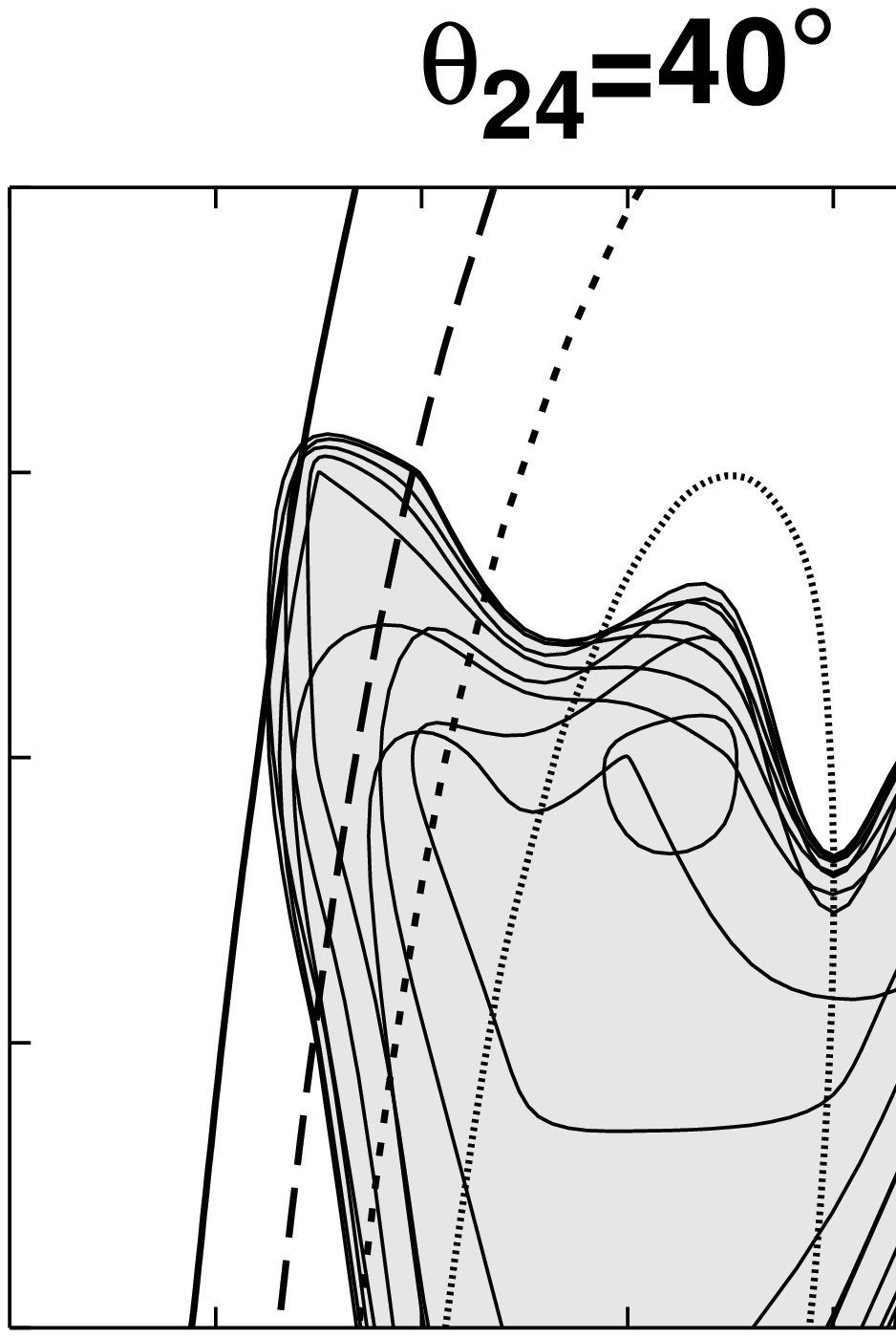,width=10cm}
\vglue -10cm \hglue 1cm \epsfig{file=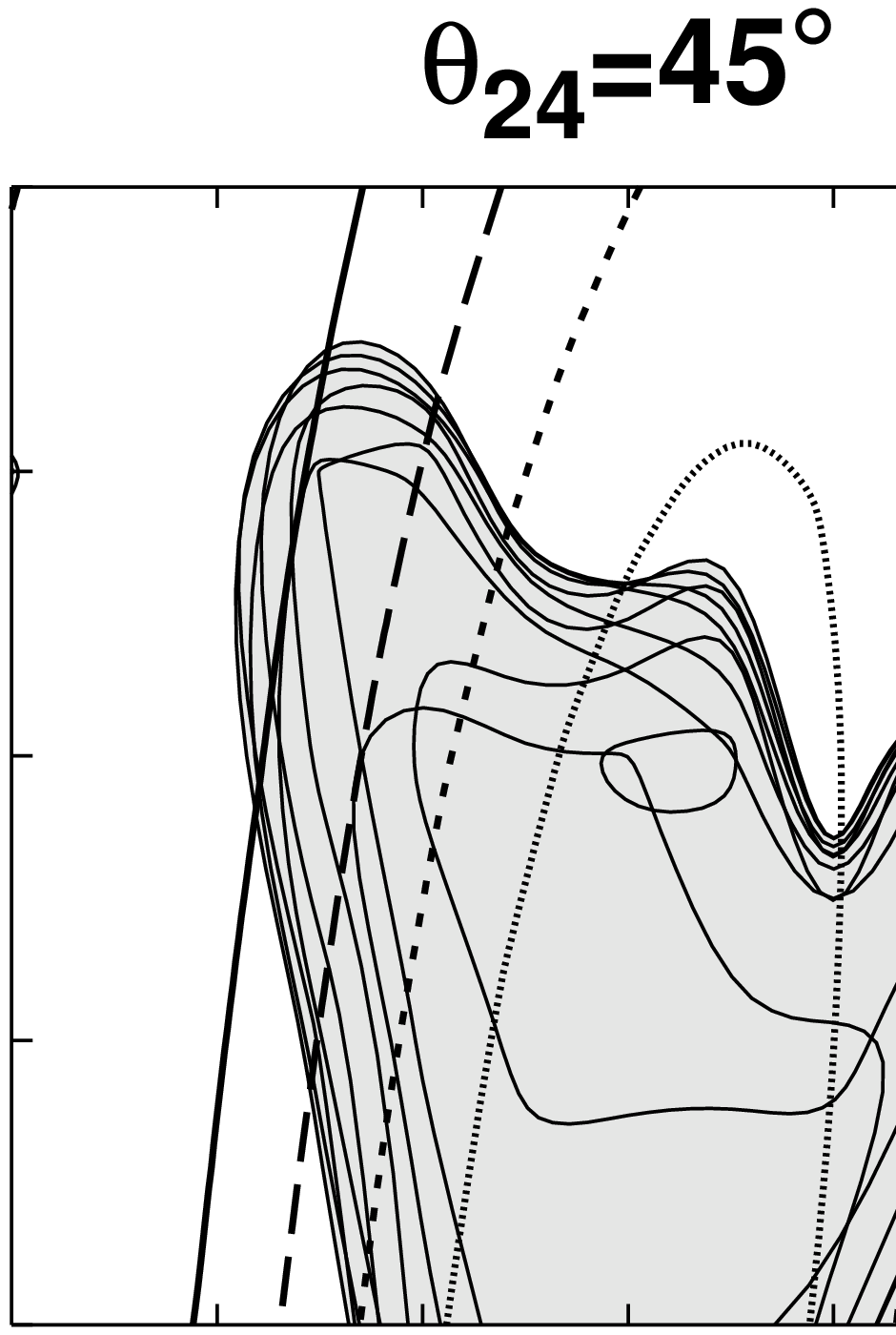,width=10cm}

\vglue -3.8cm
\hglue -6cm 
\epsfig{file=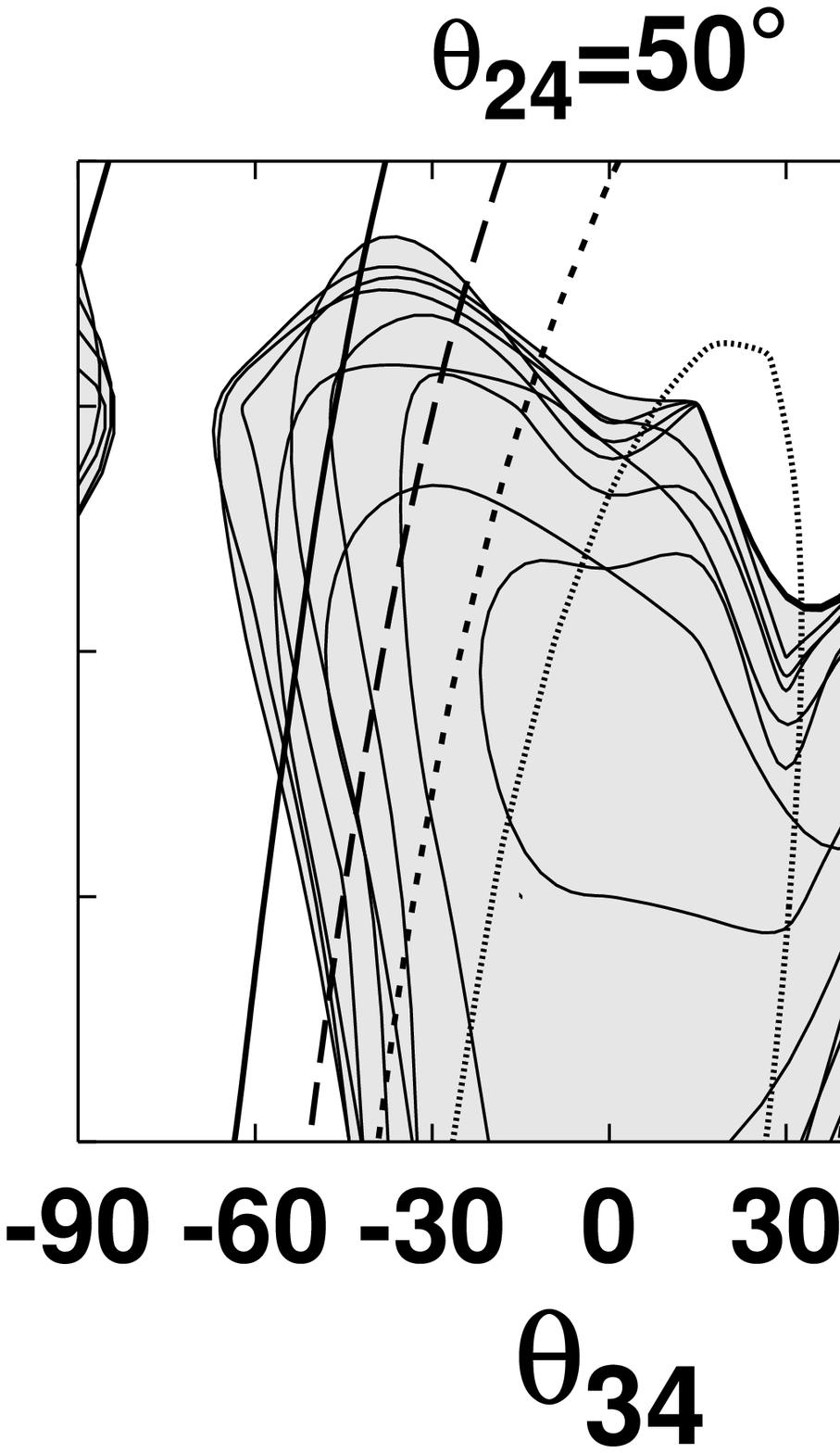,width=10cm}
\vglue -10cm \hglue 1cm \epsfig{file=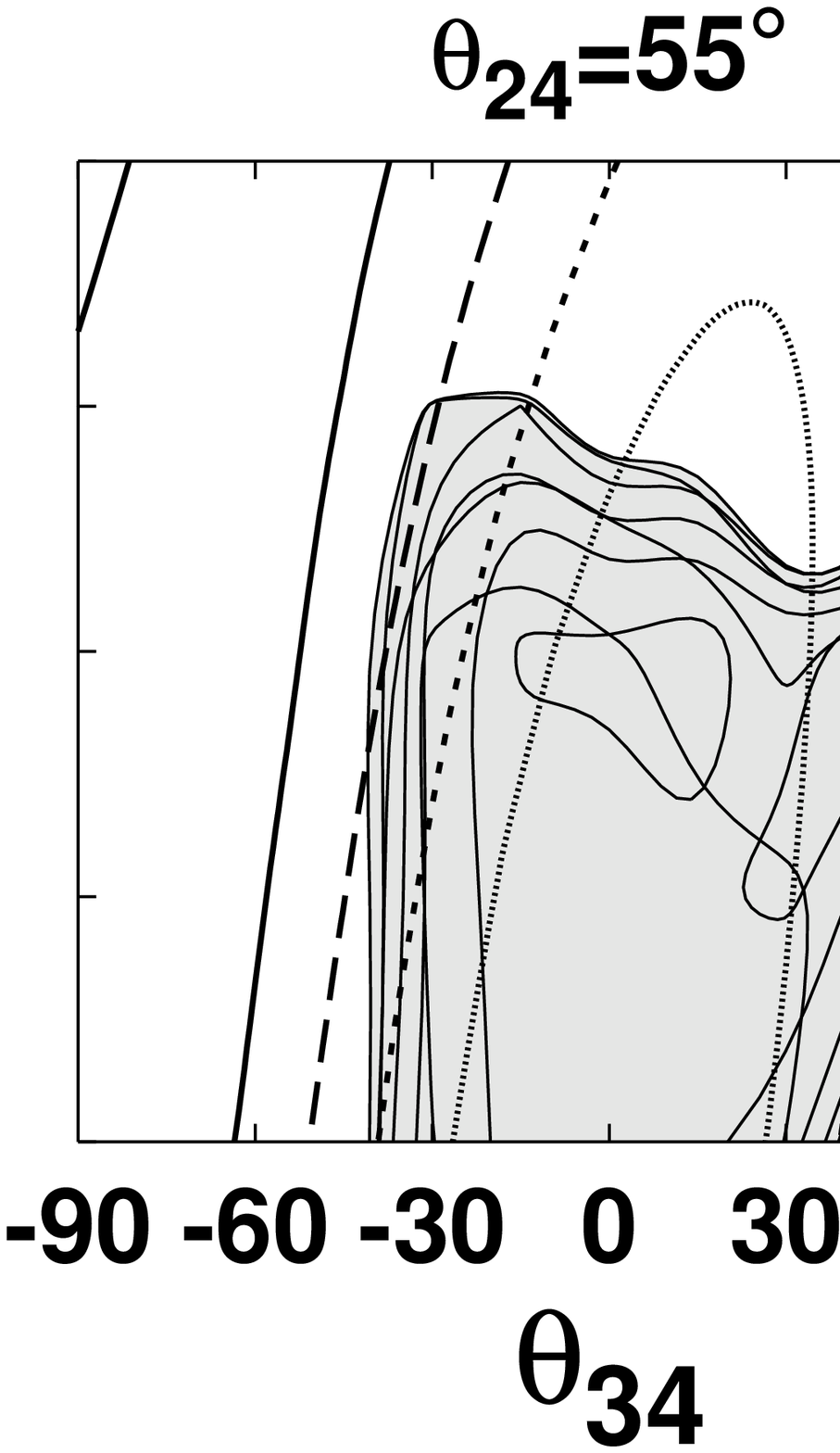,width=10cm}
\vglue -2.0cm \hglue -1.5cm \epsfig{file=cs.eps,width=10cm}

\vglue -7.cm
\hglue 6.0cm \epsfig{file=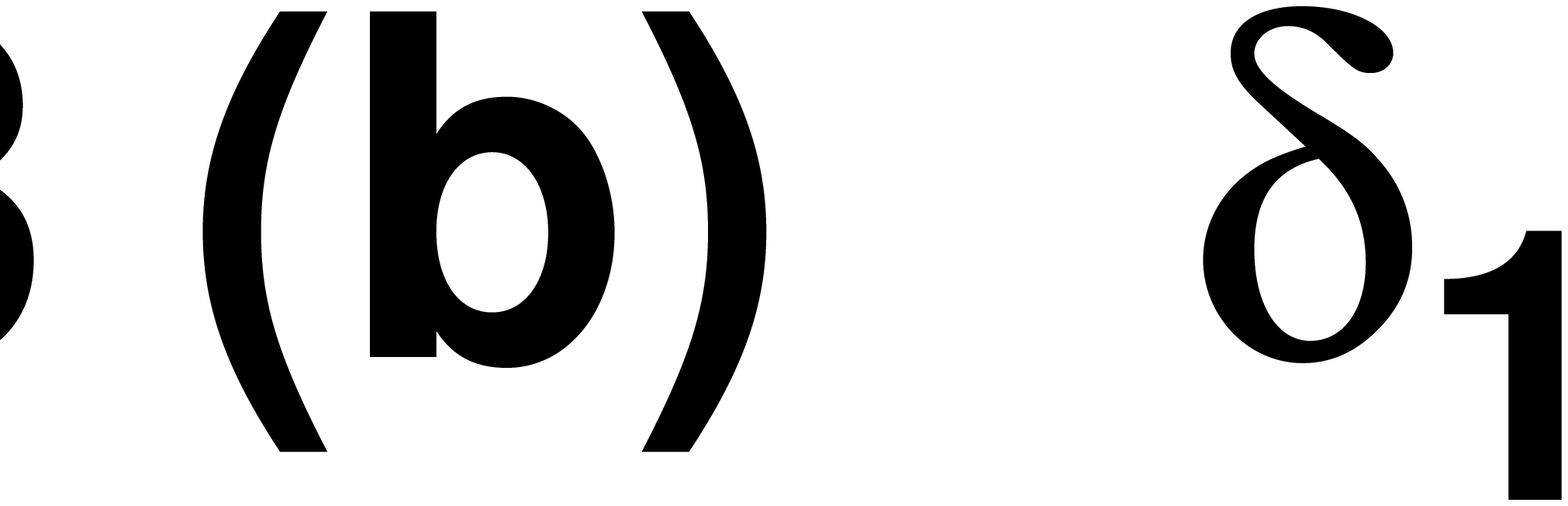,width=3cm}
\newpage
\pagestyle{empty}
\vglue -2.5cm
\hglue -6cm 
\epsfig{file=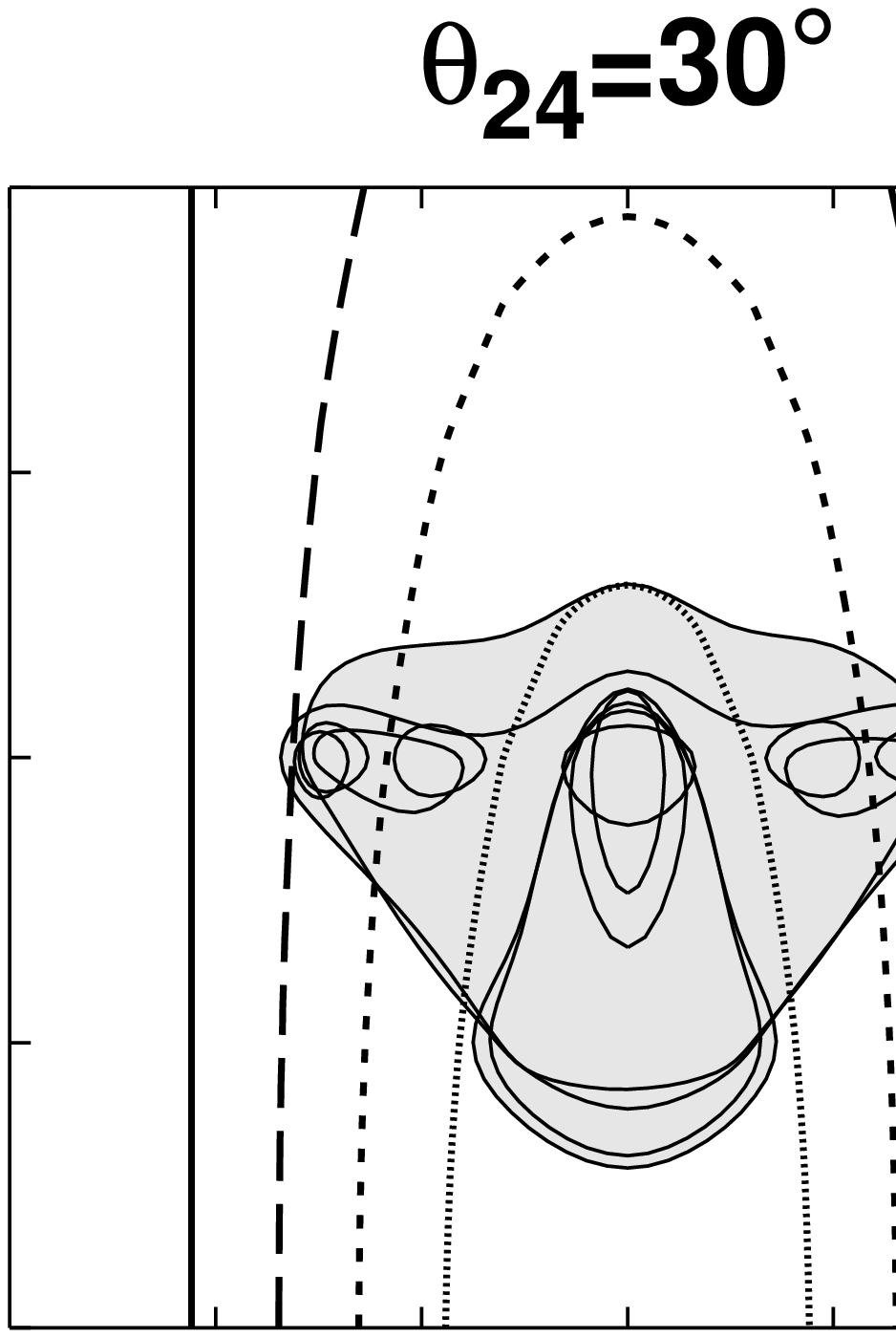,width=10cm}
\vglue -10.1cm \hglue 1cm \epsfig{file=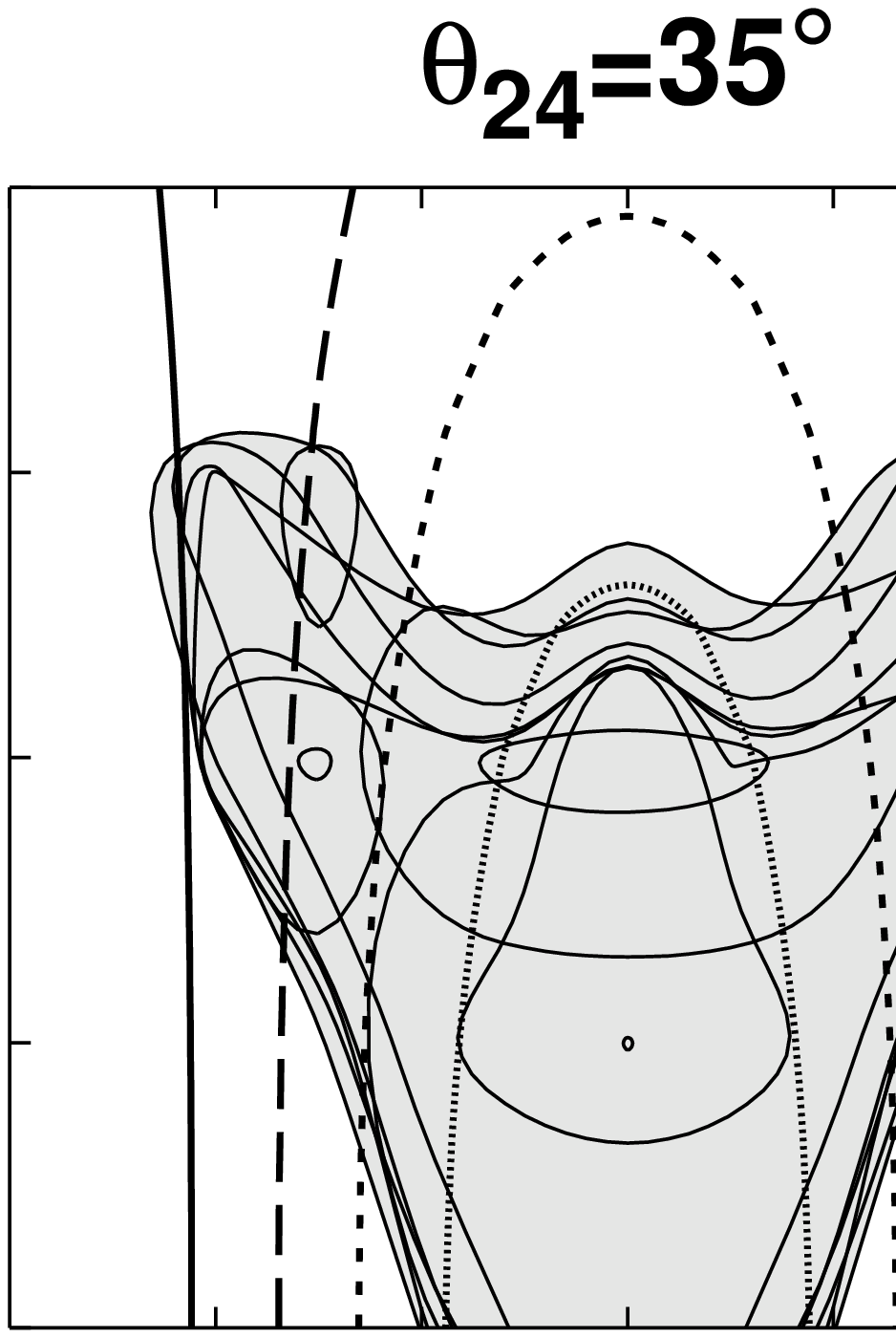,width=10cm}

\vglue -3.8cm
\hglue -6cm 
\epsfig{file=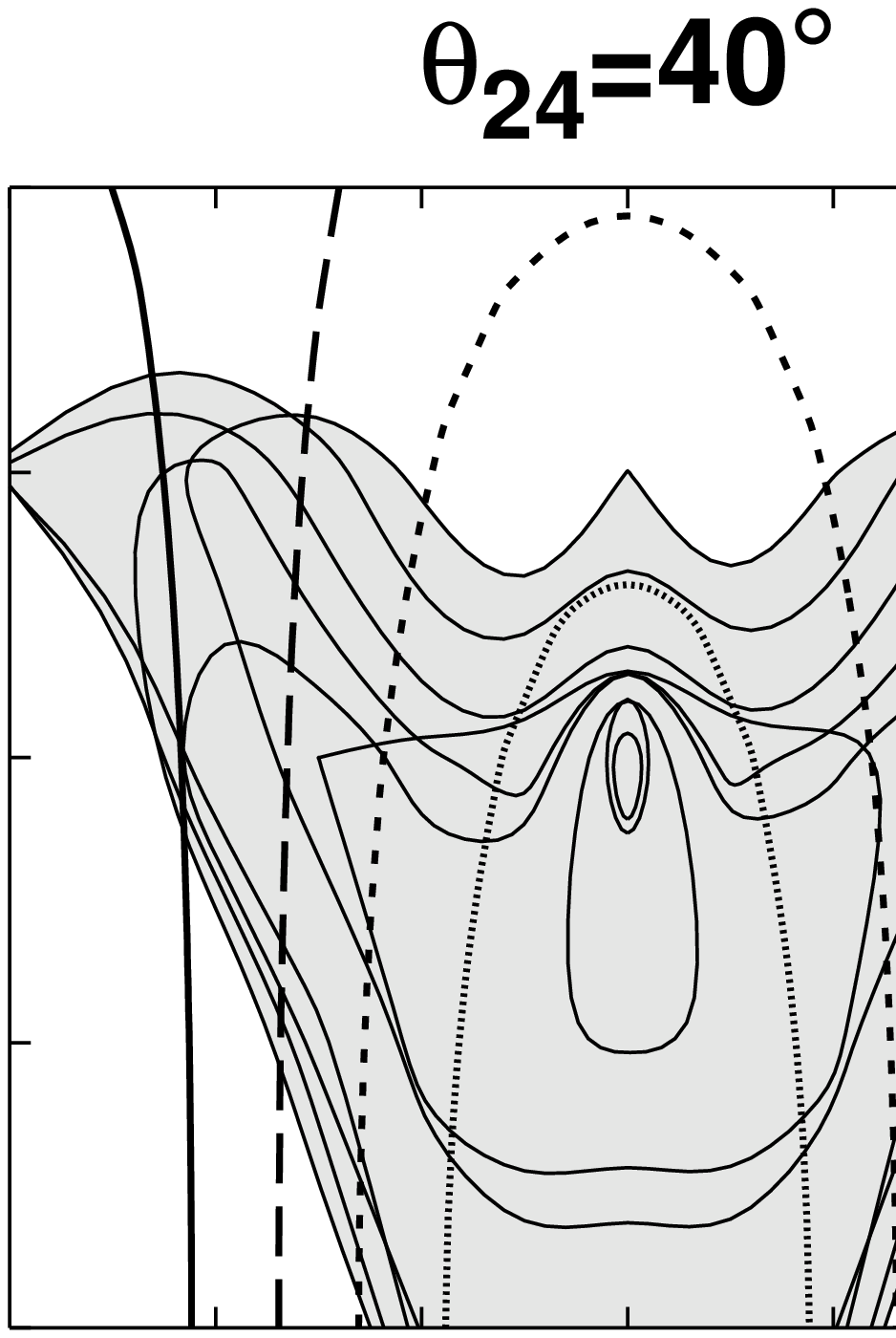,width=10cm}
\vglue -10cm \hglue 1cm \epsfig{file=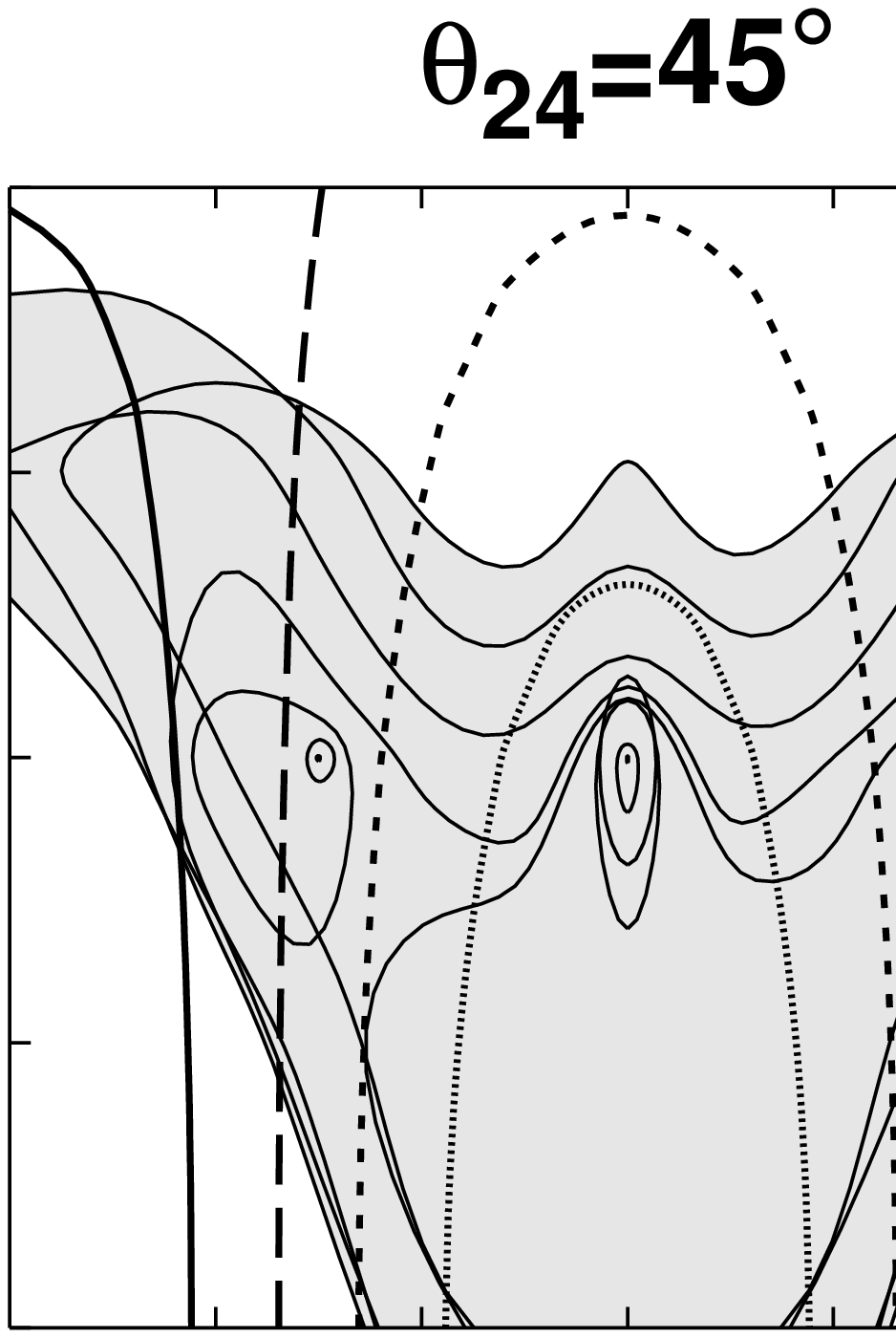,width=10cm}

\vglue -3.8cm
\hglue -6cm 
\epsfig{file=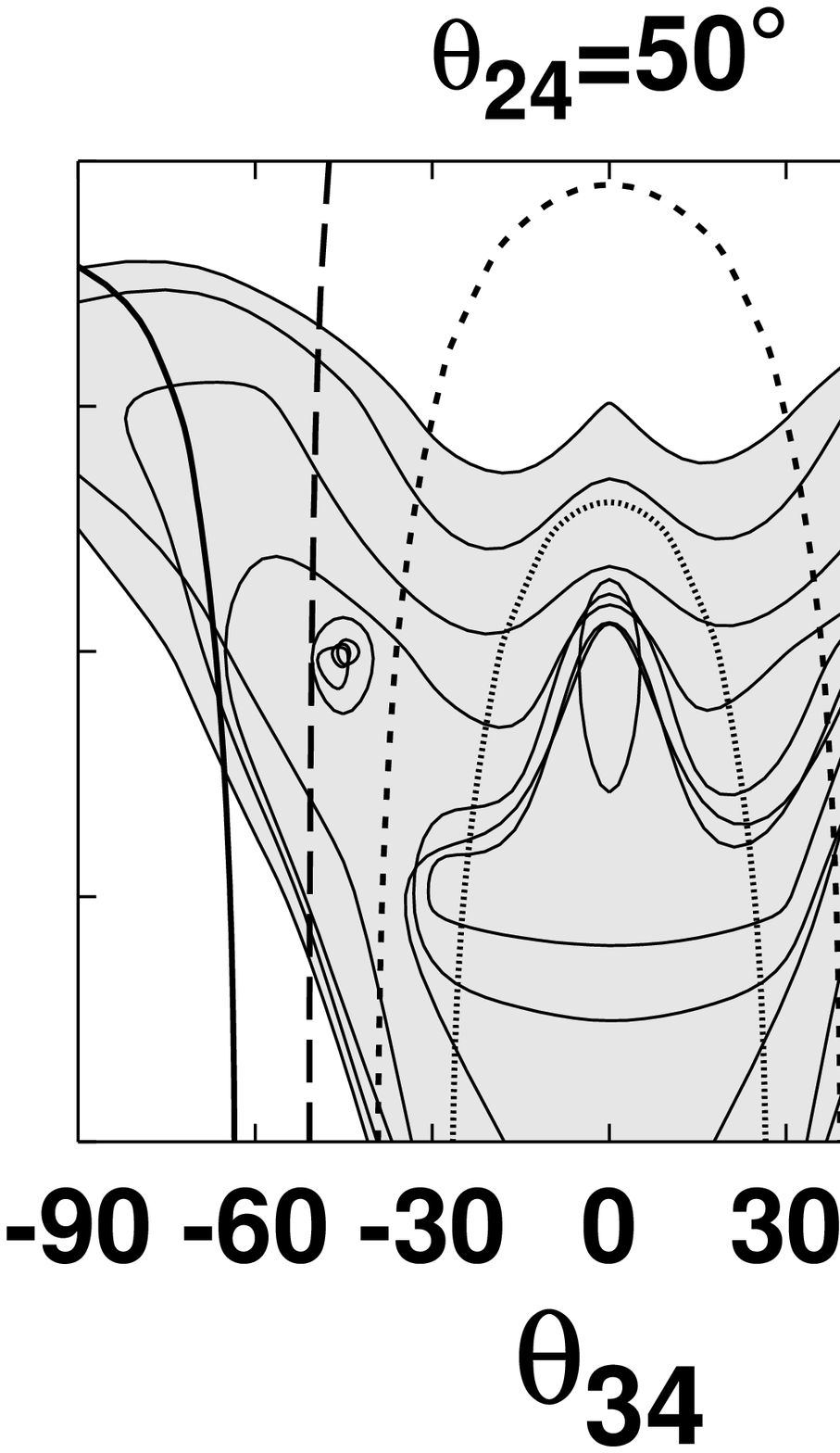,width=10cm}
\vglue -10cm \hglue 1cm \epsfig{file=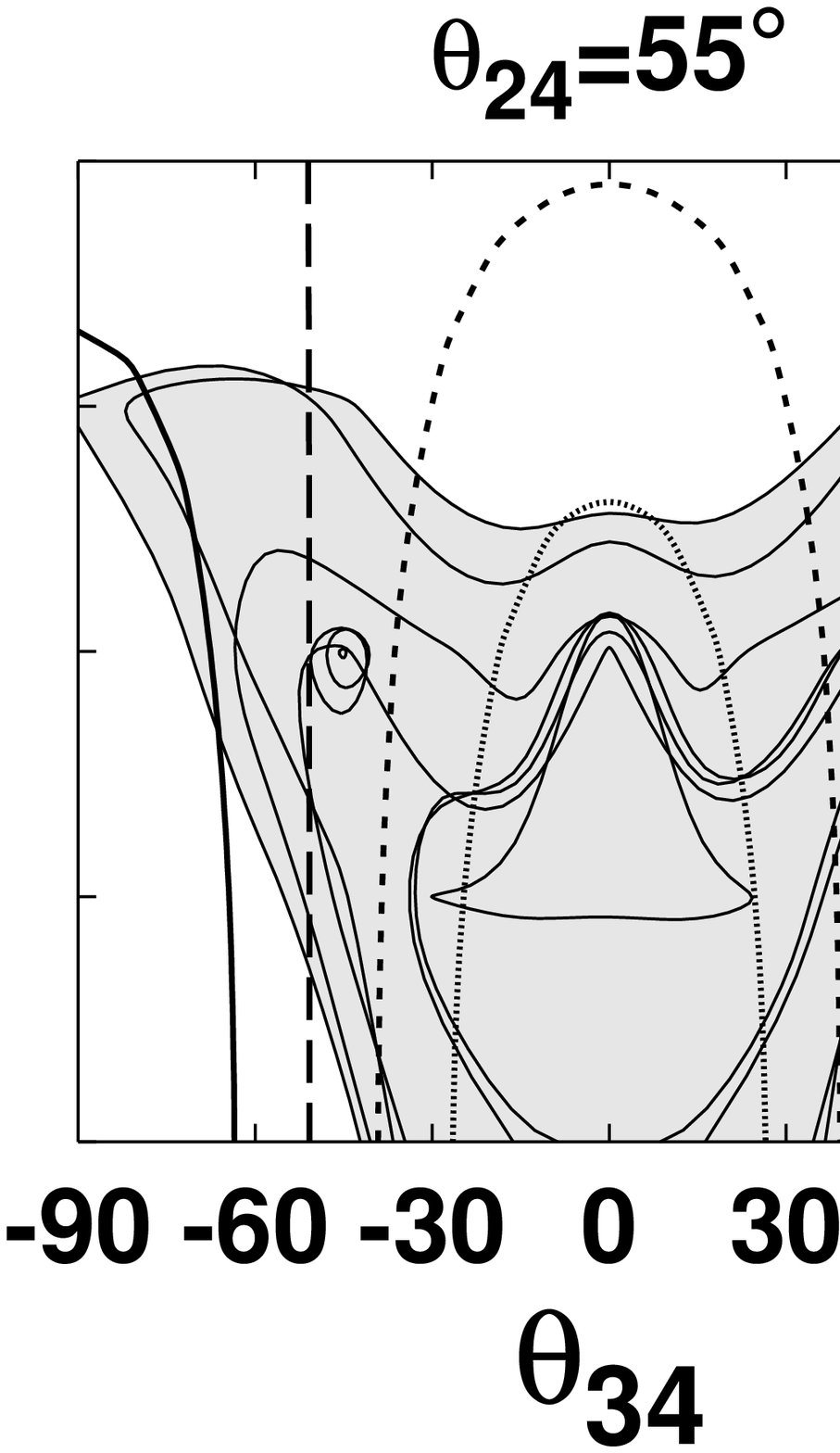,width=10cm}
\vglue -2.0cm \hglue -1.5cm \epsfig{file=cs.eps,width=10cm}

\vglue -7.cm
\hglue 6.0cm \epsfig{file=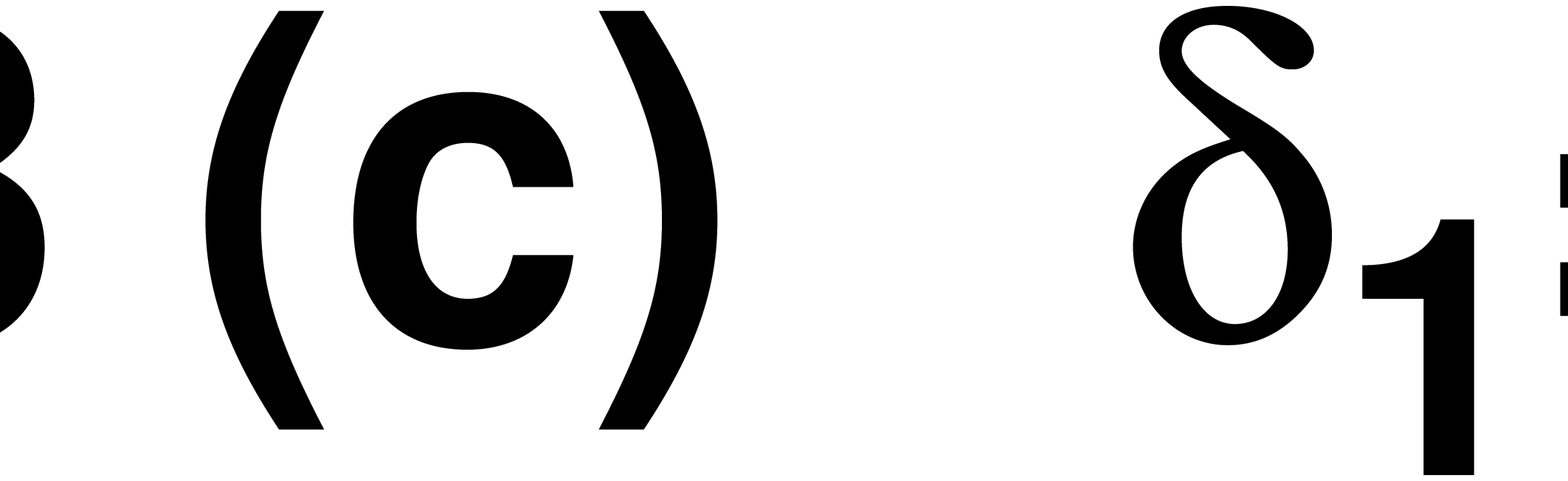,width=3cm}
\end{document}